%% file: main-arXiv.tex
\documentclass[11pt]{article}
\usepackage{fullpage}

\usepackage{calc,amsfonts,amssymb,amsmath,bm,url,color,theorem,graphicx,cite}
\usepackage{enumerate}
\usepackage{psfrag,float}
\usepackage{algorithm}
\usepackage{array}
\usepackage{algorithmic}
\usepackage{subcaption,epstopdf}
\usepackage{xr}
\usepackage{soul}
\usepackage{slashbox}

\usepackage{shortcuts_OPT}

\newcolumntype{M}[1]{>{\centering\arraybackslash}m{#1}}

\def\BibTeX{{\rm B\kern-.05em{\sc i\kern-.025em b}\kern-.08em
    T\kern-.1667em\lower.7ex\hbox{E}\kern-.125emX}}
\AtBeginDocument{\definecolor{ojcolor}{cmyk}{0.93,0.59,0.15,0.02}}

\definecolor{orange}{RGB}{255,107,0}
\def\blue{\color{blue}}

\def\revcolor{\color{black}}

\newtheorem{Fact}{Fact}

\newtheorem{Prop}{Proposition}

\newtheorem{Assumption}{Assumption}

\theorembodyfont{\rmfamily}
\newtheorem{Exa}{Example}

\input{sym_marco}

\begin{document}
\title{\papertitle}
    
    \author{{\blue Wai-Yiu Keung} and Wing-Kin Ma \\ ~ \\
    Department of Electronic Engineering, The Chinese University of Hong Kong, \\
    Hong Kong SAR of China
    }

\title{
Spatial Sigma-Delta Modulation for 
Coarsely Quantized
Massive MIMO Downlink: Flexible Designs by Convex Optimization
}

\author{
Wai-Yiu Keung${}^{\dagger, \mathsection}$ and Wing-Kin Ma${}^\dagger$
\\ ~ \\
    ${}^\dagger$Department of Electronic Engineering\\ 
    ${}^\mathsection$Department of Computer Science and Engineering \\ ~ \\  
    The Chinese University of Hong Kong, 
    Hong Kong SAR of China\\ ~ \\
    E-mails: wykeung@cse.cuhk.edu.hk, wkma@ee.cuhk.edu.hk, 
}

\maketitle

\begin{abstract}
 This paper considers the context of multiuser massive MIMO downlink precoding with low-resolution digital-to-analog converters (DACs) at the transmitter.
This subject is motivated by the consideration that it is expensive to employ high-resolution DACs for practical massive MIMO implementations.
The challenge with using low-resolution DACs is to overcome the detrimental quantization error effects.
Recently, spatial Sigma-Delta ($\Sigma\Delta$) modulation has arisen as a viable way to put quantization errors under control.
This approach takes insight from temporal $\Sigma\Delta$ modulation in classical DAC studies.
Assuming a $1$D uniform linear transmit antenna array,
the principle is to shape the quantization errors in space such that the shaped quantization errors are pushed away from the user-serving angle sector.
In the previous studies, spatial $\Sigma\Delta$ modulation was performed by direct application of the basic first- and second-order modulators from the $\Sigma\Delta$ literature.
In this paper, we develop a general $\Sigma\Delta$ modulator design framework for any given order, for any given number of quantization levels, and for any given angle sector.
We formulate our design as a problem of maximizing the signal-to-quantization-and-noise ratios (SQNRs) experienced by the users. 
The formulated problem is convex and can be efficiently solved by available solvers.
Our proposed framework offers the alternative option of focused quantization error suppression in accordance with channel state information. 
Our framework can also be extended to $2$D planar transmit antenna arrays.
We perform numerical study under different operating conditions, and the numerical results suggest that, 
given a moderate number of quantization levels, say, $5$ to $7$ levels, our optimization-based $\Sigma\Delta$ modulation schemes can lead to bit error rate performance close to that of the unquantized counterpart.   
\end{abstract}


\maketitle

\input{wkIntro}
\section{Background}
\label{sect:bg}

This section intends to provide the background of this study. We review the basics of $\Sigma\Delta$ modulation in the first subsection, give the problem statement of  coarsely quantized MIMO precoding in the second subsection, and describe the spatial $\Sigma\Delta$ modulation approach for the precoding problem in the third subsection.

\subsection{$\Sigma\Delta$ Modulation}
\label{sect:sdmod_intro}

\begin{figure}
	\centering
	\includegraphics[width=0.7\linewidth]{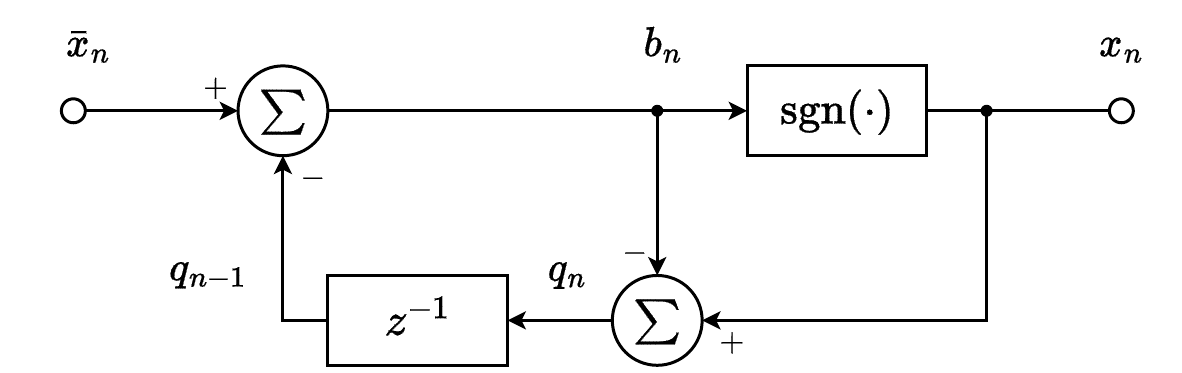}
	\caption{The one-bit first-order $\Sigma\Delta$ modulator.}
	\label{fig:1o_sd}
\end{figure}

{We introduce the basics of $\Sigma\Delta$ modulation by considering the one-bit first-order modulator, the most basic {scheme in} $\Sigma\Delta$ modulation.
The system architecture of the modulator is depicted in Fig.~\ref{fig:1o_sd}.
Let $\{ \bar{x}_n \}_{n \in \Nbb} \subset \Rbb$ be a real-valued time sequence.
Let $\sgn: \Rbb \rightarrow \{ \pm 1 \}$ be the signum function. 
The modulator takes $\{ \bar{x}_n \}_{n \in \Nbb}$ as the input and generates a binary output $\{ {x}_n \}_{n \in \Nbb} \subset \{ \pm 1 \}$ by
\begin{equation} \label{eq:io_relation}
	x_n = \sgn( \bar{x}_n - q_{n-1} )  = \bar{x}_n - q_{n-1} + q_n, ~ n \in \Nbb,
\end{equation}
where
$q_n$ is the quantization error associated with $\bar{x}_n - q_{n-1}$, for $n \in \Nbb$; 
and we have $q_{-1} = 0$.
The rationale of this process should be described.
The input $\{ \bar{x}_n \}_{n \in \Nbb}$ is a lowpass temporal signal.
We want to coarsely quantize the input in such a way that the error signal at the output is weak in the low frequency band. 
We make the following assumption which is used in nearly every $\Sigma\Delta$ literature.
\begin{Assumption} \label{assume:quant}
	Consider the modulator in Fig.~\ref{fig:1o_sd} or the system in \eqref{eq:io_relation}.
	Each quantization error $q_n$ is $[-1,1]$-supported, uniformly distributed on its support, and independent of any other random variables.	
\end{Assumption}
Let $v_n = q_n - q_{n-1}$ be the error at the output $x_n$.
The magnitude spectrum of $\{ v_n \}_{n \in \Nbb}$ equals
\[
| V(\omega) |^2 = | Q(\omega) - e^{-\jj \omega} Q(\omega) |^2 = | 1 - e^{-\jj \omega} |^2 |Q(\omega)|^2,
\]
where $| 1 - e^{-\jj \omega} |^2 = 4 | \sin(\omega/2) |^2$ is a highpass response.
Also, under Assumption~\ref{assume:quant} we can see $ |Q(\omega)|^2$ as a flat spectrum; more precisely, the power spectral density of $\{ q_n \}_{n \in \Nbb}$ is flat.
Hence, the modulator can be viewed as a quantizer that has the ability to shape the quantization error signal as highpass noise,
and by doing so we reduce the undesirable interference effects of the quantization errors on the lowpass input signal over the low frequency band.}

{In Assumption~\ref{assume:quant} we assume that every quantization error $q_n$ is bounded, lying in 
$[-1,1]$.
We want to discuss how this can be guaranteed.}
It can be easily shown that if the pre-quantized signal $b_n = \bar{x}_n - q_{n-1}$ has amplitude greater than $2$, then the associated quantization error $q_n$ will have $|q_n| > 1$---such phenomena are called overloading in the literature.
Overloading can lead to large $q_n$ in terms of the amplitude,
and mathematically one can show that there exists an input $\{ \bar{x}_n \}_{n \in \Nbb}$ such that $|q_n| \rightarrow \infty$ as $n \rightarrow \infty$~\cite{shao2019one}.
Overloading can be prevented by restricting the input to be amplitude limited:
\begin{Fact} \label{fact:no_overload1}
Consider the modulator in Fig.~\ref{fig:1o_sd} or the system in \eqref{eq:io_relation}. 
Let $A > 0$ be the maximum input amplitude, i.e., $|\bar{x}_n | \leq A$ for all $n$.
If $A \leq 1$, then $|q_n| \leq 1$ for all $n$.
\end{Fact}
The proof of Fact~\ref{fact:no_overload1} is simple:
Suppose $|q_{n-1}| \leq 1$. 
Then $|\bar{x}_n - q_{n-1}| \leq A + 1 \leq 2$, and we have $|q_n| \leq 1$.
The proof is complete.

The one-bit first-order $\Sigma\Delta$ {modulation scheme} introduced above is basic.
There are many other $\Sigma\Delta$ modulation {schemes}, as well as a variety of aspects related to the modulator designs.
We refer the reader to the literature 
(e.g., \cite{gray1990quantization,schreier2005understanding})
for details such as the multi-level and higher-order generalizations of the above $\Sigma\Delta$ modulator;
the various $\Sigma\Delta$ modulator architectures;
reasonability of the independent and identical distributed (i.i.d.) assumption in Assumption~\ref{assume:quant} in practice, and
the practical trick of dithering to try to make the quantization error more i.i.d.;
and the impact of overloading in practice.
{\revcolor We also refer the reader to the mathematical studies in \cite{daubechies2003approximating,gunturk2003one}, which analyze the reconstruction accuracy of temporal $\Sigma\Delta$ modulation schemes without Assumption~\ref{assume:quant}.}

{\revcolor 
Before we finish our review,
we should note a basic implementation aspect.
For DACs with $\Sigma\Delta$ modulation (the case of interest here), the modulators appear in digital domain and can be flexibly implemented by digital signal processing.
For ADCs (outside the scope of this study), 
$\Sigma\Delta$ modulators are implemented in analog domain and require dedicated analog and digital hardware to build.}

\subsection{Coarsely Quantized MIMO Precoding}
\label{sect:cq_intro}

Consider the following {multiuser MIMO downlink communication problem.}
The base station (BS) serves a number of $K$ users and has $N$ transmit antennas. 
The users have a single antenna.
Assuming frequency-flat time-invariant channels over a finite time frame of transmission,
{the transmit-receive relation from the BS to the users is modeled as}
\begin{equation} \label{eq:sig_mod}
y_{i,t}  = \sqrt{\rho} \bh_i^T \bx_t + \eta_{i,t}, ~ t =1,\ldots,T,
\end{equation}
where 
{ $y_{i,t}$ is the received signal of user $i$ at symbol time $t$;
$\sqrt{\rho} \bx_t \in \Cbb^N$ is the transmitted signal at symbol time $t$, with its $n$th component $\sqrt{\rho} x_{n,t}$ being the transmitted signal at the $n$th antenna;
$\bx_t$ is the transmitted signal before power amplification;
$\rho > 0$ is a power scaling factor;
$\eta_{i,t}$ is i.i.d. circular complex Gaussian noise with mean zero and variance $\sigma_\eta^2$;
$T$ is the transmission block length.}
Assume that the BS is informed of $\bh_1,\ldots,\bh_K$.
The problem, called precoding, 
is to design the transmitted signals $\{ \bx_t \}_{t=1}^T$ such that each user will { receive} its own data symbol stream with minimal distortions. 
Specifically, we want the noise-free part of $y_{i,t}$ to take the form
\begin{equation} \label{eq:sym_shape}
\bh_i^T \bx_t \approx c_i s_{i,t},
\end{equation}
where $\{ s_{i,t} \}_{t=1}^T$ is the symbol stream for user $i$; $c_i$ represents the signal gain.
{ For example, the zero-forcing (ZF) scheme performs precoding by}
\[
\bx_t = \bH^\dag \bs_t, ~ t=1,\ldots,T,
\]
where $\bH = [~ \bh_1,\ldots,\bh_K ~]^T$; $\bs_t = (s_{1,t},\ldots,s_{K,t})$.
It is easy to see that the ZF scheme leads to $y_{i,t} = \sqrt{\rho} s_{i,t} + \eta_{i,t}$.

{ In precoding,} it is common to assume that the transmitted signals $x_{n,t}$'s are continuous valued.
The problem of interest in this paper is coarsely quantized precoding,
wherein the $x_{n,t}$'s are { discrete} valued.
For example, for the one-bit case, 
the real and imaginary components of every $x_{n,t}$ are binary.
{ The motivation, as discussed in the Introduction, is to reduce massive MIMO hardware costs and power consumption by replacing high-resolution DACs with low-resolution ones.}
{ A straightforward solution to coarsely quantized precoding is to directly quantize the precoded signals.
For example, we can directly quantize the ZF scheme by $\bx_t = \setQ_c(\bH^\dag \bs_t)$, where 
{ $\setQ_c$ denotes the quantization function associated with the low-resolution DACs.}
But such a precode-then-quantized scheme can significantly suffer from quantization error effects.
Some recent studies seek a different approach, namely,
by directly optimizing the discrete variables $x_{n,t}$'s to shape symbols at the user side (cf. \eqref{eq:sym_shape}) {\cite{castaneda-jacobsson-durisi-2017, jacobsson-durisi-coldrey-2017, Sohrabi2018,li-masouros-liu-2018, jedda-mezghani-swindlehurst-2018}.}
This direct design approach was found to be able to provide promising performance by numerical experiments.
It however requires us to solve a large-scale discrete optimization problem.
Also, its optimization-oriented design principle largely disallows us from reusing precoding {concepts} in the unquantized case, such as the simple ZF scheme. 
}

\subsection{Spatial $\Sigma\Delta$ Modulation}
\label{sect:spatial_sd}

We recently proposed a spatial $\Sigma\Delta$ modulation approach for the { above stated} coarsely quantized precoding problem \cite{shao2019one}.
It falls into the precode-then-quantized scope, and
the spirit is to use $\Sigma\Delta$ modulation to shape the noise spectrum---spatially---such that users are less affected by { the} quantization error { effects}.
{ The spatial $\Sigma\Delta$ modulation approach is described as follows.}
We assume angular channels
\[
\bh_i = \alpha_i \, \ba(\theta_i),
\]
where $\alpha_i \in \Cbb$ is { a complex channel} gain; $\theta_i \in (-\pi/2,\pi/2)$ is the user angle;
\[
\ba(\theta) = ( 1, e^{-\jj \frac{2\pi d}{\lambda} \sin(\theta)}, \ldots,  e^{-\jj (N-1) \frac{2\pi d}{\lambda} \sin(\theta)})
\]
is the angular response, with $\lambda$ being the carrier wavelength, $d \leq \lambda/2$ being the inter-antenna spacing, and $\theta \in (-\pi/2,\pi/2)$ being the angle.
{ The angular channels are based on the operating assumptions that the transmit antennas are arranged as a uniform linear array, and we consider a single-path far-field channel from the BS to each user;
the reader is referred to the literature (e.g., \cite{tse2005fundamentals}) for details.}
Consider, at each symbol time $t$, that we apply the $\Sigma\Delta$ modulator in Section~\ref{sect:sdmod_intro} to the transmitted signals.
To be careful, let $\bar{\bx}_t$ and $\bx_t$ be the transmitted signals before and after $\Sigma\Delta$ modulation, respectively.
Also, as a slight abuse of notations, let $\bar{x}_{n,t}$ and $x_{n,t}$ denote the $(n+1)$th elements of $\bar{\bx}_t$ and $\bx_t$, respectively.
We apply the one-bit first-order $\Sigma\Delta$ modulator in Section~\ref{sect:sdmod_intro} to $\{ \Re(\bar{x}_{n,t}) \}_{n=0}^{N-1}$ to obtain $\{ \Re({x}_{n,t}) \}_{n=0}^{N-1}$,
and we apply another one-bit first-order $\Sigma\Delta$ modulator to $\{ \Im(\bar{x}_{n,t}) \}_{n=0}^{N-1}$ to obtain $\{ \Im({x}_{n,t}) \}_{n=0}^{N-1}$.
The appealing result goes as follows: for any $\theta \in (-\pi/2,\pi/2)$,
\begin{align}
\ba(\theta)^T \bx_t &  = \ba(\theta)^T \bar{\bx}_t + \sum_{n=0}^{N-1} ( q_{n,t} - q_{n-1,t} ) e^{-\jj n \omega} \nonumber \\
& \simeq  \ba(\theta)^T \bar{\bx}_t + (1 - e^{\jj \omega}) Q_t(\omega),
\label{eq:aTx}
\end{align}
where $\omega = \frac{2\pi d}{\lambda} \sin(\theta)$; 
{ $\{ q_{n,t} \}_{n=0}^{N-1}  \subset \Cbb$ is the quantization error sequence;} 
$Q_t(\omega)$ is the Fourier transform of $\{ q_{n,t} \}_{n=0}^{N-1}$.
In the second equation in \eqref{eq:aTx}, we assume that $N$ is large, and we will continue to assume this without explicit mentioning.
We observe from \eqref{eq:aTx} that the quantization error term is a highpass response---its magnitude is expected to be smaller if the frequency $\omega$, or its respective angle $\theta$, is closer to $0$.

The above observation suggests the following possibility: 
Consider a sectored antenna array setting wherein we serve users within a lowpass angle sector, say, { $[-30^\circ,30^\circ]$.}
Then, the spatial $\Sigma\Delta$ modulation introduced above can lead to reduced  quantization { error} effects on the users.
Specifically, by plugging \eqref{eq:aTx} into the signal model \eqref{eq:sig_mod}, we see that the received signals can be written as
\begin{equation} \label{eq:sig_mod_2}
y_{i,t} = \sqrt{\rho} \bh_i^T \bar{\bx}_t + \sqrt{\rho} \alpha_i \underbrace{(1 - e^{\jj \omega_i}) Q_t(\omega_i)}_{:= v_{i,t}} + \eta_{i,t},
\end{equation}
where $\omega_i = \frac{2\pi d}{\lambda} \sin(\theta_i)$.
By applying Assumption~\ref{assume:quant} to the real and imaginary components of $q_{n,t}$, it can be shown that the power of the quantization noise term $v_{i,t}$ is
\[
\Exp[ | v_{i,t} |^2 ]   
{\revcolor  = | 1 - e^{\jj \omega_i} |^2 \frac{2N}{3}} = 
4 \left| \sin\left(  \frac{ \revcolor \pi d}{\lambda} \sin(\theta_i) \right)  \right|^2 \frac{2N}{3} ,
\]
which reduces with $|\theta_i|$.
Note that we can also reduce the quantization noise power by reducing the inter-antenna spacing $d$, but in practice we cannot make $d$ too small { due to mutual coupling effects;} the reader is referred to our previous work \cite{shao2019one} for further discussion.

It is also necessary to describe the precoding part of the spatial $\Sigma\Delta$ modulation approach.
The idea is nothing more than treating the second and third term on the right-hand side of the received signal model \eqref{eq:sig_mod_2} as a single noise term, and then designing $\{ \bar{\bx}_t \}_{t=1}^T$ by an existing unquantized precoding scheme.
But there is a new constraint unique to spatial $\Sigma\Delta$ modulation.
To guarantee no overloading with the $\Sigma\Delta$ modulator, it is suggested by Fact~\ref{fact:no_overload1} that we should limit the amplitude of the real and imaginary components of $\bar{\bx}_t$, specifically,
\begin{equation} \label{eq:IQ_infty}
\| \bar{\bx}_t \|_{{\sf IQ}-\infty} :=
\max\{ \| \Re(\bar{\bx}_t) \|_\infty, \| \Im(\bar{\bx}_t) \|_\infty \}
\leq 1,
\end{equation}
for all $t$.
Hence, the precoding problem in spatial $\Sigma\Delta$ modulation is an amplitude-limited unquantized precoding problem,
which is still not exactly the same as the popular unquantized precoding problem which typically considers average power constraints.
But some precoding schemes can be easily modified to fit into the amplitude-limited case.
For example, for the ZF scheme, we can do normalization
\begin{equation}\label{eq:zero-forcing}
    \bar{\bx}_t =  \frac{\bH^\dag \bs_t}{C}, ~ t=1,\ldots,T,
\end{equation}
where $C= \max_{t=1,\ldots,T} \| \bH^\dag \bs_t \|_{{\sf IQ}-\infty}$,
such that the amplitude constraints $\| \bar{\bx}_t \|_{{\sf IQ}-\infty} \leq 1$ are satisfied \cite{shao2019one}.
The reader is referred to our previous work \cite{shao2019one} for more amplitude-limited precoding designs.

\section{General and Flexible Designs for Spatial $\Sigma\Delta$ Modulation}
\label{sect:main_meat}

In our previous study with the spatial $\Sigma\Delta$ modulation approach, we mainly applied an existing $\Sigma\Delta$ modulator;
we used the one-bit first-order modulator 
in \cite{shao2019one},
and later we adopted the two-bit second-order modulator in the $\Sigma\Delta$ literature \cite{shao2020multiuser}.
From this section we set our sight on designing our own $\Sigma\Delta$ modulator. 
The study to be described revolve around the following questions.
\begin{enumerate}
	\item Can we have a general and flexible design for $\Sigma\Delta$ modulation of any quantization level { number} and of any order?
	\item Can we make the designs a better fit to coarsely quantized MIMO precoding, specifically, by explicitly working on the signal-to-quantization-and-noise ratios (SQNRs)?
	\item Given an angle sector 
        { $[\theta_l,\theta_u] \subset (-\pi/2,\pi/2)$,}
 	a modulator order { $L$}, and { a quantization level number $M$}, can we design a $\Sigma\Delta$ modulator that works better than the standard $\Sigma\Delta$ modulators in the $\Sigma\Delta$ literature?
    	\item Can we lift the angle sector restriction and allow users to freely lie in  any angles? 
\end{enumerate}

\subsection{A General $\Sigma\Delta$ Modulator Structure}
\label{sect:SD_struct}

\begin{figure}[t]
	\centering
	\includegraphics[width=0.7\linewidth]{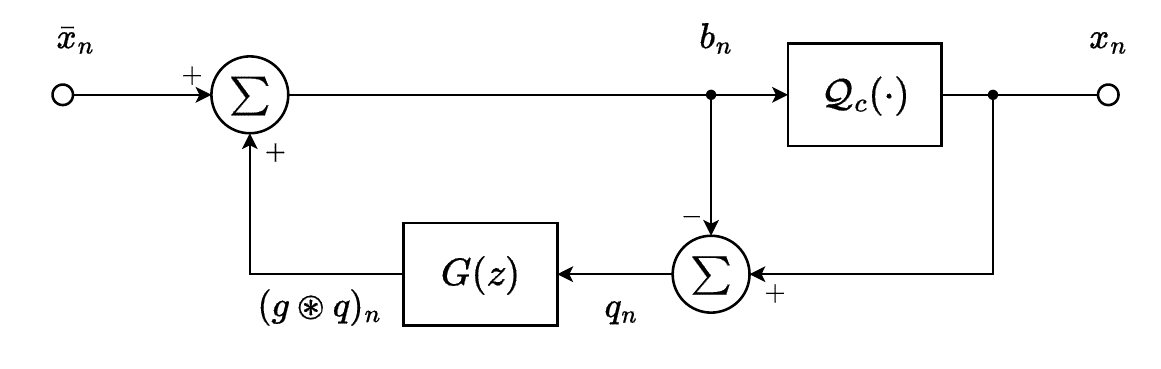}
	\caption{A general $\Sigma\Delta$ modulator.}
	\label{fig:gen_sd}
\end{figure}

{We consider a multi-level, higher-order and complex-valued generalization of the one-bit first-order $\Sigma\Delta$ modulator in Section~\ref{sect:sdmod_intro}. 
The system architecture is depicted in Fig.~\ref{fig:gen_sd}.
It should be noted that this generalized structure was mentioned or considered in the literature~\cite{schreier2005understanding,nagahara2012frequency}, often for the real-valued case.}
{ The rationale of this modulator} 
is identical to that of its predecessor in 
Section~\ref{sect:sdmod_intro},
and we shall be concise with our description.
The input $\{ \bar{x}_n \}_{n \in \Nbb}$ is a complex-valued sequence.
{ The function $\setQ_c$  applies $M$-level quantization to the real and imaginary components.}
To be specific, let
\begin{equation} \label{eq:setX}
\setX = \left\{
\begin{array}{ll}
\{ \pm 1, \pm 3, \ldots, \pm (M-1) \}, & \text{$M$ is even} \\
\{ 0, \pm 2, \ldots, \pm (M-1) \}, & \text{$M$ is odd}
\end{array} 
\right. 
\end{equation}
be the multi-level signal set,
and let $\setQ: \Rbb \rightarrow \setX$ be the quantizer associated with $\setX$.
The quantizer $\setQ_c$ is given by $\setQ_c(x) = \setQ( \Re(x)) + \jj \, \setQ(\Im(x))$.
The error feedback is given by
\[
( g \circledast q )_n = \sum_{l=1}^L g_l q_{n-l},
\]
which is the convolution of the quantization error { sequence} $\{ q_n \}_{n \in \Nbb}$ and an impulse response $\{ g_l \}_{l=1}^L$ of a filter.
The filter coefficients $g_1,\ldots,g_L$ are complex-valued and are to be designed.
The input-output relation of the modulator is 
\begin{equation} \label{eq:io_relation_gen}
x_n = \setQ_c( \bar{x}_n + ( g \circledast q )_n ) = \bar{x}_n + ( g \circledast q )_n + q_n,
~ n \in \Nbb.
\end{equation}
We assume that 
\begin{Assumption} \label{assume:quant_gen}
	Consider the modulator in Fig.~\ref{fig:gen_sd} or the system in \eqref{eq:io_relation_gen}.
	Each quantization error component $\Re(q_n)$ or $\Im(q_n)$ is $[-1,1]$-supported, uniformly distributed on its support, and independent of any other random variables.	
\end{Assumption}
Let $v_n = ( g \circledast q )_n + q_n$ be the quantization noise term at the output.
Its magnitude spectrum is
\[
|V(\omega)|^2 = | 1 + G(\omega) |^2 | Q(\omega) |^2,
\]
where the response $1+G(\omega)$ plays the { key} role of shaping the noise magnitude spectrum.

Furthermore,
we are concerned with overloading.
We will adopt the following no-overload condition.
\begin{Fact} \label{fact:no_overload_gen}
	Consider the modulator in Fig.~\ref{fig:gen_sd} or the system in \eqref{eq:io_relation_gen}. 
	Let $A > 0$ be the maximum input amplitude, specifically, $|\bar{x}_n |_{{\sf IQ}-\infty} := \max\{ |\Re(\bar{x}_n)|, |\Im(\bar{x}_n)| \} \leq A$ for all $n$. Let $\bg= (g_1,\ldots,g_L)$, 
    and $\| \bg \|_{{\sf IQ}-1} = \sum_{i=1}^L | \Re(g_i)| + | \Im(g_i) |$.
	If 
	\[
	A + \| \bg \|_{{\sf IQ}-1}  \leq M,
	\]
	then $|q_n|_{{\sf IQ}-\infty} \leq 1$ for all $n$.
\end{Fact}
Fact~\ref{fact:no_overload_gen} is the multi-level higher-order complex-valued counterpart of  Fact~\ref{fact:no_overload1}.
The result is considered known in the literature; see e.g.,
\cite{schreier1991stability},
\cite[Section 4.2.2]{schreier2005understanding}, \cite{gunturk2003one,nagahara2012frequency}
for the real-valued case.
We provide the proof of Fact~\ref{fact:no_overload_gen} in Appendix~\ref{app:no_overload_gen} for the reader's reference.

To give the reader some insight,
we show some examples covered by the  general $\Sigma\Delta$ modulator.

\begin{Exa}[{first-order modulator}]\label{exa:1st}
	Consider $L= 1$, $g_1= -1$.
	This is the previously studied first-order modulator, which has a noise shaping response $1+ G(\omega) = 1 - e^{-\jj \omega}$.
	According to Fact~\ref{fact:no_overload_gen}, the modulator is guaranteed to have no overload if $A \leq M-1$.
\end{Exa}

\begin{Exa}[{second-order modulator}] \label{exa:2nd}
	Consider $L= 2$, $g_1= -2$, $g_2 = 1$.
	This modulator is called the second-order modulator {in the $\Sigma\Delta$ literature.}  
    {It has} a shaping response $1+ G(\omega) = (1 - e^{-\jj \omega})^2$, which is a stronger highpass response than the first-order.
	By Fact~\ref{fact:no_overload_gen}, the modulator has no overload if $A \leq M-3$. This further implies that the second-order modulator requires at least $M= 4$, or two bits, to achieve the no-overload condition.
\end{Exa}

\begin{Exa}
[{frequency-shifted modulator}]
\label{exa:f_shift} 
	Consider $L= 1$, $g_1= -e^{\jj \omega_c}$ for some given frequency $\omega_c$.
	The shaping response is $1+ G(\omega) = 1 - e^{-\jj (\omega - \omega_c)}$, which is a band-stop response centered at $\omega_c$ {\cite{shao2019one}}.
	By Fact~\ref{fact:no_overload_gen}, the modulator has no overload if $A \leq M-|\sin(\omega_c)| - |\cos(\omega_c)|$,
	or, more conservatively, if $A \leq M - \sqrt{2}$.
\end{Exa}

\subsection{Spatial $\Sigma\Delta$ Modulation by the General Structure}

Consider the spatial $\Sigma\Delta$ modulation for coarsely quantized MIMO precoding in Sections~\ref{sect:cq_intro} and \ref{sect:spatial_sd}.
We want to replace the previous one-bit first-order $\Sigma\Delta$ modulator by the general $\Sigma\Delta$ modulator in the last subsection,
specifically, by applying the general $\Sigma\Delta$ modulator to $\{ \bar{x}_{n,t} \}_{n=0}^{N-1}$ to yield $\{ {x}_{n,t} \}_{n=0}^{N-1}$.
Following the same derivations in Sections~\ref{sect:cq_intro} and \ref{sect:spatial_sd}, we can show that the received signals can be modeled as
\begin{equation} \label{eq:sig_mod_gen}
y_{i,t} = \sqrt{\rho} \bh_i^T \bar{\bx}_t + \sqrt{\rho} \alpha_i \underbrace{(1 + G(\omega_i)) Q_t(\omega_i)}_{:= v_{i,t}} + \eta_{i,t},
\end{equation}
where we recall { $\omega_i = \frac{2\pi d}{\lambda} \sin(\theta_i).$}
Also, by applying Assumption~\ref{assume:quant_gen} to $\{ q_{n,t} \}_{n=0}^{N-1}$, the quantization noise power is
\begin{equation} \label{eqn:effective_noise_power}
  \Exp[ | v_{i,t} |^2 ] = | 1 + G(\omega_i) |^2 \frac{2N}{3}.  
\end{equation}
Our problems, to be studied in the subsequent subsections, 
{ is to}
design the filter coefficients $g_1,\ldots,g_L$ such that the quantization noise powers of the users are mitigated, while, at the same time, the no-overload condition in Fact~\ref{fact:no_overload_gen} is satisfied.

\subsection{Zero Quantization Noise?}

It is natural to question this: Can we have zero quantization noise for all the users?
In raising this question, we allow  the angle $\theta_i$ of each user to lie freely in the admissible 
{ angle region $(-\pi/2,\pi/2)$.}
Achieving zero quantization noise means that $1+ G(\omega_i) = 0$ for all $i$, and this can be made possible by setting the shaping response as 
\begin{equation} \label{eq:zero_q_err_resp}
1+ G(\omega) = \prod_{k=1}^K ( 1 - e^{-\jj (\omega - \omega_k)} ).
\end{equation}
It should be noted that $1+G(\omega)$ produces a multiple notch filter response, with nulls placed at the $\omega_k$'s.
The above shaping response corresponds to a $K$-th order $\Sigma\Delta$ modulator with coefficients
\begin{equation} \label{eq:zero_q_err_coeff}
g_k = \sum_{1 \leq i_1 < \cdots < i_k \leq K} (-e^{\jj \omega_{i_1}}) \cdots (-e^{\jj \omega_{i_k}}),
\end{equation}
for $k=1,\ldots,K$.
This zero quantization noise design, however, has a serious limitation.
\begin{Prop} \label{prop:zero_qe}
	Consider the $\Sigma\Delta$ modulator with coefficients given by \eqref{eq:zero_q_err_coeff}, which achieves zero quantization noise with the users.
	We have the following results.
	\begin{enumerate}[(a)]		
		\item It holds that $$\| \bg \|_{{\sf IQ}-1} \leq \sqrt{2} ( 2^K - 1 ),$$
		and equality is attained when $\omega_1 = \cdots = \omega_K \in \{ \pi/4, 3\pi/4, 5\pi/4, 7\pi/4 \}$.
		\item As a simplifying assumption, assume each $\omega_i$ to be i.i.d. uniformly distributed on $(-\pi,\pi)$. Then,
		\[
		2^{(K-1)/2} \leq \sqrt{ \Exp [ \| \bg \|_{{\sf IQ}-1}^2  ]} \leq 2^K.
		\]
	\end{enumerate}
\end{Prop}
We show the proof of Proposition~\ref{prop:zero_qe} in Appendix~\ref{app:props}.
Proposition~\ref{prop:zero_qe} suggests that $\| \bg \|_{{\sf IQ}-1}$ may increase exponentially with $K$.
To give some idea, Table~\ref{tab:g_IQ} shows some empirical evaluation results for $\| \bg \|_{{\sf IQ}-1}$. 
\begin{table*}[tbp!]
	\centering
	\caption{The minimum, mean, root mean square (RMS), and maximum values of $\| \bg \|_{{\sf IQ}-1}$ for \eqref{eq:zero_q_err_coeff} and for a number of randomly generated $\omega_i$'s. The $\omega_i$'s are i.i.d. $(-\pi,\pi)$-uniform distributed.}
	\label{tab:g_IQ}
	\renewcommand*{\arraystretch}{1.5}
	\begin{tabular}{c|c|c|c|c|c|c|c }%
		\hline  %
		\backslashbox{$\|\bg\|_{\sf IQ-1}$}{$K$}          & $2$ & $3$        & $4$ &$5$ & $6$& $7$ & $8$\\
		\hline \hline %
		min. & 1.00006 & 1.078  & 1.23   & 1.26 & 1.86  & 2.53   & 3.09   \\
		\hline
		mean &2.89  & 5.28  & 8.37  & 12.85  & 18.83  & 27.17   & 38.06    \\
		\hline %
		
		RMS &3.01  & 5.60   & 9.27   & 14.72   & 22.53  & 33.78  & 50.37  \\
		\hline %

		max. & 4.09  & 9.46  & 20.23 & 41.33  & 81.96  & 160.34   & 315.88  \\
		\hline

		\hline
	\end{tabular}
	
\end{table*}
{ We observe that the empirical results are in}
agreement with our theoretical prediction.
By also considering the no-overload requirement in Fact~\ref{fact:no_overload_gen},
we { see} the following implication: the quantization level { number} $M$ may need to increase exponentially with the number of users $K$ to achieve zero quantization noise at the user side.

\subsection{SQNR Maximization in a User Targeted Fashion}
\label{sect:SQNR_max}

Since zero quantization noise is practically infeasible even for a moderate number of users, we turn to the alternative of maximizing the SQNRs experienced by the users.
Note that, as in the previous problem, we allow the user angles $\theta_i$'s to freely lie in $(-\pi/2,\pi/2)$.
Our tasks are divided into three parts: define a suitable SQNR for the problem at hand, properly formulate the $\Sigma\Delta$ modulator design as an optimization problem, and develop a solution.

We start with the SQNR.
From the received signal model \eqref{eq:sig_mod_gen}, we see that the signal part $\sqrt{\rho} \bh_i^T \bar{\bx}_t$ scales with $\sqrt{\rho} |\alpha_i| A$.
Here, it is important to note that $A$ describes the {maximum input signal amplitude,}
{ i.e., $\| \bar{\bx}_t \|_{{\sf IQ}-\infty} \leq A$ for all $t$.}
We define the SQNR of user $i$ as the ratio of the square of the received signal scale factor $\sqrt{\rho} |\alpha_i| A$ to the quantization and noise power, which can be shown to be
\begin{equation} \label{eq:SQNR_def}
{\sf SQNR}_i = \frac{ \rho |\alpha_i |^2 A^2 }{ \frac{2N \rho |\alpha_i |^2}{3} | 1 + G(\omega_i) |^2  + \sigma_\eta^2  }. 
\end{equation}

Next, we formulate the $\Sigma\Delta$ modulator design.
Our underlying assumption is that the BS is informed of the channels $\bh_i$'s, or, the complex gains $\alpha_i$'s and angles $\theta_i$'s of all the users. 
The BS is assumed to know the background noise power $\sigma_\eta^2$, too.
Also, the modulator order $L$ and the quantization level number $M$ of the $\Sigma\Delta$ modulator are prespecified.
We design the $\Sigma\Delta$ filter coefficients by the max-min-fair criterion, subject to the no-overload condition in Fact~\ref{fact:no_overload_gen}:
\begin{equation} \label{eq:MMF}
\begin{aligned}
\max_{ \bg \in \Cbb^L, A \in \Rbb} & ~ \min_{i=1,\ldots,K} {\sf SQNR}_i  \\
{\rm s.t.} & ~ A + \| \bg \|_{{\sf IQ}-1} \leq M, ~ A > 0.
\end{aligned}
\end{equation}
Here, fairness is achieved by maximizing the weakest user's SQNR, thereby sacrificing no one in the interest of others.
It is worth noting that we also optimize the {maximum} input signal amplitude $A$, rather than {prefixing} it, to give the design more degrees of freedom.

The max-min-fair design \eqref{eq:MMF} can be converted to a convex problem and can be efficiently solved. 
To see how this is done, we substitute \eqref{eq:SQNR_def} into problem \eqref{eq:MMF} and rewrite the problem as
\begin{equation} \label{eq:MMF2}
\begin{aligned}
\min_{ \bg \in \Cbb^L, A \in \Rbb} & ~ \max_{i=1,\ldots,K} \frac{\sqrt{|1+G(\omega_i)|^2 + \gamma_i}}{A}  \\
{\rm s.t.} & ~ A + \| \bg \|_{{\sf IQ}-1} \leq M, ~ A > 0,
\end{aligned}
\end{equation}
where $\gamma_i =  3 \sigma_\eta^2 / (2 N \rho |\alpha_i |^2 )$.
Problem~\eqref{eq:MMF2} is quasi-convex, but not convex.
Consider the following transformation
\begin{equation} \label{eq:cct}
{\revcolor \bnu} = \bg / A, ~  \xi = 1/A,
\end{equation}
which is known as the Charnes-Cooper transformation in optimization \cite{charnes1962programming}.
The transformation \eqref{eq:cct} is one-to-one if $A$ and $\xi$ are positive.
Using \eqref{eq:cct}, problem~\eqref{eq:MMF2} can be transformed as
\begin{equation} \label{eq:MMF3}
\begin{aligned}
\min_{ {\revcolor \bnu} \in \Cbb^L, \xi \in \Rbb} & ~ \max_{i=1,\ldots,K} \sqrt{|\xi + \ba(\omega_i)^T  {\revcolor \bnu} |^2 + \gamma_i \xi^2 }  \\
{\rm s.t.} & ~ 1 + \| {\revcolor \bnu} \|_{{\sf IQ}-1} \leq M \xi, ~ \xi > 0,
\end{aligned}
\end{equation}
where we redefine $\ba(\omega) = (1, e^{-\jj \omega}, \ldots, e^{-\jj (N-1) \omega})$.
Moreover, problem~\eqref{eq:MMF3} is equivalent to
\begin{equation} \label{eq:MMF4}
\begin{aligned}
\min_{ {\revcolor \bnu} \in \Cbb^L, \xi \in \Rbb} & ~ \max_{i=1,\ldots,K} \sqrt{|\xi + \ba(\omega_i)^T  {\revcolor \bnu} |^2 + \gamma_i \xi^2 }  \\
{\rm s.t.} & ~ 1 + \| {\revcolor \bnu} \|_{{\sf IQ}-1} \leq M \xi, ~ \xi \geq 0,
\end{aligned}
\end{equation}
where we replace $\xi > 0$ with $\xi \geq 0$.
This is without loss, because the first constraint in \eqref{eq:MMF4} implies $1 \leq M \xi$, and with the second constraint $\xi \geq 0$ we further get $\xi \geq 1/M > 0$.
Problem \eqref{eq:MMF4} is convex,
and its solution can be conveniently and efficiently obtained by using a convex optimization software, such as the widely-used {\tt CVX} \cite{cvx}.

\input{example-UT}

\subsection{SQNR Maximization for a Fixed Angle Sector}
\label{sect:SQNR_max_sect}

The {user-targeted} SQNR maximization design in the last subsection assumes that we can change the $\Sigma\Delta$ modulator whenever the user angles $\theta_i$'s and channel gains $|\alpha_i|^2$ change.
Suppose { that} we are prohibited to do so due to implementation reasons,
and we can only re-design  the $\Sigma\Delta$ modulator once in a while.
We hence return to the angle sector setting wherein we serve users in a prespecified angle sector $[\theta_l, \theta_u] \subset (-\pi/2,\pi/2)$ { (which can be lowpass or bandpass).}
Our problem is to adapt the preceding $\Sigma\Delta$ modulator design to { this fixed sector setting.}

We start with off-the-shelf designs from the $\Sigma\Delta$ literature.
Consider the following shaping response
\beq \label{eq:bandpass0}
1 + G(\omega) = ( 1 - e^{-\jj (\omega - \omega_c )} )^L
\eeq 
for a given positive integer $L$,
where $\omega_c = ( \omega_l + \omega_u )/2$,  $\omega_l = \frac{2\pi d}{\lambda} \sin(\theta_l),$ $\omega_u = \frac{2\pi d}{\lambda} \sin(\theta_u).$
This is a band-stop response with center frequency $\omega_c$.
The corresponding coefficients are 
\begin{equation} \label{eq:bandpass}
g_l = {L \choose l} (- e^{\jj  \omega_c })^l, ~l =1,\ldots,L.
\end{equation}
This modulator is essentially the combination of the standard $L$-th order modulator and the frequency-shifted modulator; see Examples~\ref{exa:2nd}--\ref{exa:f_shift}.
Increasing the order $L$ makes the band-stop response sharper, but this comes with a limitation.
\begin{Prop} \label{prop:zero_qe_2}
	Consider the $\Sigma\Delta$ modulator with coefficients given by \eqref{eq:bandpass},
    { and with the shaping response given by \eqref{eq:bandpass0}.}
 We have
	\[
	2^L - 1  \leq \| \bg \|_{{\sf IQ}-1} \leq \sqrt{2} ( 2^L - 1 ).
	\]
\end{Prop}
The proof of Proposition~\ref{prop:zero_qe_2} is shown in Appendix~\ref{app:props}.
Proposition~\ref{prop:zero_qe_2}, together with Fact~\ref{fact:no_overload_gen}, indicate that the quantization level number $M$ needs to increase exponentially with $L$ to achieve the no-overload condition.

Alternatively, we can repurpose the SQNR-based design in Section~\ref{sect:SQNR_max}.
Suppose { that} the channel gains $|\alpha_i|$'s are known to lie in a range $[r_{\rm min}, r_{\rm max}]$.
{ Our goal is to design the $\Sigma\Delta$ modulator in accordance with the {prespecified} angle sector $[\theta_l, \theta_u]$ and the channel gain range $[r_{\rm min}, r_{\rm max}]$.}
Following the SQNR definition in \eqref{eq:SQNR_def}, a user with angle $\theta \in [\theta_l, \theta_u]$ and channel gain $|\alpha| \in [r_{\rm min}, r_{\rm max}]$ will experience an SQNR
\begin{align*}
	{\sf SQNR} & = \frac{ \rho |\alpha |^2 A^2 }{ \frac{2N \rho |\alpha |^2}{3} | 1 + G(\omega) |^2  + \sigma_\eta^2  } \\
	& \geq \frac{ \rho r_{\rm min}^2 A^2 }{ \frac{2N \rho r_{\rm max}^2}{3} | 1 + G(\omega) |^2  + \sigma_\eta^2  } \\
	& := \widetilde{\sf SQNR}(\omega),
\end{align*}
where $\omega = \frac{2\pi d}{\lambda} \sin(\theta).$
With the above expression, we consider the following adaptation of the max-min-fair design in \eqref{eq:MMF} to the angle sector setting:
\begin{equation} \label{eq:MMF_alt}
	\begin{aligned}
		\max_{ \bg \in \Cbb^L, A \in \Rbb} & ~ \min_{\omega \in [\omega_l,\omega_u]}  \widetilde{\sf SQNR}(\omega)  \\
		{\rm s.t.} & ~ A + \| \bg \|_{{\sf IQ}-1} \leq M, ~ A > 0,
	\end{aligned}
\end{equation}
where we  maximize the worst SQNR lower bound over the angle sector; recall that 
$\omega_l = \frac{2\pi d}{\lambda} \sin(\theta_l),$ $\omega_u = \frac{2\pi d}{\lambda} \sin(\theta_u).$
We deal with problem \eqref{eq:MMF_alt} by discretization:
\begin{equation} \label{eq:MMF_alt2}
	\begin{aligned}
		\max_{ \bg \in \Cbb^L, A \in \Rbb} & ~ \min_{i=1,\ldots,I}  \widetilde{\sf SQNR}(\omega_i)  \\
		{\rm s.t.} & ~ A + \| \bg \|_{{\sf IQ}-1} \leq M, ~ A > 0,
	\end{aligned}
\end{equation}
where, with an abuse of notations, we redefine $\omega_l \leq \omega_1 < \omega_2 < \cdots < \omega_I \leq \omega_u$ as sample points of $[\omega_l,\omega_u]$ (e.g., by uniform sampling).
Problem \eqref{eq:MMF_alt2} takes the same form as problem \eqref{eq:MMF}, and the same method in Section~\ref{sect:SQNR_max} can be used to solve problem \eqref{eq:MMF_alt2}.
We shall not repeat the details.
\input{example-FS}

\subsection{Comparison with Existing Temporal $\Sigma\Delta$ Modulator Designs}

{ It is interesting to compare our optimization-based spatial $\Sigma\Delta$ modulator designs with relevant designs in the temporal $\Sigma\Delta$ literature.}
To put { this} into perspective, let us write down the user-targeted and fixed-sector designs, shown in \eqref{eq:MMF} and \eqref{eq:MMF_alt}, respectively, as a single formulation:
\begin{equation}  \label{eq:design_explain}
	\begin{aligned}
		\min_{ \bg \in \Cbb^L, A \in \Rbb} & ~ \max_{\omega \in \Omega} \frac{\sqrt{|1+G(\omega)|^2 + \gamma_i}}{A}  \\
		{\rm s.t.} & ~ A + \| \bg \|_{{\sf IQ}-1} \leq M, ~ A > 0,
	\end{aligned}
\end{equation}
where $\Omega= \{ \omega_1,\ldots,\omega_K \}$ for the user-targeted case and {$\Omega= [ \omega_l,\omega_u]$} for the fixed-sector case; 
{note that} the constants $\gamma_i$'s scale with the background noise power $\sigma_\eta^2$.
Suppose we prefix {the maximum input signal amplitude} $A$ and set $\gamma_i = 0$ for all $i$.
The above problem then reduces to
\begin{equation}  \label{eq:design_explain2}
	\begin{aligned}
		\min_{ \bg \in \Cbb^L} & ~ \max_{\omega \in \Omega} |1+G(\omega)|  \\
		{\rm s.t.} & ~ \| \bg \|_{{\sf IQ}-1} \leq M - A
	\end{aligned}
\end{equation}
which is a multiple notch filter design for the user-targeted case, and a {band-stop} filter design for the fixed-sector case.
In fact, we have seen that in the illustrations in Figures~\ref{fig:FiltDes-UsrAdp-8-usr} and \ref{fig:FiltDes}.
In this connection, 
a formulation similar to \eqref{eq:design_explain2} was considered by Nagahara and Yamaoto \cite{nagahara2012frequency} to design temporal  $\Sigma\Delta$ modulators for {lowpass or bandpass signals.}
{There are subtle differences;
e.g., Nagahara and Yamaoto do not use the no-overload constraint in problem~\eqref{eq:design_explain2}, and they replace it with a sufficient condition in the form of a linear matrix inequality.
The distinctive difference with our designs, apart from being for a different application, 
is that we also optimize the input amplitude $A$ to maximize the users' SQNRs.
}
\section{Two-Dimensional Spatial $\Sigma\Delta$ Modulation}
\label{sect:2D_sigdel}

{ The spatial $\Sigma\Delta$ modulator designs developed in the preceding sections can be extended to the case of two-dimensional (2D) uniform planar arrays.
It should be noted that, to the best of our knowledge, spatial $\Sigma\Delta$ modulation for coarsely quantized MIMO precoding with 2D uniform planar arrays has not been considered before.
In the following subsections we will concisely describe how this is done.}

\subsection{A 2D $\Sigma\Delta$ Modulator}
\label{sect:2D_mod}

{Before we proceed, we should mention that 2D $\Sigma\Delta$ modulation was considered in, and finds important applications to, image half-toning \cite{kite1997digital}.}
{Here, we first}
consider the 2D extension of the general $\Sigma\Delta$ modulator in Section~\ref{sect:SD_struct}.
The input-output relation of the 2D modulator is
\begin{align*}
x_{n_1,n_2} & = \setQ_c( \bar{x}_{n_1,n_2} + ( g \circledast q )_{n_1,n_2} ) \\
& = \bar{x}_{n_1,n_2} + ( g \circledast q )_{n_1,n_2} + q_{n_1,n_2}, \\
( g \circledast q )_{n_1,n_2} & = \sum_{l_1=0}^{L_1} \sum_{l_2=0}^{L_2} g_{l_1,l_2} g_{n_1 - l_1, n_2 - l_2},
\end{align*}
where { $\{ \bar{x}_{n_1,n_2} \}_{n_1,n_2 \in \Nbb} \subset \Cbb$} is the input;
{ $\{ x_{n_1,n_2} \}_{n_1,n_2 \in \Nbb} \subset \setX + \jj \setX$} is the output;
{ each} $q_{n_1, n_2} \in \Cbb$ is a quantization error and is assumed to be follow the i.i.d. assumption in Assumption~\ref{assume:quant_gen};
the $g_{l_1,l_2}$'s, $l_1 =0,\ldots,L_1$,  $l_2 =0,\ldots,L_2$, with $g_{0,0}= 0$, are the filter coefficients.
The filter plays the role of shaping the noise magnitude spectrum according to $| 1 + G(\omega_1,\omega_2)|^2$,
where $$G(\omega_1,\omega_2) = \sum_{n_1=0}^{L_1} \sum_{n_2=0}^{L_2} g_{n_1,n_2} e^{-\jj (n_1 \omega_1+ n_2 \omega_2)}$$ is the 2D Fourier transform of $\{ g_{l_1,l_2} \}$.
Let $\bG \in \Cbb^{(L_1 + 1) \times (L_2 + 1)}$ be a matrix with its $(i,j)$th element given by $g_{i-1,j-1}$.
As the 2D extension of the no-overload condition in Fact~\ref{fact:no_overload_gen}, the modulator has no overload if 
\[
A + \| \bG \|_{{\sf IQ}-1} \leq M,
\]
where
$\| \bG \|_{{\sf IQ}-1} = \sum_{l_1=0}^{L_1} \sum_{l_2=0}^{L_2} |\Re( g_{l_1,l_2} )| + |\Im( g_{l_1,l_2} )|$;
$A > 0$ is the maximum input amplitude, i.e., $| x_{n_1,n_2} |_{{\sf IQ}-\infty} \leq 1$ for all $n_1,n_2$.

\subsection{Uniform Planar Array}

\begin{figure}[tp!]
	\centering
\includegraphics[width=0.65\linewidth]{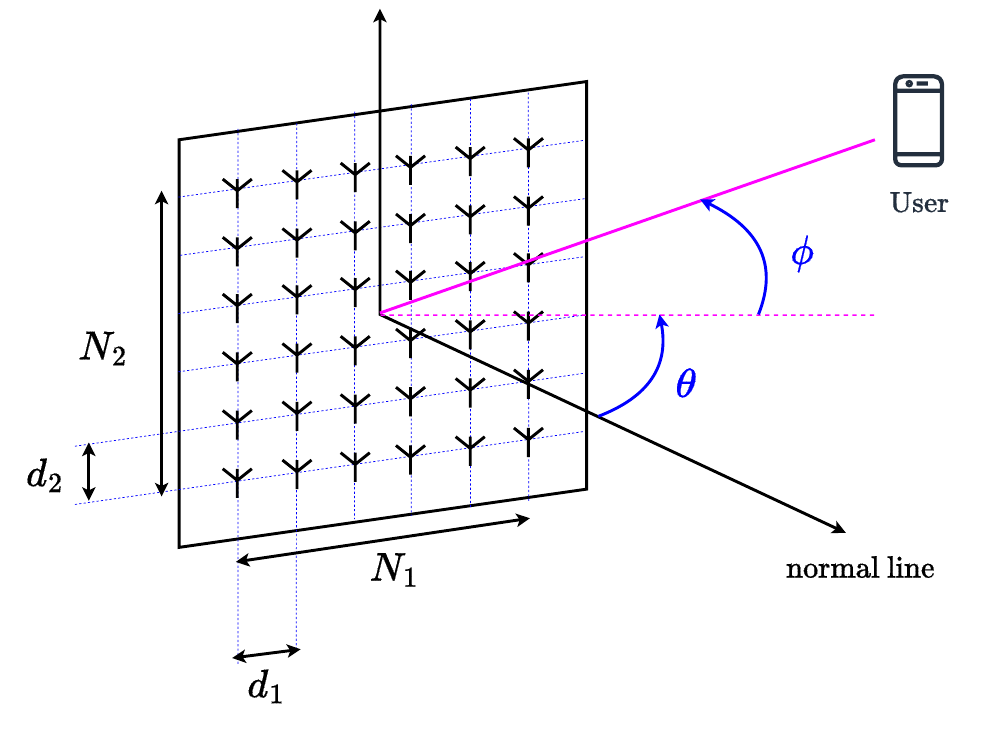}
	\caption{The uniform planar array.}
	\label{fig:upa}
\end{figure}

Second, we review some concepts with the uniform planar array.
As illustrated in Fig.~\ref{fig:upa}, a uniform planar array has the antennas 
arranged in a equi-spaced rectangular fashion~\cite{balanis2015antenna}.
It has $N_1$ and $N_2$ antennas in the horizontal and vertical directions, respectively.
Under the same set of operating assumptions as uniform linear arrays, the uniform planar array has an array response
\[
\bA(\theta,\phi) = \ba_1(\theta,\phi) \ba_2^T(\phi) \in \Cbb^{N_1 \times N_2},
\]
where $\theta \in (-\pi/2,\pi/2)$ and $\phi \in (-\pi/2,\pi/2)$ are the azimuth and elevation angles, respectively; we have
\begin{align*}
	\ba_1(\theta,\phi) & = ( 1, e^{-\jj \omega_1}, \ldots,  e^{-\jj (N_1 - 1 )\omega_1}), \\
	\ba_2(\phi) & = ( 1, e^{-\jj \omega_2}, \ldots,  e^{-\jj (N_2 - 1 )\omega_2}), \\
	\omega_1 & = \frac{2\pi d_1}{\lambda} \cos(\phi) \sin(\theta),
	~ \omega_2  = \frac{2\pi d_2}{\lambda} \sin(\phi);
\end{align*}
$d_1 \leq \lambda/2$ and $d_2 \leq \lambda/2$ are horizontal and vertical inter-antenna spacings, respectively;
$\lambda$ is the carrier wavelength.
Let $x_{n_1,n_2}$ be the transmitted signal from the $(n_1+1,n_2+1)$th antenna of the array, 
and let $\bX \in \Cbb^{N_1 \times N_2}$ be a matrix with its $(i,j)$th element given by $x_{i-1,j-1}$.
The array exhibits a transmit directional pattern
\begin{align*}
{\rm tr}( \bA^T(\theta,\phi) \bX ) & = \sum_{n_1=0}^{N_1 - 1} \sum_{n_2 = 0}^{N_2 - 1 }
x_{n_1,n_2} e^{-\jj (n_1 \omega_1 + n_2 \omega_2) } \\
& = X(\omega_1,\omega_2).
\end{align*}

\subsection{$\Sigma\Delta$ MIMO Precoding for Uniform Planar Arrays}
\label{sect:2D_sigdel_Precode}

Third, we consider spatial $\Sigma\Delta$ modulation for coarsely quantized MIMO precoding in  Sections~\ref{sect:spatial_sd} and \ref{sect:main_meat} when 
the 1D uniform linear array is replaced by the 2D uniform planar array.
{Under the 2D uniform planar array setting,}
the basic signal model \eqref{eq:sig_mod} is modified as
\[
y_{i,t} = \sqrt{\rho} \, {\rm tr}( \bH_i^T  \bX_t ) + \eta_{i,t},
\]
where 
$\bX_t \in \Cbb^{N_1 \times N_2}$ is the transmitted signal;
$\bH_i \in \Cbb^{N_1 \times N_2}$ is the channel of user $i$ and is modeled as
\[
\bH_i = \alpha_i \bA(\theta_i,\phi_i),
\]
in which $\alpha_i, \theta_i, \phi_i$ are the complex channel gain, azimuth angle and elevation angle of user $i$, respectively.
Also, the $\Sigma\Delta$ modulator is replaced by the 2D modulator in Section~\ref{sect:2D_mod}. 
Let $\bar{\bX}_t \in \Cbb^{N_1 \times N_2}$ be the 2D transmitted signal before $\Sigma\Delta$ modulation.
It can be shown that
\[
y_{i,t} \simeq  \sqrt{\rho} \, {\rm tr}( \bH_i^T  \bar{\bX}_t ) + 
\sqrt{\rho} \alpha_i (1+G(\omega_1,\omega_2))Q_t(\omega_1,\omega_2) + \eta_{i,t},
\]
which, as before, is the sum of signal, quantization noise, and background noise components; here, $Q_t(\omega_1,\omega_2)$ is the 2D Fourier transform of the quantization error $\{ q_{n_1,n_2,t} \}_{n_1,n_2}$.
Subsequently, 
we can further show that the SQNR, under the definition in \eqref{eq:SQNR_def}, is
\begin{equation*} \label{eq:SQNR_2D}
{\sf SQNR}_i = \frac{ \rho |\alpha_i |^2 A^2 }{ \frac{2N_1 N_2 \rho |\alpha_i |^2}{3} | 1 + G(\omega_{1,i},\omega_{2,i}) |^2  + \sigma_\eta^2  }. 
\end{equation*}
where $\omega_{1,i} = \frac{2\pi d}{\lambda} \cos(\phi_i) \sin(\theta_i)$,
$\omega_{2,i} = \frac{2\pi d}{\lambda} \sin(\phi_i) $.

Let us describe the modulator designs.
We can follow the user-targeted $\Sigma\Delta$ modulator design in problem~\eqref{eq:MMF} in Section~\ref{sect:SQNR_max}, 
which maximizes the users' SQNRs in the max-min-fair fashion and in a user targeted fashion.
The 2D extension of the design  is 
\begin{equation*} \label{eq:MMF_2D}
\begin{aligned}
\max_{ \bG, A \in \Rbb}
& ~ \min_{i=1,\ldots,K} {\sf SQNR}_i  \\
{\rm s.t.} & ~ A + \| \bG \|_{{\sf IQ}-1} \leq M, ~ A > 0,  
\end{aligned}
\end{equation*}
where the domain of $\bG$ is $\Cbb^{(L_1+1) \times (L_2+1)}$, $g_{0,0}= 0$.
The above problem takes the same form as its predecessor, problem~\eqref{eq:MMF},
and it can be solved by the exactly same way as in Section~\ref{sect:SQNR_max}. 
We can also adopt the fixed-sector design in problem~\eqref{eq:MMF_alt} in  Section~\ref{sect:SQNR_max_sect}, which designs a fixed modulator for an angle sector by maximizing the worst SQNR lower bound over that sector.
Let $[\theta_l,\theta_u] \times [\phi_l,\phi_u]$ be the angle sector of interest.
The 2D extension of the design is 
\begin{equation} \label{eq:MMF_fixed_2D}
\begin{aligned}
\max_{ \bG, A \in \Rbb}
& ~ \min_{(\omega_1,\omega_2) \in \Omega} \widetilde{\sf SQNR}(\omega_1,\omega_2)  \\
{\rm s.t.} & ~ A + \| \bG \|_{{\sf IQ}-1} \leq M, ~ A > 0, 
\end{aligned}
\end{equation}
where 
\begin{align*} 
&\Omega = \left \{ 
    \left[\begin{array}{c}
        \omega_1\\ \omega_2
    \end{array}\right]       
                            =       \left[\begin{array}{c}
                                    \tfrac{2\pi d}{\lambda} \cos(\phi) \sin(\theta)\\
                                     \tfrac{2\pi d}{\lambda} \sin(\phi) 
                                    \end{array}
                                    \right]
    \middle| 
    \begin{array}{l}
        \theta \in [\theta_l,\theta_u]  \\
        \phi \in [\phi_l,\phi_u]
    \end{array} \right \}; \\
&\widetilde{\sf SQNR}(\omega_1,\omega_2) = 
\frac{ \rho r_{\rm min}^2 A^2 }{ \frac{2N_1 N_2 \rho r_{\rm max}^2}{3} | 1 + G(\omega_1,\omega_2) |^2  + \sigma_\eta^2  }.
\end{align*}
The above problem can be handled by the same way as in Section~\ref{sect:SQNR_max_sect}.
\input{example-2D}

\input{simResult}


\appendix
\section{Proof of Fact~\ref{fact:no_overload_gen}}
\label{app:no_overload_gen}

Suppose $| q_{n-l} |_{{\sf IQ}-\infty} \leq 1$ for all $l \geq 1$. 
For convenience, let $b_n = \bar{x}_n + ( g\circledast q )_n$.
We have
\begin{align*}
	| \Re( b_n ) | & \leq | \Re( \bar{x}_n ) | + \left|\Re\left( \sum_{l=1}^L g_l q_{n-l} \right) \right| \\
	& \leq | \Re( \bar{x}_n ) | + \sum_{l=1}^L ( | \Re(g)_l | | \Re(q_{n-l}) | + | \Im(g)_l | | \Im(q_{n-l}) | ) \\
	& \leq A + \| \bg \|_{{\sf IQ}-1}.
\end{align*}
Recall that $\setQ$ is the quantizer associated with $\setX$ in \eqref{eq:setX},
and note $\Re(q_n) = \setQ( \Re(b_n) ) - \Re(b_n)$.
It can be verified that $| \setQ(y) - y | \leq 1$ if $| y | \leq M$.
Hence, if  $ A + \| \bg \|_{{\sf IQ}-1} \leq M$ holds, we have $|\Re(q_n)| \leq 1$.
Similarly, one can show that if $ A + \| \bg \|_{{\sf IQ}-1} \leq M$, then $|\Im(q_n)| \leq 1$.
The proof is done.

\section{Proof of Propositions~\ref{prop:zero_qe} and \ref{prop:zero_qe_2}}
\label{app:props}

First we show  Proposition~\ref{prop:zero_qe}.(a) and  Proposition~\ref{prop:zero_qe_2}.
For convenience, rewrite the coefficients $g_k$'s in \eqref{eq:zero_q_err_coeff} as
\[
g_k = \sum_{1 \leq i_1 < \cdots < i_k \leq K} \beta_{i_1} \cdots \beta_{i_k},
\]
where $\beta_i = -e^{\jj \omega_i}$.
Note that the coefficients $g_k$'s in \eqref{eq:bandpass} is a special case of the above where $\beta_1 = \cdots = \beta_K = e^{\jj \omega_c}$, $K= L$.
It can be shown that, for $x \in \Cbb$,
\begin{align}
	| x |_{{\sf IQ}-1} & := |\Re(x)| + |\Im(x)|  \geq |x|, \label{eq:proof_2prop_t1} \\
	| x |_{{\sf IQ}-1} & \leq\sqrt{2} |x|, \label{eq:proof_2prop_t2}
\end{align}
where equality in \eqref{eq:proof_2prop_t2} is attained if $x$ takes the form $x = |x| e^{j\omega}$, $\omega \in \{ \pi/4, 3\pi/4, 5\pi/4, 7\pi/4 \}$.
This leads to 
\begin{align*}
	| g_k |_{{\sf IQ}-1} & \leq \sqrt{2} \left| \sum_{1 \leq i_1 < \cdots < i_k \leq K} \beta_{i_1} \cdots \beta_{i_k}  \right| \\
	& \leq \sqrt{2} \sum_{1 \leq i_1 < \cdots < i_k \leq K} | \beta_{i_1} \cdots \beta_{i_k} | \\
	& = \sqrt{2} {K \choose k},
\end{align*}
where equality above is attained if $\beta_1 = \cdots = \beta_K = e^{\jj \omega}$, $\omega \in \{ \pi/4, 3\pi/4, 5\pi/4, 7\pi/4 \}$.
Hence we have 
\[
\| \bg \|_{{\sf IQ}-1} \leq \sqrt{2} \sum_{k=1}^K {K \choose k} = \sqrt{2} (2^K - 1),
\]
which is the inequality in Proposition~\ref{prop:zero_qe}.(a) and the upper bound inequality in Proposition~\ref{prop:zero_qe_2}.
Furthermore, for the case of $\beta_1 = \cdots = \beta_K := \beta$, we use \eqref{eq:proof_2prop_t1} to obtain
\begin{align*}
	| g_k |_{{\sf IQ}-1} & \geq \left|  \sum_{1 \leq i_1 < \cdots < i_k \leq K} \beta^k \right| 
	 = {K \choose k}.
\end{align*}
Consequently we have $\| \bg \|_{{\sf IQ}-1} \geq 2^K - 1$, the lower bound inequality in Proposition~\ref{prop:zero_qe_2}.

Second we show  Proposition~\ref{prop:zero_qe}.(b).
If $\omega_1,\ldots,\omega_K$ are i.i.d. and $(-\pi,\pi)$-uniform distributed, one can show the following: given $1 \leq i_1 < \cdots < i_k \leq K$, $1 \leq j_1 < \cdots < j_k \leq K$,
\[
\Exp[ \beta_{i_1} \cdots \beta_{i_k} \beta_{j_1}^* \cdots \beta_{j_k}^* ] =
\left\{
\begin{array}{ll}
	1, & \text{$i_l = j_l$ for all $l$} \\
	0, & \text{otherwise}
\end{array}
\right.
\]
Subsequently we have 
\begin{align*} 
	& \Exp[ | g_k |^2 ] = \\  &\textstyle\sum_{  1 \leq i_1 < \cdots < i_k \leq K } \sum_{ 1 \leq j_1 < \cdots < j_k \leq K } \Exp[ \beta_{i_1} \cdots \beta_{i_k} \beta_{j_1}^* \cdots \beta_{j_k}^* ] \\
	& = \textstyle\sum_{  1 \leq i_1 < \cdots < i_k \leq K }  1 \\ 
        &=\textstyle {K \choose k }.
\end{align*}
Also, it can be shown that, for $\bx \in \Cbb^K$, $\| \bx \|_2 \leq \| \bx \|_{{\sf IQ}-1} \leq \sqrt{2K} \| \bx \|_2$.
This leads to 
\begin{equation}  \label{eq:proof_2prop_t3}
	2^K -1 \leq \Exp[ \| \bg \|_{{\sf IQ}-1}^2 ] \leq 2K (2^K- 1).
\end{equation}
Our final step is to polish the above bounds to a simpler form.
Consider the inequalities below:
\begin{align*}
	2K & = 2^{\log(K) + \log(2)} \leq 2^{K-1+\log(2)} \leq 2^K \\
	2^K -1 & \leq 2^K \\
	2^K - 1 & \geq 2^K - 2^{K-1} = 2^{K-1}
\end{align*}
where, in the first equation, we have used $\log(x) \leq x - 1$ for $x > 0$.
Applying the above inequalities to \eqref{eq:proof_2prop_t3} gives the result in Proposition~\ref{prop:zero_qe}.(b).


\bibliographystyle{IEEEtran}
\bibliography{refs_lastime,refs_thistime,refs_wk}
\end{document}

%% file: sym_marco.tex
\newcommand\bx{\ensuremath{{\bm x}}}

\newcommand\bG{\ensuremath{{\bm G}}}
\newcommand\bh{\ensuremath{{\bm h}}}
\newcommand\bH{\ensuremath{{\bm H}}}

\newcommand\bX{\ensuremath{{\bm X}}}

\newcommand\ba{\ensuremath{{\bm a}}}

\newcommand\bA{\ensuremath{{\bm A}}}

\newcommand\bg{\ensuremath{{\bm g}}}

\newcommand\bD{\ensuremath{{\bm D}}}

\newcommand\bnu{\ensuremath{{\bm \nu}}}

\newcommand\bs{\ensuremath{{\bm s}}}

\newcommand{\Nbb}{\mathbb{N}}
\newcommand{\Rbb}{\mathbb{R}}
\newcommand{\Cbb}{\mathbb{C}}

\newcommand{\setQ}{\mathcal{Q}}

\newcommand{\setX}{\mathcal{X}}

\newcommand{\setN}{\mathcal{N}}

\newcommand{\Exp}{\mathbb{E}}
\newcommand{\jj}{\mathfrak{j}}

\newcommand{\Diag}{\mathrm{Diag}}

\newcommand\sgn{\ensuremath{{\rm sgn}}}



%% file: wkintro.tex
\section{Introduction}

Physical-layer or signal-level transceiver techniques have been playing a key part in massive multi-input multi-output (MIMO) communications.
They serve the crucial role of physically realizing the promise of massive MIMO, such as substantial gains in spectral efficiency and greatly improved spatial degrees of freedom for serving multiple users \cite{marzetta_larsson_yang_ngo_2016}.
Recent research has focused on how MIMO transceiver techniques can allow us to better cope with practical limitations with 
the radio frequency (RF) front ends, specifically,  issues with the energy efficiency and hardware cost of power amplifiers and analog-to-digital/digital-to-analog converters (ADCs/DACs).
Let us narrow down our scope to 
the ADCs/DACs.
We want fine signal resolution to support currently-used  transceiver techniques. 
This calls for high resolution ADCs/DACs being employed at the receiver/transmitter,
and a higher resolution means a higher hardware cost and energy consumption. 
{\revcolor Employing high-resolution ADCs/DACs would not be a serious issue if the MIMO scale (the number of antennas) is small.}
But, for massive MIMO, the total hardware cost and energy consumption required by the high-resolution ADCs/DACs will be a burden.
One solution  is to replace the high-precision converters 
with lower precision ones \cite{risi2014massive,choi-mo-heath-2016, MollenCLH17,li-tao-gonzalo-2017, saxena-fijalkow-swindlehurst-2017,swindlehurst-saxena-mezghani-2017,jacobsson-durisi-coldrey-2017}.

The challenge with using low-resolution ADCs/DACs is that we need to deal with the undesirable error effects caused by coarse quantization. 
In this paper we are interested in the context of multiuser massive MIMO downlink with low-resolution DACs at the transmitter.
It is important to mention that  massive MIMO uplink with low-resolution ADCs at the receiver is another key topic; the reader is referred to the literature, such as  \cite{risi2014massive,choi-mo-heath-2016,MollenCLH17,li-tao-gonzalo-2017,studer2016quantized,shao2022accelerated} and the references therein, for details.
In coarsely quantized MIMO downlink precoding, the existing studies can be taxonomized into two types, namely, the precode-then-quantize type and the direct signal design type.
The precode-then-quantize type takes a precoding scheme in the {unquantized case,} such as the {popularly-used} zero-forcing scheme, and then quantizes the precoded signals to produce the few-bit transmitted signals.
This approach is straightforward, but the precoding schemes are not designed to resist the adverse effects of quantization errors.
Performance analysis for the precode-then-quantize approach has been a subject of interest, helping us better understand the nature of coarsely quantized MIMO; see, e.g., \cite{li-tao-swindlehurst-2017, saxena-fijalkow-swindlehurst-2017, mezghani-ghiat-nossek-2009}.
The direct signal design type seeks to directly manipulate the few-bit signals by optimization, with the aim to optimize some symbol-level performance metric such as mean square error \cite{jacobsson-durisi-coldrey-2017} and symbol error probability 
\cite{Sohrabi2018,shao2018framework}.
Doing so requires us to handle a large-scale discrete optimization problem, which may not be easy.
Also, this optimization-oriented approach is, by its nature, unable to leverage our community's rich understanding of MIMO precoding in the unquantized case.  
That being said, direct signal designs have been empirically found to provide significantly better performance than the precode-then-quantize methods \cite{castaneda-jacobsson-durisi-2017, jacobsson-durisi-coldrey-2017, Sohrabi2018,shao2018framework,li-masouros-liu-2018, jedda-mezghani-swindlehurst-2018}.
The advances of direct signal designs are mostly with the one-bit case and with the related context of constant envelope precoding  \cite{kazemi-aghaeinia-2017, jedda-mezghani-swindlehurst-2018,shao2018framework, wu-liu-jiang-dai-2023}.
So far we have not seen direct signal designs for the general multi-bit case,
due possibly to the difficulty of such optimization.

The traditional precode-then-quantize approach, which directly quantizes the precoded signals, has no control with the quantization noise.
Lately, spatial Sigma-Delta ($\Sigma\Delta$) modulation has arisen as a new precode-then-quantize approach that features  quantization noise control or containment \cite{shao2019one, shao2020multiuser}.
Spatial $\Sigma\Delta$ modulation draws inspiration from temporal $\Sigma\Delta$ modulation in the classical ADC/DAC literature \cite{schreier2005understanding}. 
The basic idea is to add an error feedback loop to the quantizer so that the quantization noise is shaped toward the high frequency band.
Consequently, given a low-pass temporal signal, we can convert it to a few-bit signal whose frequency domain sees the signal and quantization noise well separated.
In spatial $\Sigma\Delta$ modulation, we turn such noise shaping idea to space.
To be specific, we consider a uniform linear transmit antenna array at the base station (BS). 
We pass the quantization noise of each antenna to the adjacent antenna, thereby forming a spatial $\Sigma\Delta$ feedback loop.
This leads to the quantization noise being pushed toward high spatial frequencies, or angles.
Consequently we can use a low angle sector to serve users, who will experience reduced quantization noise effects compared to the direct quantization case.
While this means that we cannot use the high angle sectors, it is common in practice to consider an angle sector, rather than the full angle range, due to the directivity of antennas.
As a precode-then-quantize approach, spatial $\Sigma\Delta$ modulation allows us to use precoding techniques established for the unquantized case---which is a merit.
It is worth noting that, recently, spatial $\Sigma\Delta$ modulation has also been considered for MIMO uplink 
\cite{baracspatial, Corey_Sig, pirzadeh2020spectral, rao-seco-pirzadeh-2021, sankar-chepuri-2022}.

In the previous study of $\Sigma\Delta$ MIMO downlink {\cite{shao2019one,shao2020multiuser}},
the basic first- and second-order modulators from the temporal $\Sigma\Delta$ literature
were directly applied to perform spatial $\Sigma\Delta$ modulation.
An interesting question is whether we can build $\Sigma\Delta$ modulators that are general, flexible and specifically designed for the context of multiuser massive MIMO downlink precoding. 
In this paper, we develop a $\Sigma\Delta$ modulator design framework for such a purpose. 
Our framework considers a general $\Sigma\Delta$ error-feedback  structure for any given modulator order and for any given number of quantization levels (or bits).
We design $\Sigma\Delta$ modulators by optimization.
By characterizing the signal-to-quantization-and-noise ratio (SQNR) experienced by the users, 
we formulate the $\Sigma\Delta$ modulator designs as some form of SQNR maximization problems.
The formulated problems are convex and can be conveniently solved by calling available solvers.
Our designs offer two options with quantization noise suppression, namely,
(i) quantization noise suppression over a prescribed angle sector;
and 
(ii) focused quantization noise suppression at the user angles, based on the instantaneous channel state information available at the BS.
In particular, option (ii) is a new idea.
Our framework can also be extended 
to the $2$D uniform planar antenna array setting.

We should describe the relationship of this study to the prior studies in the temporal $\Sigma\Delta$ literature.
We commonly see closed-form modulator designs in the temporal $\Sigma\Delta$ literature.
While optimization-based modulator designs do not seem to be commonplace in the ADC/DAC literature, our background research found that, curiously, optimization-based modulator designs were considered in the signal processing literature;  see \cite{nagahara2012frequency}
and the references therein.
In particular, the work by Nagahara and Yamamoto  \cite{nagahara2012frequency} is worth noting, as it provides a convex optimization framework for Chebyshev-type filter designs for $\Sigma\Delta$ noise shaping.
As we will elaborate upon in this paper, our spatial  $\Sigma\Delta$ modulator designs happen to share some similarities with the temporal $\Sigma\Delta$ modulator designs by Nagahara and Yamamoto.
We should however emphasize that, to the best of our knowledge, optimization-based designs have not been previously considered in spatial $\Sigma\Delta$ modulation for coarsely quantized MIMO precoding.
Furthermore, our design philosophy differs in that we aim at maximization of the SQNRs experienced by the users, with the MIMO application aspects taken into consideration,
while Nagahara and Yamamoto consider noise shaping.

The organization of this paper is as follows.
Section~\ref{sect:bg} reviews the background of spatial $\Sigma\Delta$ modulation for coarsely quantized massive MIMO precoding.
Section~\ref{sect:main_meat}
presents our $\Sigma\Delta$ modulator design framework. 
Section~\ref{sect:2D_sigdel}
describes the extension of our framework to the $2$D uniform planar array case.
Section~\ref{sect:sim}
provides numerical results to show how the $\Sigma\Delta$ modulators designed under our framework perform.
Section~\ref{sect:con}
concludes this work.

Our notations are as follows.
The symbols $\Rbb$, $\Cbb$ and $\mathbb{N}$ denote the sets of real numbers, complex numbers and non-negative integers, respectively.
A scalar, a column vector and a matrix are  represented by a lowercase normal letter, a lowercase boldfaced letter and a capital boldfaced letter,  respectively; e.g., $a$, $\ba$ and $\bA$, respectively.
The real and imaginary parts of a given vector $\ba$ are denoted by $\Re(\ba)$ and $\Im(\ba)$, respectively.
The transpose of a vector $\ba$ is denoted by $\ba^T$, and the same convention applies to matrices. 
The trace, inverse and pseudo-inverse of a matrix $\bA$ are denoted by ${\rm tr}(\bA)$, $\bA^{-1}$ and $\bA^\dag$, respectively.
Given a vector $\ba$, the notation $\Diag(\ba)$ denotes a diagonal matrix with the $(i,i)$th component given by the $i$th component of $\ba$.
Given a collection of scalars $a_1,\ldots,a_n$,
the notation $(a_1,\ldots,a_n)$ denotes the concatenation of the $a_i$'s as a vector, i.e., $(a_1,\ldots,a_n)= [~ a_1,\ldots,a_n ~]^T$.
The same convention applies when the $a_i$'s are vectors.
We denote $\jj= \sqrt{-1}$.
Given a sequence $\{ a_n \}_{n \in \setN}$,
where $\setN$ equals either $\mathbb{N}$ or $\{0,1,\ldots,N-1\}$ for some positive integer $N$,
the Fourier transform of $\{ a_n \}_{n \in \setN}$ is denoted by $A(\omega)= \sum_{n\in\setN} a_n e^{-\jj n \omega}$.
Given a vector $\ba$,
the notations $\| \ba \|_1$, $\| \ba \|_2$ and $\| \ba \|_\infty$ denote the $1$-norm, Euclidean norm and $\infty$-norm of $\ba$, respectively. 
Given a complex vector $\ba$, 
the notations $\| \ba \|_{{\sf IQ}-1}$ and $\| \ba \|_{{\sf IQ}-\infty}$ denote the $1$-norm and $\infty$-norm with respect to $(\Re(\ba),\Im(\ba))$, respectively; i.e.,
$\| \ba \|_{{\sf IQ}-1} = \| (\Re(\ba),\Im(\ba)) \|_1$ and $\| \ba \|_{{\sf IQ}-\infty} = \| (\Re(\ba),\Im(\ba)) \|_\infty$.
The same conventions apply to matrices.

%% file: example-UT.tex
\begin{figure*}[t!] 
\begin{minipage}[c]{\linewidth}
\centering
\begin{subfigure}[t]{0.33\textwidth}
       \centerline{\includegraphics[width=\textwidth]{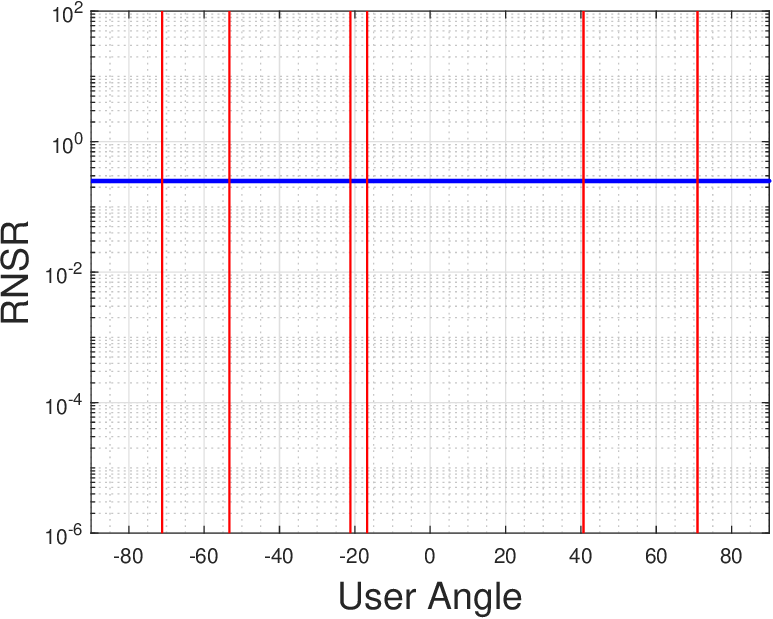}}\vspace{-0pt}
       \caption{$M = 2$}
\end{subfigure}\hfill
\begin{subfigure}[t]{0.33\textwidth}
       \centerline{\includegraphics[width=\textwidth]{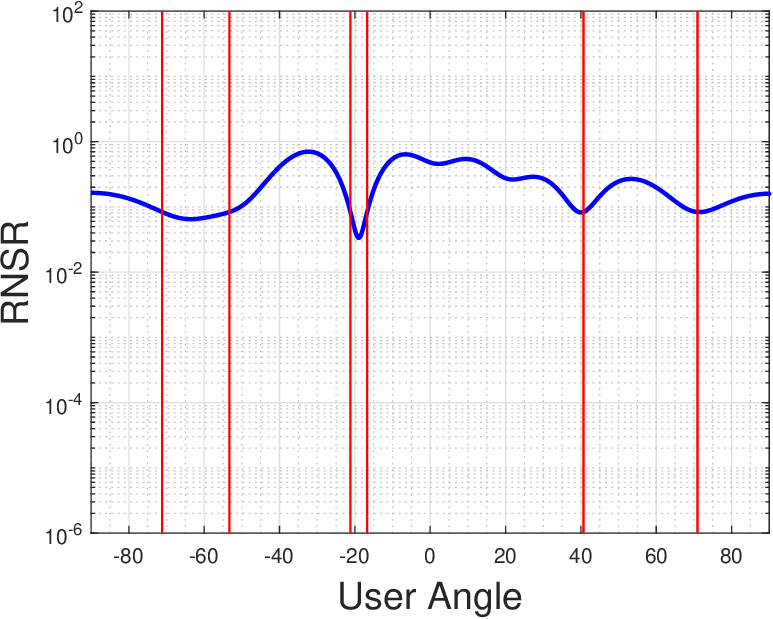}}\vspace{-0pt}
       \caption{$M = 3$}
\end{subfigure}\hfill
\begin{subfigure}[t]{0.33\textwidth}
       \centerline{\includegraphics[width=\textwidth]{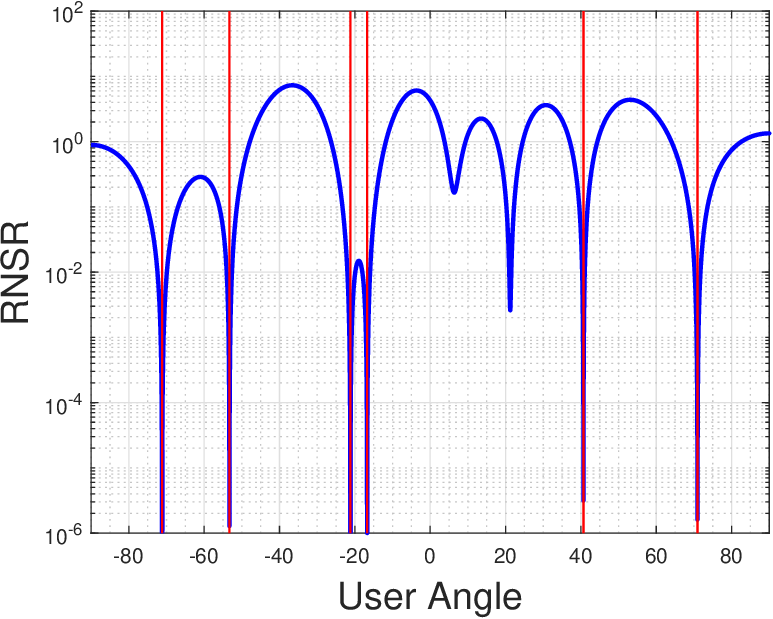}}\vspace{-0pt}
       \caption{$M = 4$}
\end{subfigure}\\
\begin{subfigure}[t]{0.33\textwidth}
       \centerline{\includegraphics[width=\textwidth]{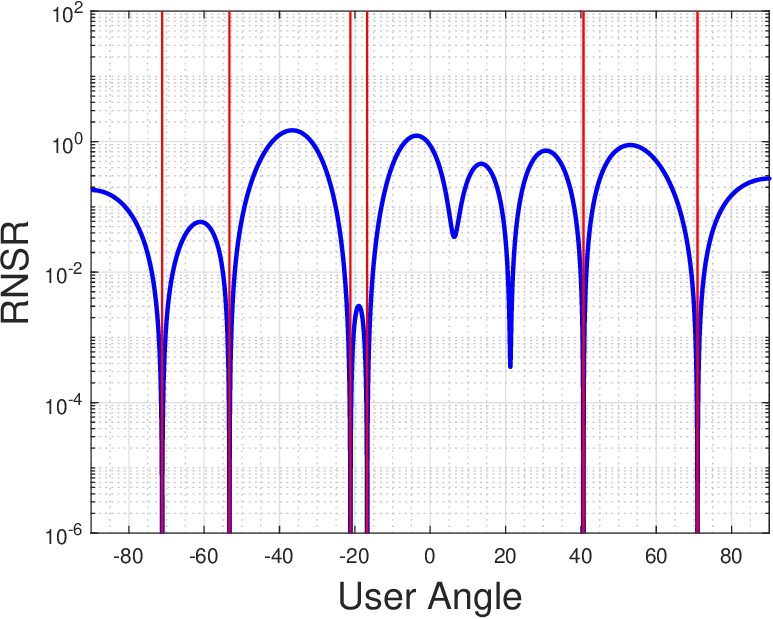}}\vspace{-0pt}
       \caption{$M = 5$}
\end{subfigure}\hfill
\begin{subfigure}[t]{0.33\textwidth}
       \centerline{\includegraphics[width=\textwidth]{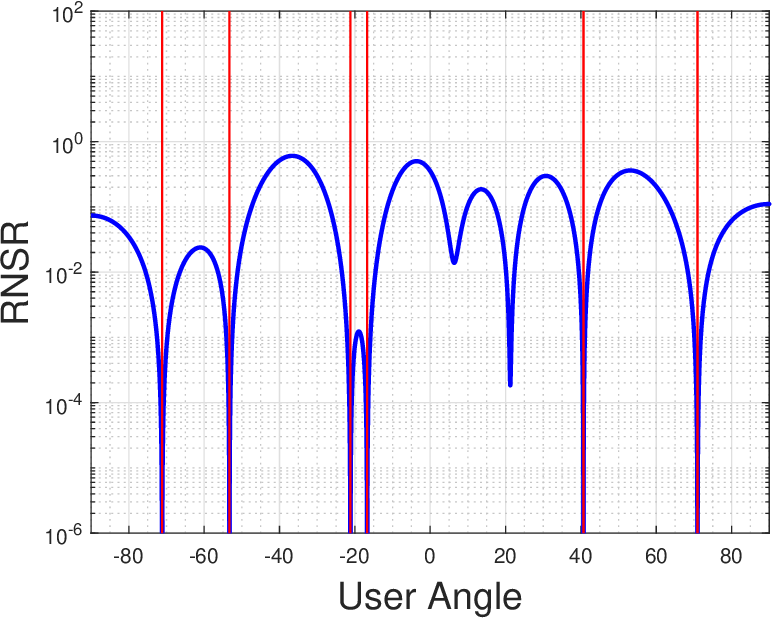}}\vspace{-0pt}
       \caption{$M = 6$}
\end{subfigure}\hfill
\begin{subfigure}[t]{0.33\textwidth}
       \centerline{\includegraphics[width=\textwidth]{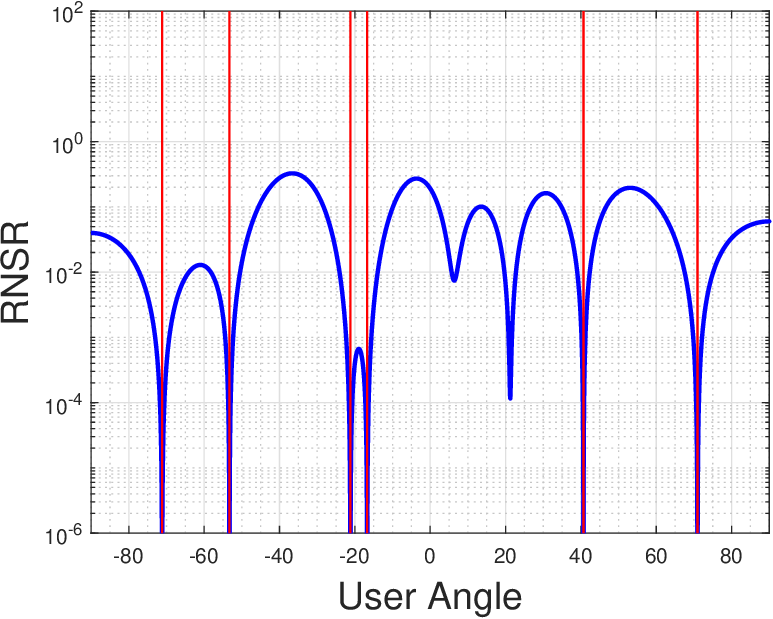}}\vspace{-0pt}
       \caption{$M = 7$}
\end{subfigure}\\
\end{minipage}
\caption{Illustration of the relative noise-shaping responses of the $\Sigma\Delta$ modulators designed by the user-targeted formulation \eqref{eq:MMF}. Red line: user angles.}\label{fig:FiltDes-UsrAdp-8-usr}
\end{figure*}

{Let us provide some numerical illustration.
Figure~\ref{fig:FiltDes-UsrAdp-8-usr} shows the relative noise shaping responses, defined as  
\begin{equation}\label{eqn:RNSR}
    {\rm RNSR}(\theta) = \frac{|1+G(\frac{2\pi d}{\lambda} \sin(\theta))|^2}{A^2},
\end{equation}
of the user-targeted design \eqref{eq:MMF}.
The red vertical lines in the figure indicate the user angles, and the system settings are $N = 1024, K = 6, L = 16, d = \lambda/4, |\alpha_i| = 1$ for all $i$, and $\sigma_\eta^2 = 0$.
We see that, as the quantization level number $M$ increases, the user-targeted design provides sharper notches, and therefore better quantization noise suppression, at the user angles. 
}

%% file: example-FS.tex
\begin{figure*}[t!] 
\centering
\begin{subfigure}[t]{0.33\textwidth}
       \centerline{\includegraphics[width=\textwidth]{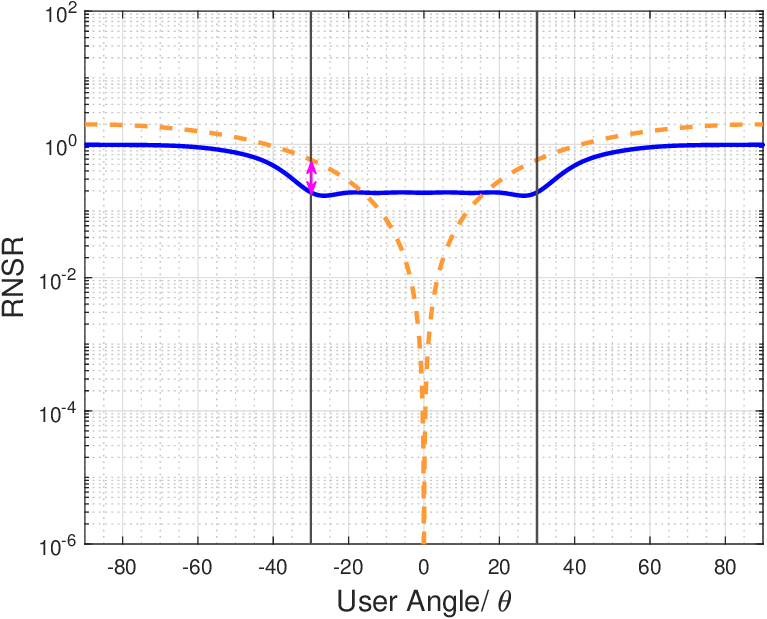}}\vspace{-0pt}
       \caption{$M = 2$}
\end{subfigure}\hfill
\begin{subfigure}[t]{0.33\textwidth}
       \centerline{\includegraphics[width=\textwidth]{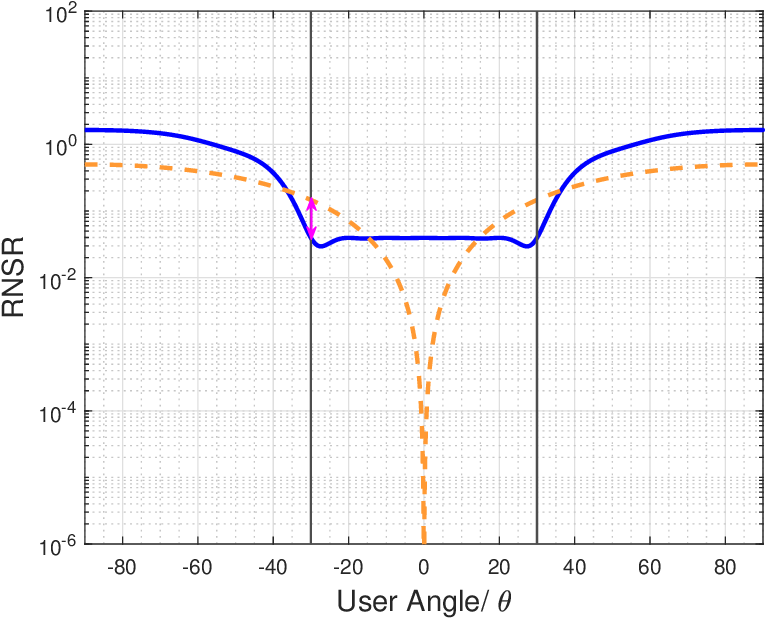}}\vspace{-0pt}
       \caption{$M = 3$}
\end{subfigure}\hfill
\begin{subfigure}[t]{0.33\textwidth}
       \centerline{\includegraphics[width=\textwidth]{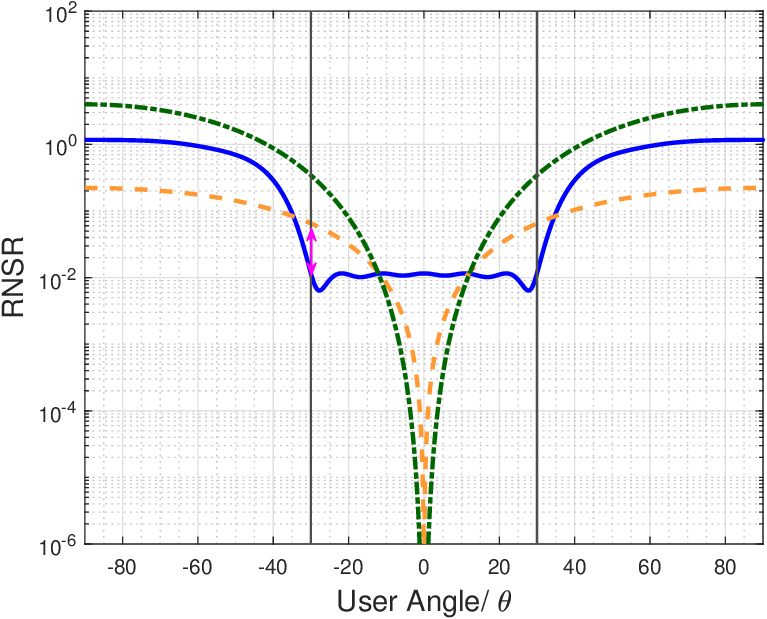}}\vspace{-0pt}
       \caption{$M = 4$}
\end{subfigure}\\
\begin{subfigure}[t]{0.33\textwidth}
       \centerline{\includegraphics[width=\textwidth]{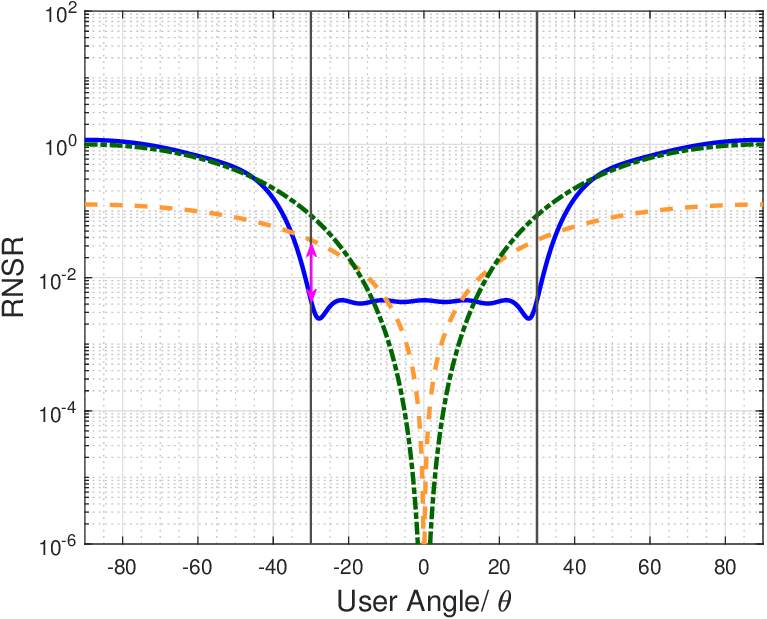}}\vspace{-0pt}
       \caption{$M = 5$}
\end{subfigure}\hfill
\begin{subfigure}[t]{0.33\textwidth}
       \centerline{\includegraphics[width=\textwidth]{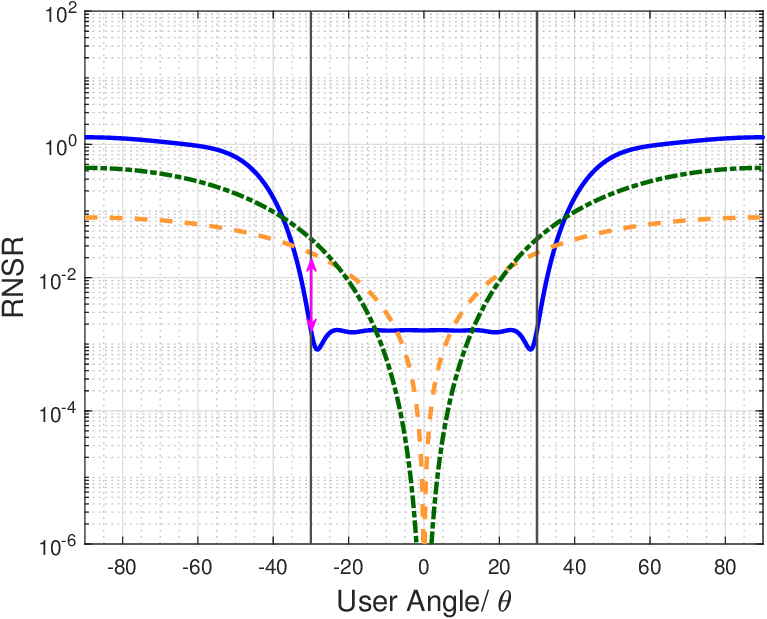}}\vspace{-0pt}
       \caption{$M = 6$}
\end{subfigure}\hfill
\begin{subfigure}[t]{0.33\textwidth}
       \centerline{\includegraphics[width=\textwidth]{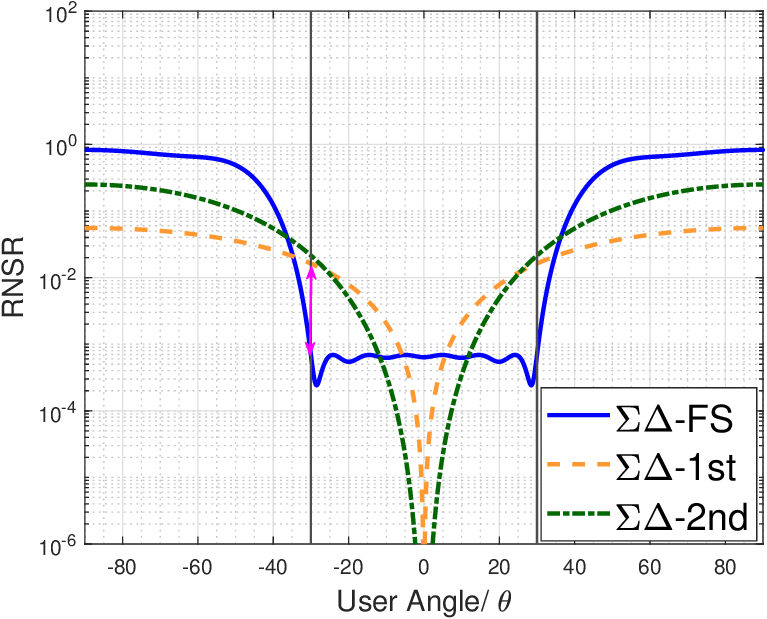}}\vspace{-0pt}
       \caption{$M = 7$}
\end{subfigure}\\
\caption{Illustration of the relative noise-shaping responses of the $\Sigma\Delta$ modulators designed by the fixed sector formulation \eqref{eq:MMF_alt2}. Black line: the boundary of the angle sector $[\theta_l,\theta_u]$. $\Sigma\Delta$-FS: the fixed-sector design \eqref{eq:MMF_alt2}. $\Sigma\Delta$-1st: the standard first-order $\Sigma\Delta$ modulator. $\Sigma\Delta$-2nd: the standard second-order $\Sigma\Delta$ modulator.  }\label{fig:FiltDes}
\end{figure*}

We give a numerical illustration by plotting the relative noise shaping responses of the fixed-sector design \eqref{eq:MMF_alt2} in Figure~\ref{fig:FiltDes}. 
To benchmark, we also plot the relative noise shaping responses of the first-order and second-order $\Sigma\Delta$ modulators in Examples~\ref{exa:1st} and \ref{exa:2nd}. The settings are $N = 1024, L = 16, d = \lambda/4, \sigma_\eta^2 = 0$, $r_{\rm min} = r_{\rm max} = 1$, and $[\theta_l, \theta_u] = [-30^\circ, 30^\circ]$. 
The first- and second-order $\Sigma\Delta$ modulators have the {maximum} input amplitude $A$ set to be the largest under the no-overload condition, i.e. $A = M- 1$ and $A = M - 3$, respectively (see Examples \ref{exa:1st} and \ref{exa:2nd}).
In the plots in Figure~\ref{fig:FiltDes}, the vertical black lines indicate the angle sector. The magenta double-headed arrows indicate the gap between the worst-case relative noise-shaping response, $\max_{\theta\in[\theta_l, \theta_u]} {\rm RNSR}(\theta)$, of the fixed-sector design and the worst-case relative noise-shaping response of the first-order and second-order $\Sigma\Delta$ modulators. 
We see that the fixed-sector design provides a uniform quantization noise suppression over the angle sector of interest. We also see that, for larger quantization level numbers $M$'s, the fixed-sector design provides considerably improved quantization noise suppression in an angle-sector uniform sense.

%% file: example-2D.tex
\begin{figure*}[tp!] 
\begin{minipage}[c]{\linewidth}
 \begin{subfigure}[t]{0.49\textwidth}
       \centerline{\includegraphics[width=\textwidth]{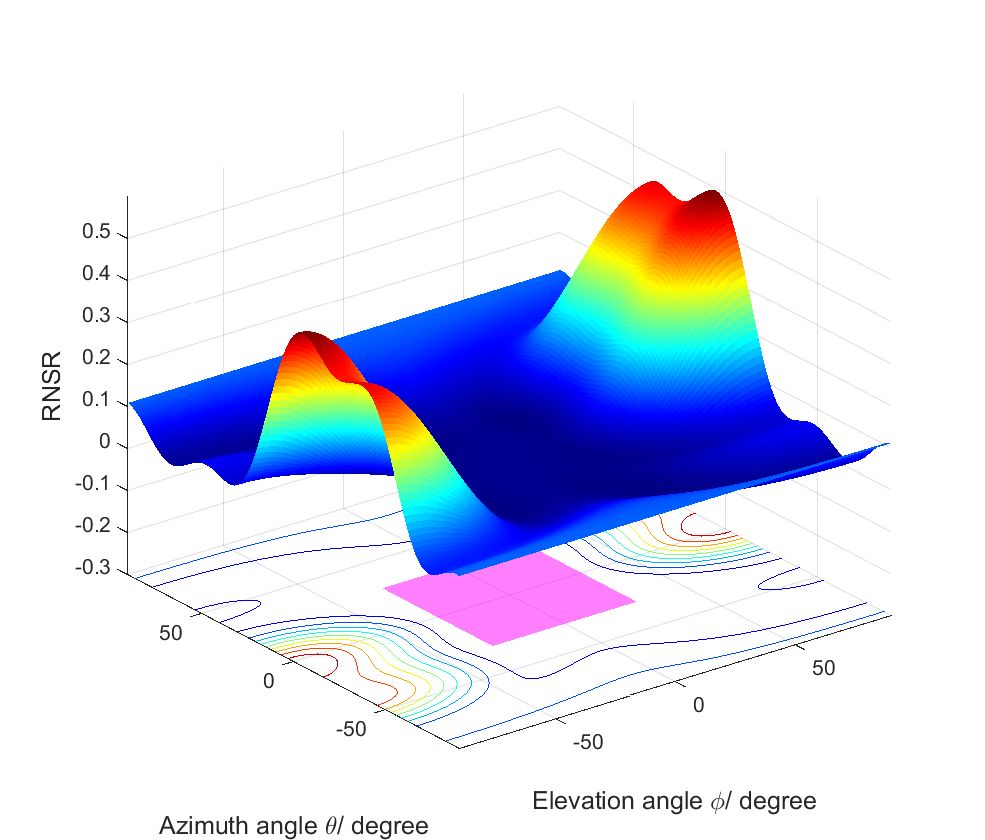}}\vspace{-0pt}
       \caption{The fixed-sector design \eqref{eq:MMF_fixed_2D}.}
\end{subfigure}\hfill
\begin{subfigure}[t]{0.49\textwidth}
       \centerline{\includegraphics[width=\textwidth]{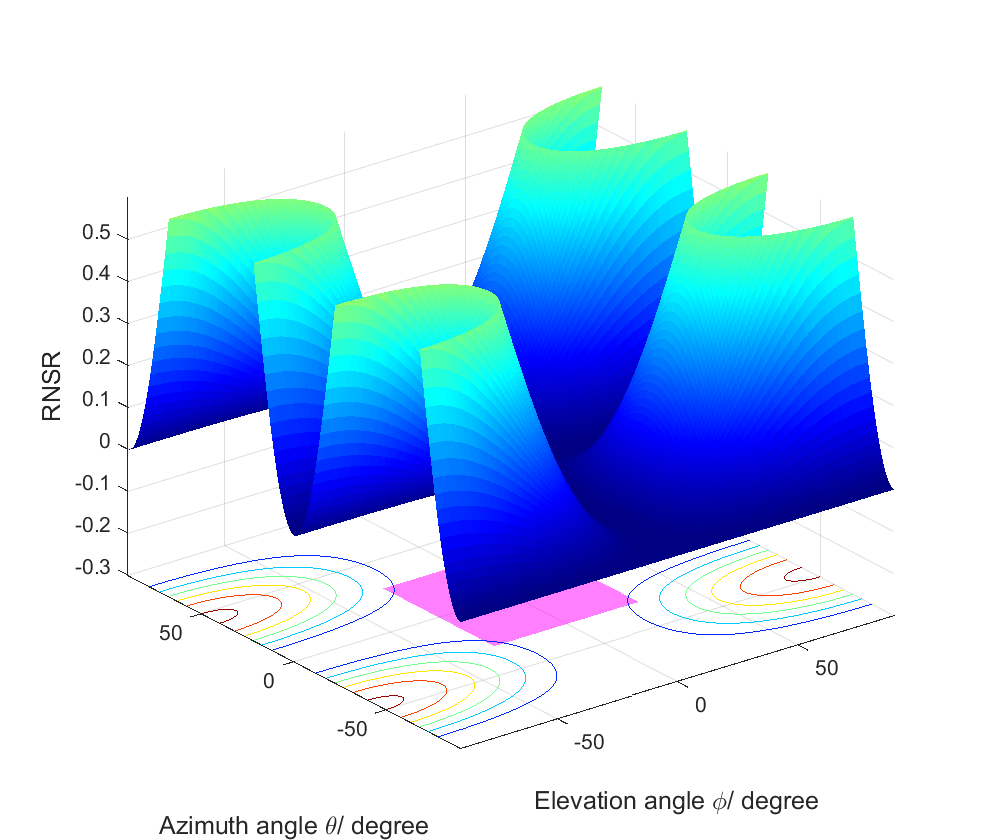}}\vspace{-0pt}
       \caption{The first-order $\Sigma\Delta$ modulator.}
\end{subfigure}
\end{minipage}
\caption{Illustration of the relative noise-shaping response in the 2D case.} \label{fig:FDFilter}
\end{figure*}

{As an illustration, Figure~\ref{fig:FDFilter}(a) plots the relative noise shaping response of the fixed-sector design \eqref{eq:MMF_fixed_2D}.}
The settings are $(N_1, N_2)=(60,60), d_1 = d_2 = \lambda/4, (L_1, L_2) = (4,4)$, $r_{\rm max} = r_{\rm min} = 1, \sigma_\eta^2 = 0, [\theta_l, \theta_u]\times[\phi_l, \phi_u] = [-30^\circ, 30^\circ] \times  [-30^\circ, 30^\circ], M = 4$.
{To benchmark, we also consider} 
a 2D first-order $\Sigma\Delta$ modulator whose shaping response is 
\begin{equation}\label{eq:1st-2D-noise-shaping}
    1+G(\omega_1, \omega_2) = (1-e^{-\jj\omega_1})(1-e^{-\jj\omega_2}),
\end{equation}
and whose coefficients are $(g_{01}, g_{10}, g_{11}) = (-1,-1,1)$; we set $A=M-3$, the maximum under the no-overload condition. 
{The relative noise shaping response of this first-order modulator is plotted in Figure~\ref{fig:FDFilter}(b).
Comparing Figures~\ref{fig:FDFilter}(a) and \ref{fig:FDFilter}(b), the fixed-sector design \eqref{eq:MMF_fixed_2D} appears to provide better quantization noise suppression over the given angle sector than the first-order modulator.}

%% file: simResult.tex
\section{Numerical Results}
\label{sect:sim}

In this section we provide numerical results.
We simulate both spatial $\Sigma\Delta$ modulation and precoding at the signal level,
and we evaluate  users' bit error rates (BERs) as our way to assess the performance of our method.
The symbol stream $\{ s_{i,t} \}_{t=1}^T$ of each user is drawn from the $64$-QAM constellation, with symbol stream length $T= 500$.
The precoding scheme is the ZF scheme. 
To be specific, for a given spatial $\Sigma\Delta$ modulator, the ZF precoded signals are given by 
\beq \label{eq:sd_zf}
\bar{\bx}_t = \frac{A}{C} \bH^\dag \bD \bs_t,
~ t=1,\ldots,T,
\eeq 
where $C= \max_{t=1,\ldots,T} \| \bH^\dag \bD \bs_t \|_{{\sf IQ}-\infty}$, $\bD= \Diag(\sigma_{w,1},\ldots,\sigma_{w,K})$,
$\sigma_{w,i}^2 = \rho |\alpha_i|^2 \Exp[ |v_{i,t} |^2] + \sigma_\eta^2$, and $\Exp[ |v_{i,t} |^2] = 2N |1+G(\omega_i)|^2/3$;
see \cite{shao2019one}.
Note that we scale the symbol streams such that the post-precoding SNRs of all the users are equal, and that the normalization with $C$ is to enforce the peak signal amplitude constraint $\| \bar{\bx}_t \|_{{\sf IQ}-\infty} \leq A$ for all $t$.
For each symbol time $t$, the precoded signal $\bar{\bx}_t$ is fed to the $\Sigma\Delta$ modulator to generate the transmitted signal $\bx_t$.
Our $\Sigma\Delta$ modulation scheme is designed either by the fixed-sector design in Section~\ref{sect:main_meat}-\ref{sect:SQNR_max_sect} or by the user-targeted design in  Section~\ref{sect:main_meat}-\ref{sect:SQNR_max}.
As benchmarks, we also consider the first- and second-order $\Sigma\Delta$ modulation schemes in Examples~\ref{exa:1st} and \ref{exa:2nd}, respectively, which are standard modulators in the $\Sigma\Delta$ literature.
{The maximum input signal amplitude} $A$ of the first- and second-order  modulators is set to be the largest under the no-overload condition, which are $A=M-1$ and $A=M-3$ for the first- and second-order modulators, respectively.
In addition we benchmark the direct quantization method.
We employ the ZF precoding scheme, to be consistent with our benchmarking, and the transmitted signals for the directly quantized ZF scheme are given by
\[
\bx_t = \mathcal{Q}_c \left(M \frac{1}{C} \bH^\dag  \bs_t \right),
~ t=1,\ldots,T,
\]
where $C= \max_{t=1,\ldots,T} \| \bH^\dag \bD \bs_t \|_{{\sf IQ}-\infty}$.
Furthermore we provide a performance baseline by evaluating the BER performance of the following unquantized ZF scheme:
\beq \label{eq:nq_zf}
{\bx}_{t} = \frac{M-1}{C} \bH^\dag \bs_t,
~ t=1,\ldots,T,
\eeq 
where $C= \max_{t=1,\ldots,T} \| \bH^\dag \bs_t \|_{{\sf IQ}-\infty}$.
Note that this unquantized scheme satisfies $\| \bx_t \|_{{\sf IQ}-\infty} \leq M-1$ for all $t$, which complies with the peak signal amplitude constraint for the coarsely quantized case.
We define the  SNR as ${\sf SNR} = (M-1)^2\rho/ \sigma_{\eta}^2$, which is the ratio of the per-antenna peak power to the background noise power.

The BER performance to be reported was obtained by Monte-Carlo simulations with 1,000 trials.
At each trial, the user angles $\theta_i$'s and the complex channnel gains $\alpha_i$'s are generated by the following way.
The user angles $\theta_i$'s are randomly drawn from a prespecified angle sector $[\theta_l, \theta_u]$, and they are separated by no less than $1^\circ$.
The phases of $\alpha_i$'s are uniformly drawn from $[-\pi, \pi]$.
The amplitudes of $\alpha_i$'s are  generated by $|\alpha_i| = r_0/r_1$, where $r_0 = 30$ and $r_1$ is randomly drawn from $[20, 100]$.

\subsection{Fixed-Sector Design}

\begin{figure*}[tp!]
     \begin{minipage}[c]{\linewidth}
    \begin{subfigure}[t]{0.245\textwidth}
       \centerline{\includegraphics[width=\textwidth]{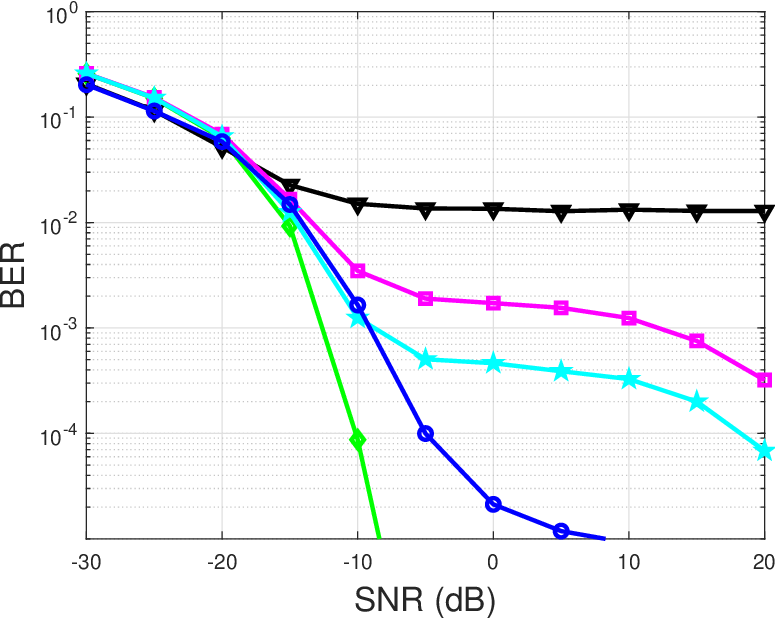}}
       \caption{ $M\!=\!4, |\theta|\leq 30^\circ$}
    \end{subfigure}\hfill 
\begin{subfigure}[t]{0.245\textwidth}
       \centerline{\includegraphics[width=\textwidth]{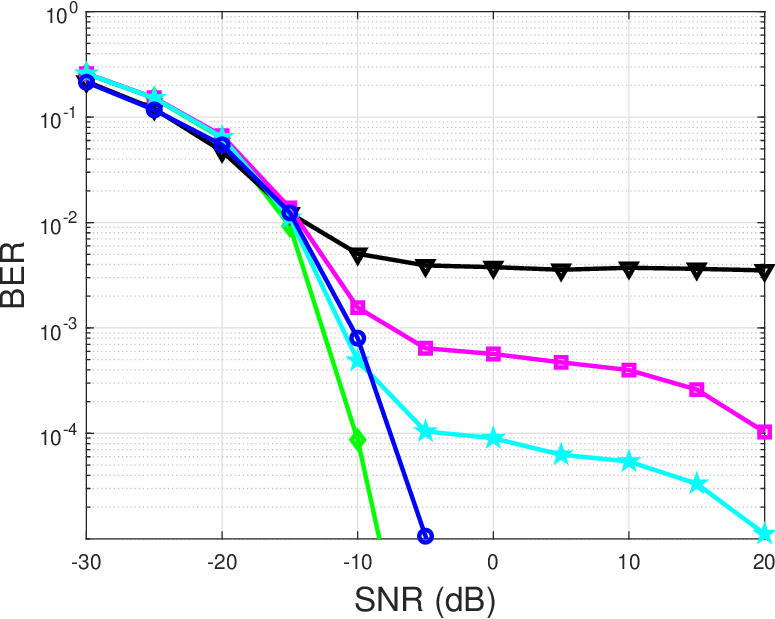}}
       \caption{$M\!=\!5, |\theta|\leq 30^\circ$}
    \end{subfigure}\hfill
       \begin{subfigure}[t]{0.245\textwidth}
       \centerline{\includegraphics[width=\textwidth]{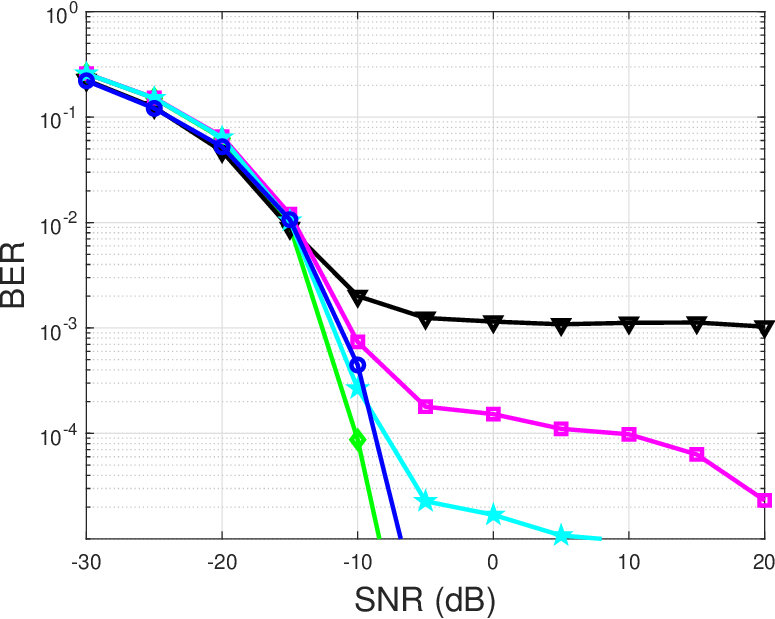}}
       \caption{$M\!=\!6, |\theta|\leq 30^\circ$}
    \end{subfigure}\hfill
          \begin{subfigure}[t]{0.245\textwidth}
       \centerline{\includegraphics[width=\textwidth]{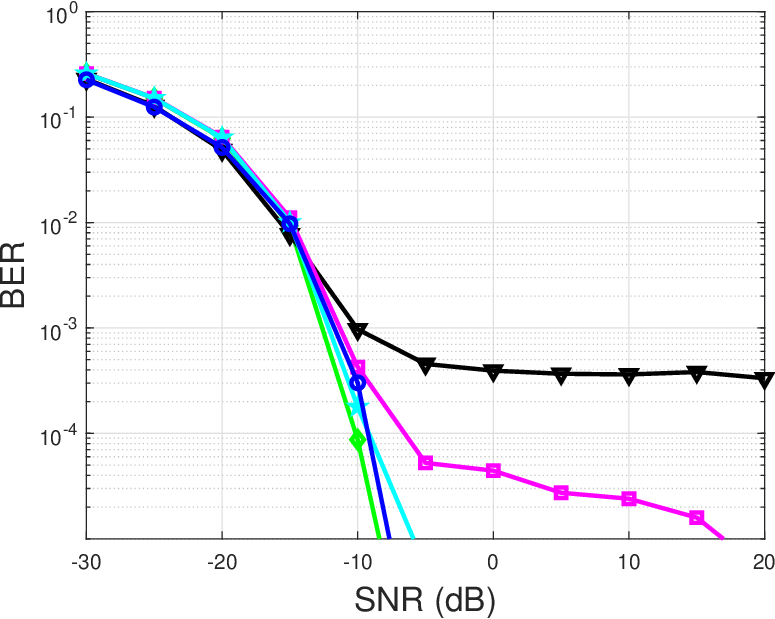}}
       \caption{$M\!=\!7, |\theta|\leq 30^\circ$}
    \end{subfigure}  \\\vspace{10pt}\\
    \begin{subfigure}[t]{0.245\textwidth}
       \centerline{\includegraphics[width=\textwidth]{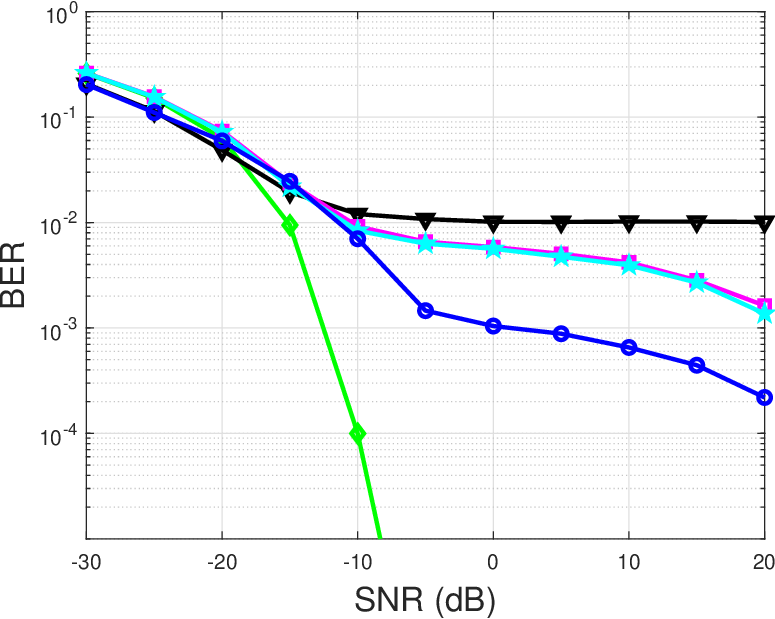}}
       \caption{ $M\!=\!4, |\theta|\leq 45^\circ$}
    \end{subfigure}\hfill 
\begin{subfigure}[t]{0.245\textwidth}
       \centerline{\includegraphics[width=\textwidth]{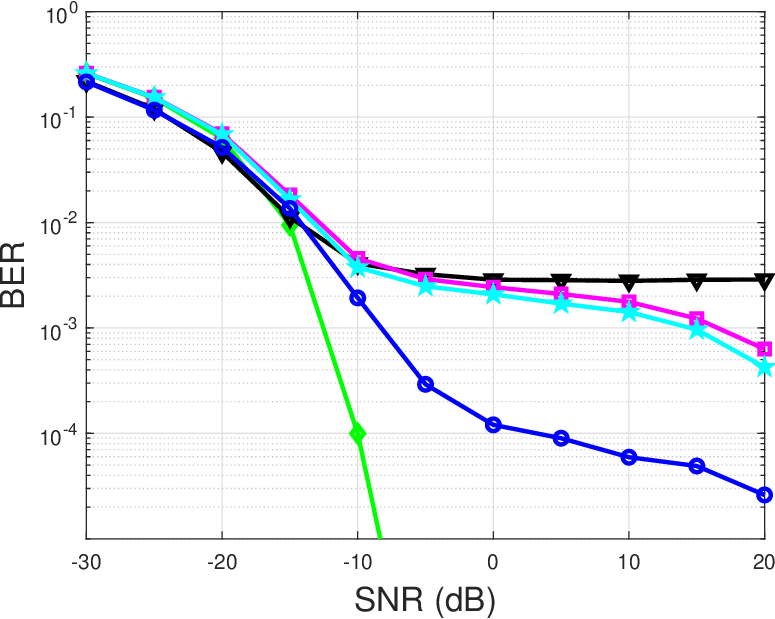}}
       \caption{$M\!=\!5, |\theta|\leq 45^\circ$}
    \end{subfigure}\hfill
       \begin{subfigure}[t]{0.245\textwidth}
       \centerline{\includegraphics[width=\textwidth]{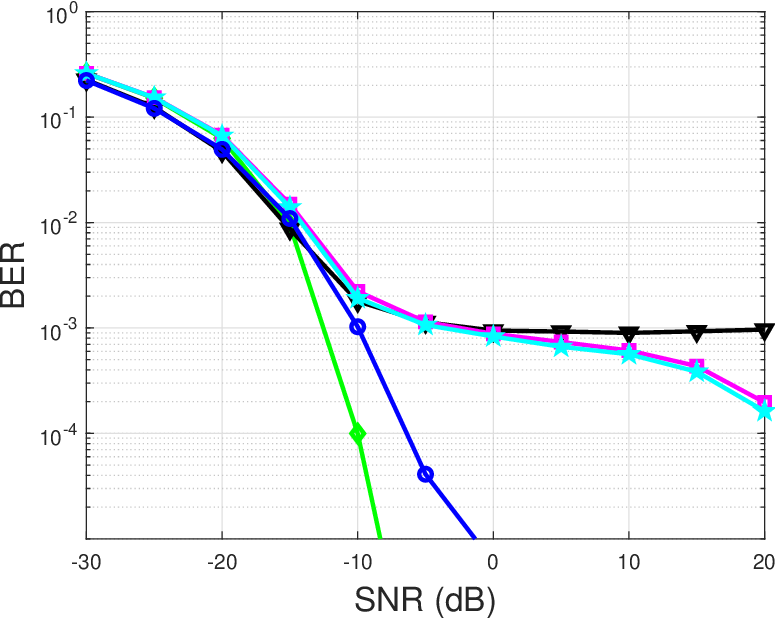}}
       \caption{$M\!=\!6,|\theta|\leq 45^\circ$}
    \end{subfigure}\hfill
          \begin{subfigure}[t]{0.245\textwidth}
       \centerline{\includegraphics[width=\textwidth]{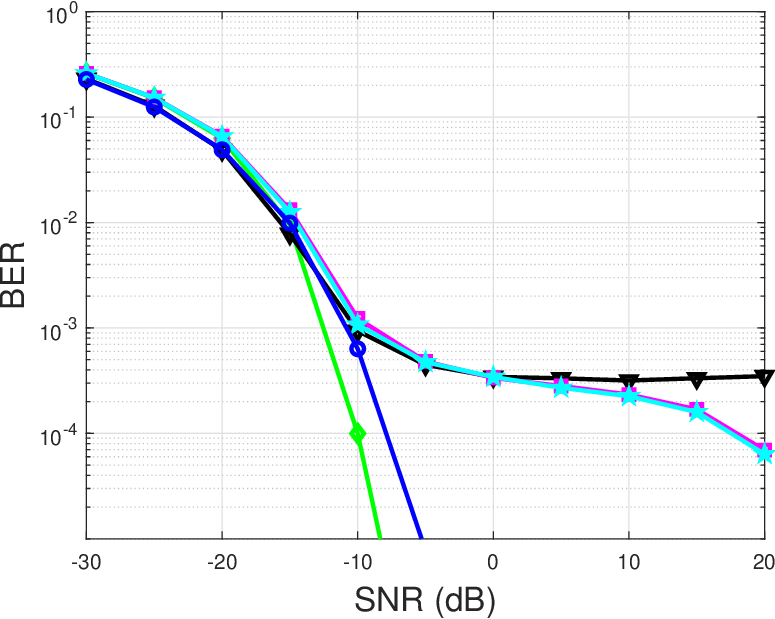}}
       \caption{$M\!=\!7, |\theta|\leq 45^\circ$}
    \end{subfigure}  \\\vspace{10pt}\\
    \begin{subfigure}[t]{0.245\textwidth}
       \centerline{\includegraphics[width=\textwidth]{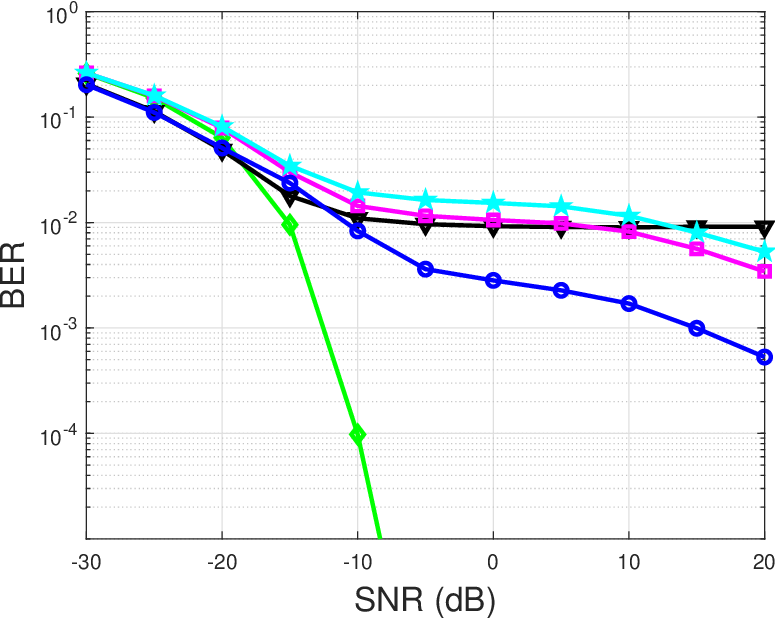}}
       \caption{$M\!=\!4, |\theta|\leq 60^\circ$}
    \end{subfigure}\hfill
        \begin{subfigure}[t]{0.245\textwidth}
       \centerline{\includegraphics[width=\textwidth]{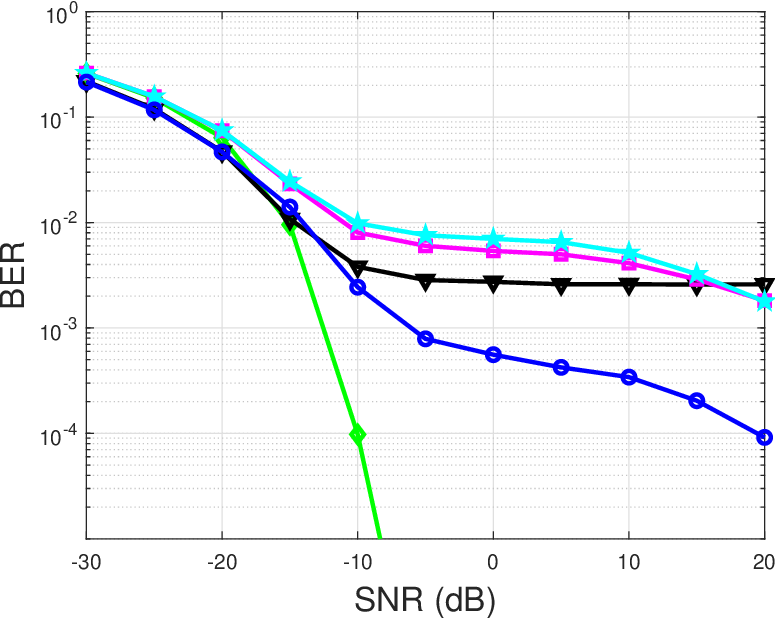}}
       \caption{$M\!=\!5, |\theta|\leq 60^\circ$}
    \end{subfigure}\hfill
    \begin{subfigure}[t]{0.245\textwidth}
       \centerline{\includegraphics[width=\textwidth]{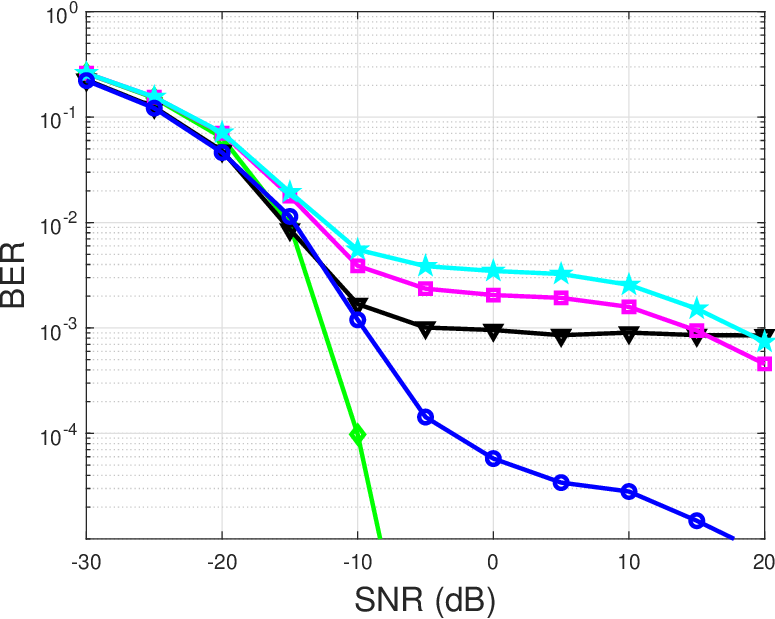}}
       \caption{$M\!=\!6, |\theta|\leq 60^\circ$}
    \end{subfigure}\hfill
    \begin{subfigure}[t]{0.245\textwidth}
       \centerline{\includegraphics[width=\textwidth]{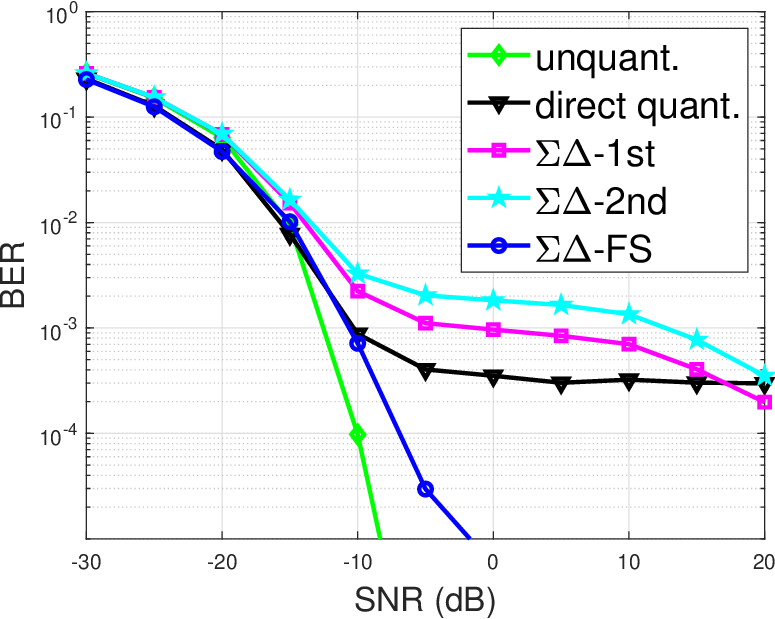}}
       \caption{$M\!=\!7, |\theta|\leq 60^\circ$}
    \end{subfigure}
\end{minipage}
\caption{BER performance of the fixed-sector optimized $\Sigma\Delta$ modulation scheme for various settings of the angle sector $[\theta_l,\theta_u]$. $N = 1024$, $d = \lambda/4$, $K=8$, $L = 16$, $|\theta| \leq \vartheta$ means that the angle sector is $[\theta_l,\theta_u]= [-\vartheta,\vartheta]$.
$\Sigma\Delta$-FS: the fixed-sector $\Sigma\Delta$ modulation scheme,
$\Sigma\Delta$-1st: the first-order $\Sigma\Delta$ modulation scheme,
$\Sigma\Delta$-2nd: the second-order $\Sigma\Delta$ modulation scheme,
direct quant.: the direct quantization scheme, 
unquant.: the unquantized performance baseline.
}\label{fig:sectorpass-baseband}
\end{figure*}

We consider the fixed-sector design in Section~\ref{sect:main_meat}-\ref{sect:SQNR_max_sect}.
Here are the settings:
The number of transmit antennas is $N=1024$;
the inter-antenna spacing is $d= \lambda/4$ {\revcolor (we remind the reader that a small $d$ leads to a small quantization noise power, as described in Section~\ref{sect:bg}-\ref{sect:spatial_sd})};
the number of users is $K= 8$;
the filter order of our fixed-sector optimized $\Sigma\Delta$ modulator is $L=16$.
Figure \ref{fig:sectorpass-baseband} shows the BER-versus-SNR plots of our scheme and the benchmarked schemes for various settings of the quantization level number $M$ and the angle sector  $[\theta_l, \theta_u]$.
In particular, Figures \ref{fig:sectorpass-baseband}(a)-(d) consider $[\theta_l, \theta_u] = [-30^\circ,30^\circ]$,
Figures \ref{fig:sectorpass-baseband}(e)-(h) consider  $[\theta_l, \theta_u] = [-45^\circ, 45^\circ]$,
and Figures \ref{fig:sectorpass-baseband}(i)-(l) consider
$[\theta_l, \theta_u] = [-60^\circ, 60^\circ]$.
We see that our fixed-sector optimized $\Sigma\Delta$ modulation scheme generally leads to better BER performance than the benchmarked schemes.
We also see that, as the width of the angle sector increases, we need a larger quantization level number $M$ to provide the same or similar BER performance level.

\begin{figure*}[t] 
\begin{minipage}[c]{\linewidth}
  \centering
   \begin{subfigure}[t]{0.245\textwidth}
       \centerline{\includegraphics[width=\textwidth]{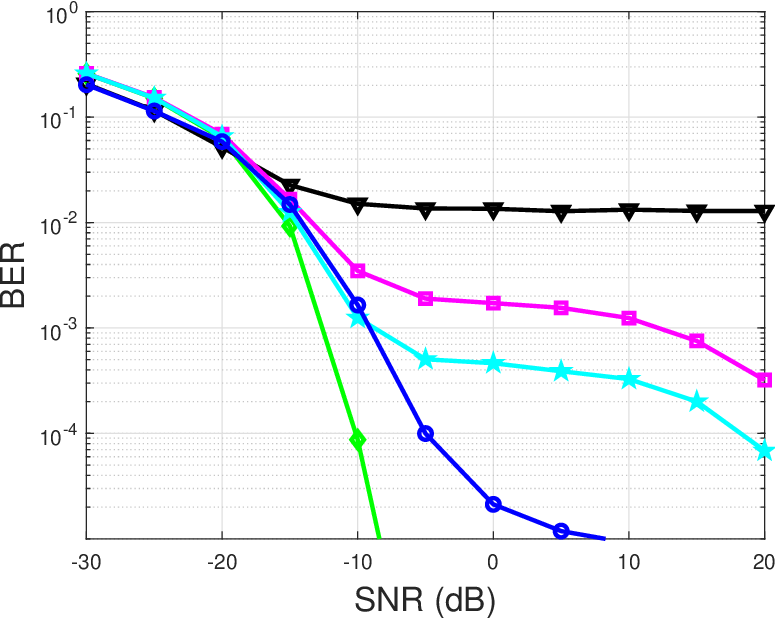}}
       \caption{ $M\!=\!4, K\!=\!8$}
    \end{subfigure}\hfill 
\begin{subfigure}[t]{0.245\textwidth}
       \centerline{\includegraphics[width=\textwidth]{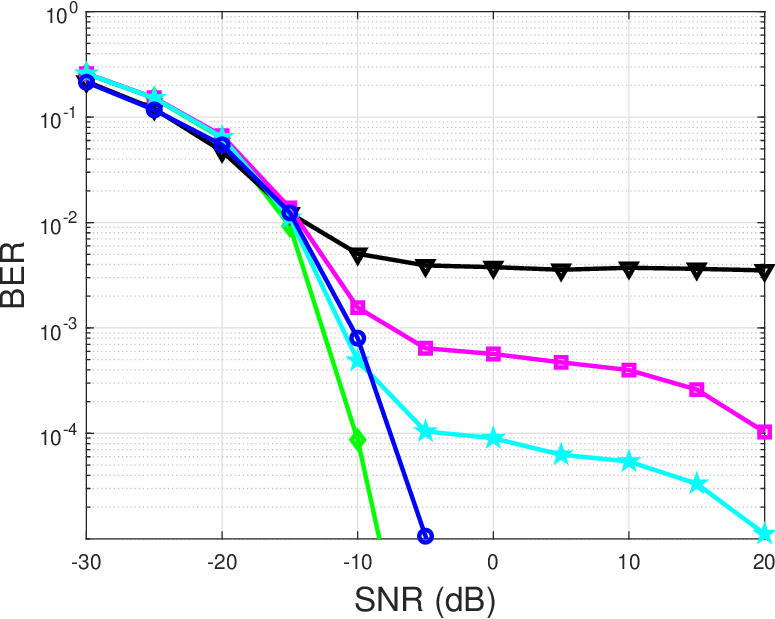}}
       \caption{$M\!=\!5, K\!=\!8$}
    \end{subfigure}\hfill
       \begin{subfigure}[t]{0.245\textwidth}
       \centerline{\includegraphics[width=\textwidth]{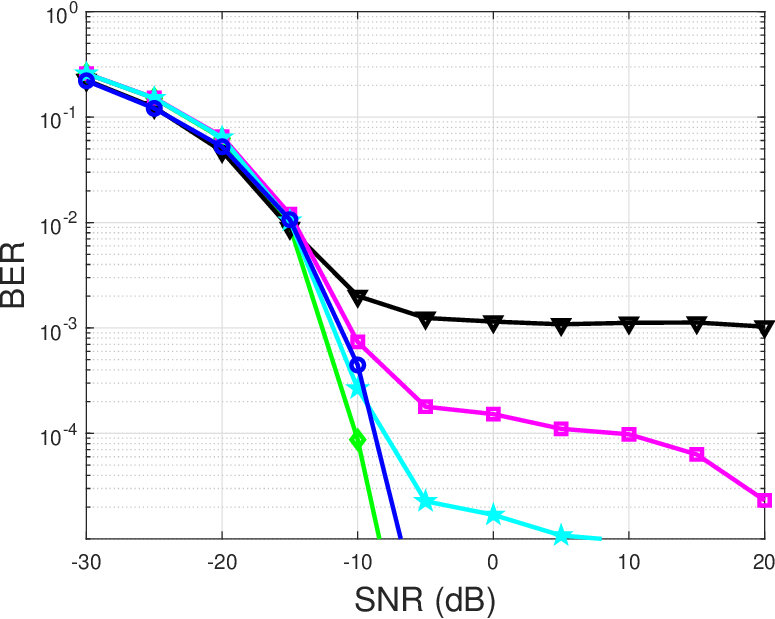}}
       \caption{$M\!=\!6, K\!=\!8$}
    \end{subfigure}\hfill
          \begin{subfigure}[t]{0.245\textwidth}
       \centerline{\includegraphics[width=\textwidth]{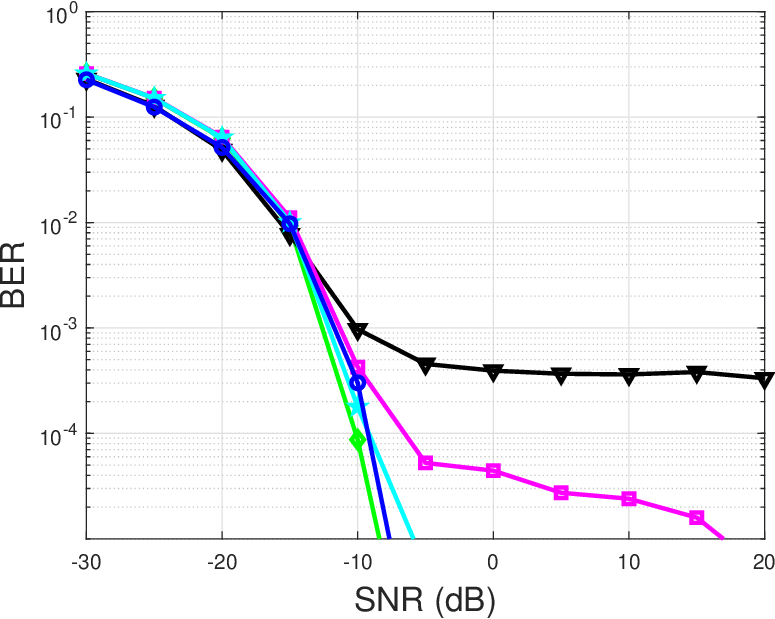}}
       \caption{$M\!=\!7, K\!=\!8$}
    \end{subfigure} \\ \vspace{10pt} 
    \begin{subfigure}[t]{0.245\textwidth}
       \centerline{\includegraphics[width=\textwidth]{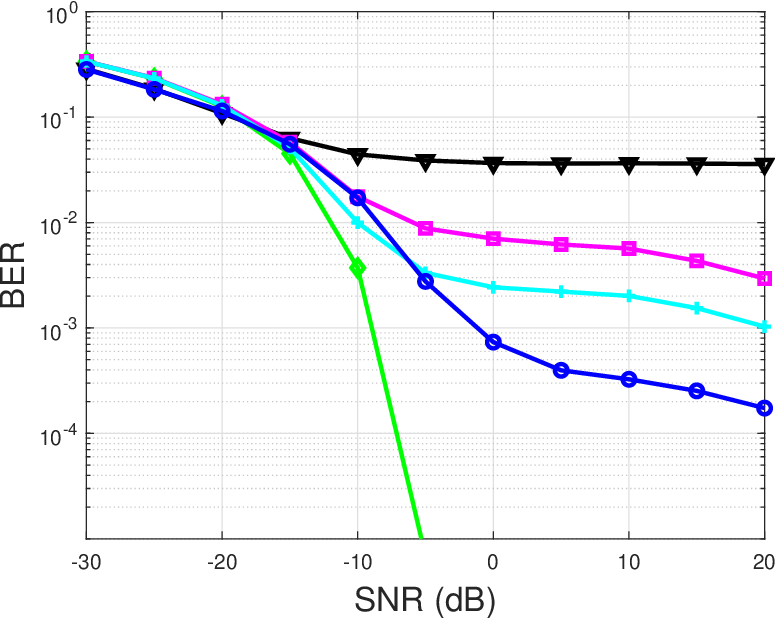}}
       \caption{$M\!=\!4, K\!=\!16 $}
    \end{subfigure}\hfill
        \begin{subfigure}[t]{0.245\textwidth}
       \centerline{\includegraphics[width=\textwidth]{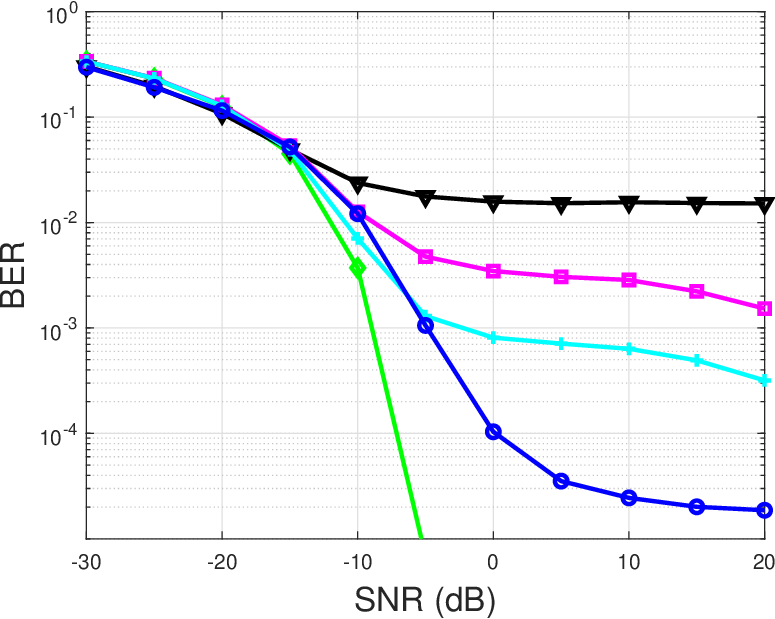}}
       \caption{$M\!=\!5,K\!=\!16 $}
    \end{subfigure}\hfill
    \begin{subfigure}[t]{0.245\textwidth}
       \centerline{\includegraphics[width=\textwidth]{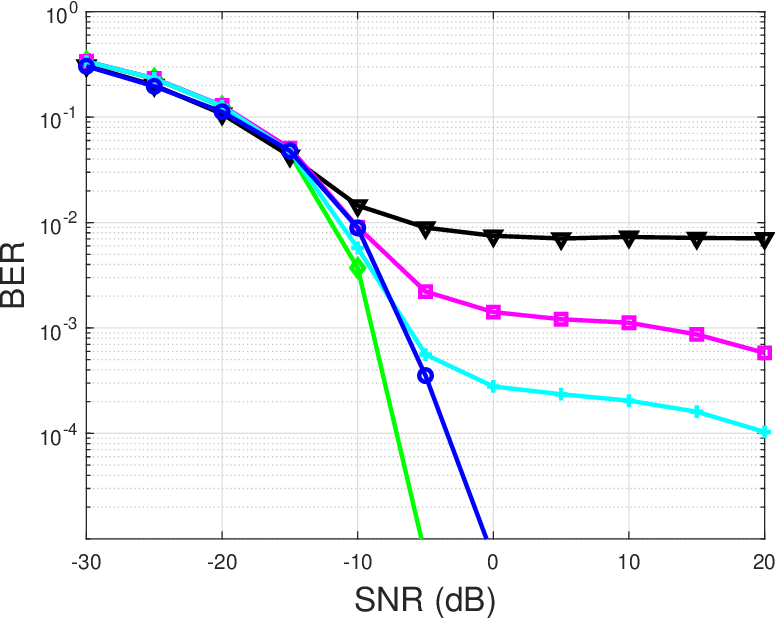}}
       \caption{$M\!=\!6,K\!=\!16 $}
    \end{subfigure}\hfill
    \begin{subfigure}[t]{0.245\textwidth}
       \centerline{\includegraphics[width=\textwidth]{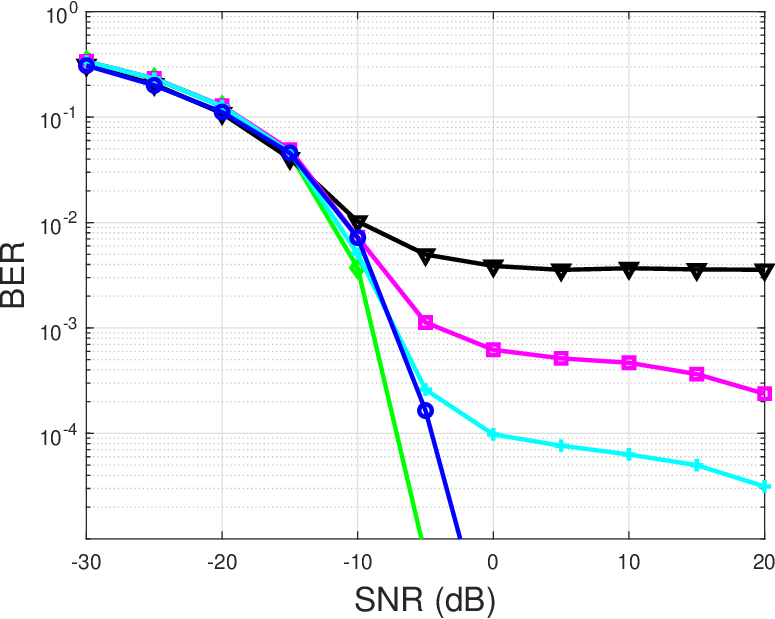}}
       \caption{$M\!=\!7,K\!=\!16 $}
    \end{subfigure} \\ \vspace{10pt}
    \begin{subfigure}[t]{0.245\textwidth}
       \centerline{\includegraphics[width=\textwidth]{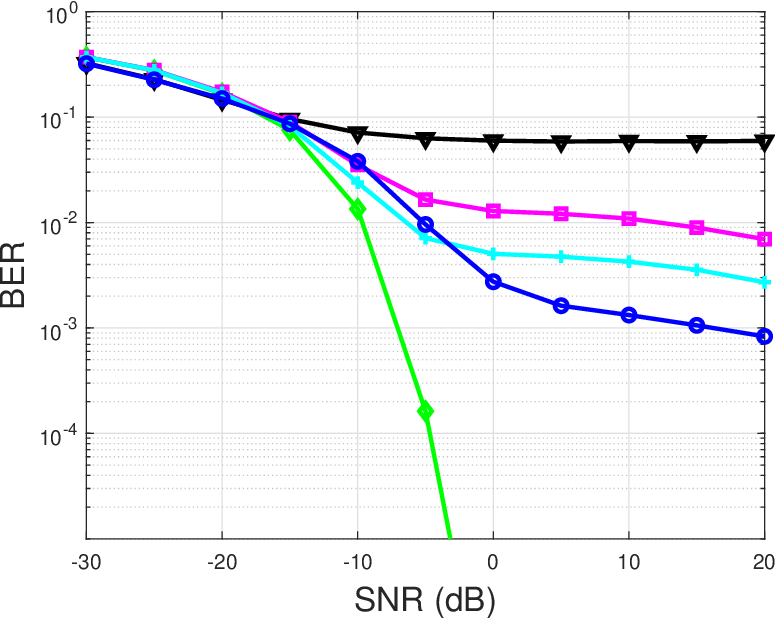}}
       \caption{$M\!=\!4, K\!=\!24$}
    \end{subfigure}\hfill
        \begin{subfigure}[t]{0.245\textwidth}
       \centerline{\includegraphics[width=\textwidth]{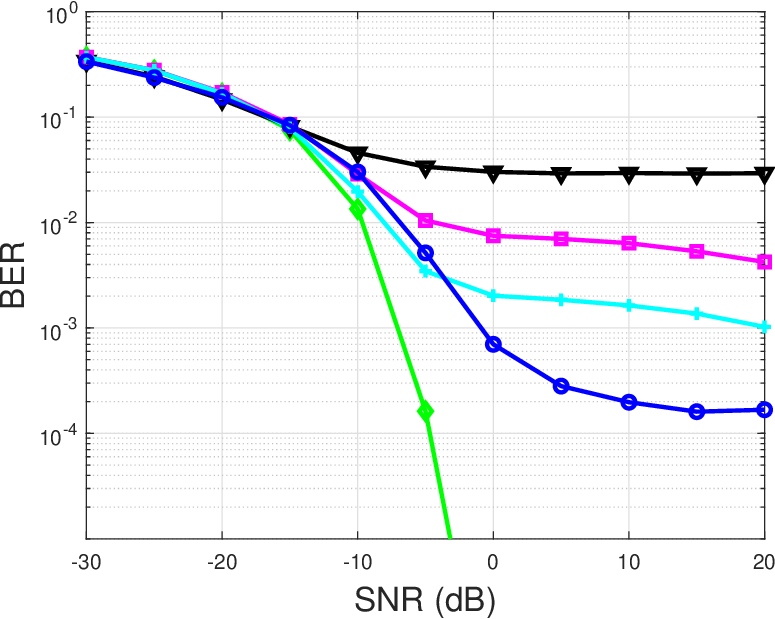}}
       \caption{$M\!=\!5, K\!=\!24$}
    \end{subfigure}\hfill
    \begin{subfigure}[t]{0.245\textwidth}
       \centerline{\includegraphics[width=\textwidth]{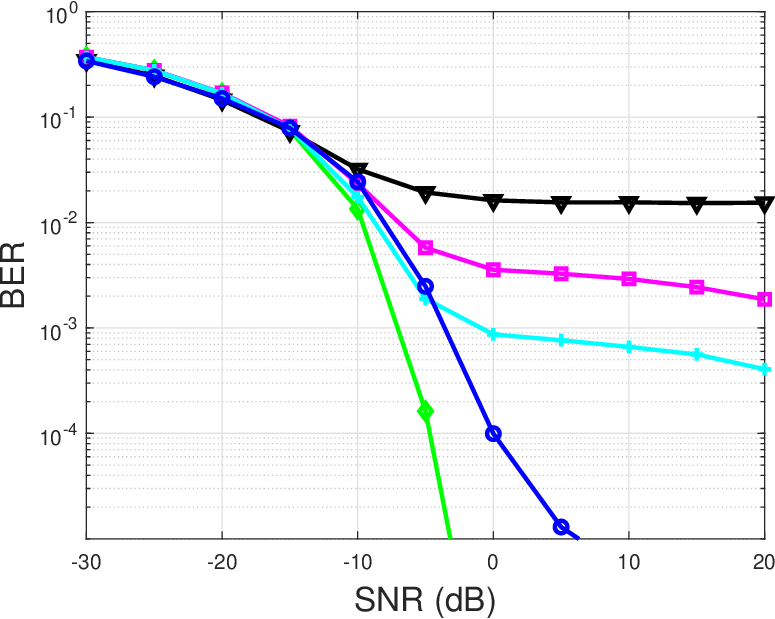}}
       \caption{$M\!=\!6, K\!=\!24$}
    \end{subfigure}\hfill
    \begin{subfigure}[t]{0.245\textwidth}
       \centerline{\includegraphics[width=\textwidth]{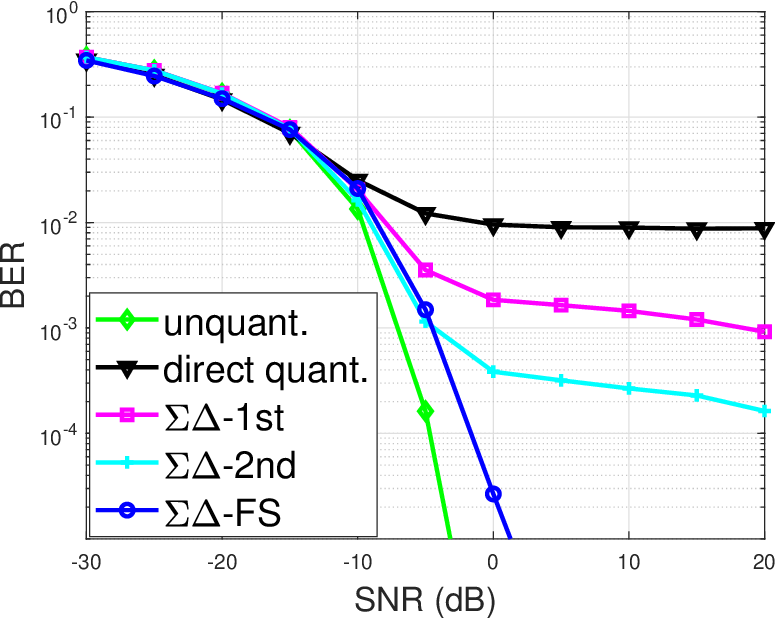}}
       \caption{$M\!=\!7, K\!=\!24$}
    \end{subfigure}
\end{minipage}
\caption{BER performance of the fixed-sector optimized $\Sigma\Delta$ modulation scheme for various values of the user number $K$. $N = 1024$, $d = \lambda/4$, $L = 16$, $[\theta_l, \theta_u] = [-30^\circ, 30^\circ]$.
See the caption of Figure \ref{fig:sectorpass-baseband} for a description of the legend labels. 
}
\label{fig:diff-no-usr}
\end{figure*}

The next simulation result has the challenge raised by increasing the user number $K$.
Figure \ref{fig:diff-no-usr} displays a set of BER-versus-SNR plots for various values of $K$, wherein we fix the angle sector as $[\theta_l, \theta_u] = [-30^\circ,30^\circ]$.
Once again, the fixed-sector optimized $\Sigma\Delta$ modulation scheme is seen to lead to better BER performance than the benchmarked schemes.
It is also noticed that, as the user number $K$ increases, we need a larger quantization level number $M$  to get close to the unquantized performance baseline.

\begin{figure*}[tp!]
\begin{minipage}[c]{\linewidth}
  \centering\vspace{6pt}
    \begin{subfigure}[t]{0.245\textwidth}
        \centerline{\includegraphics[width=\textwidth]{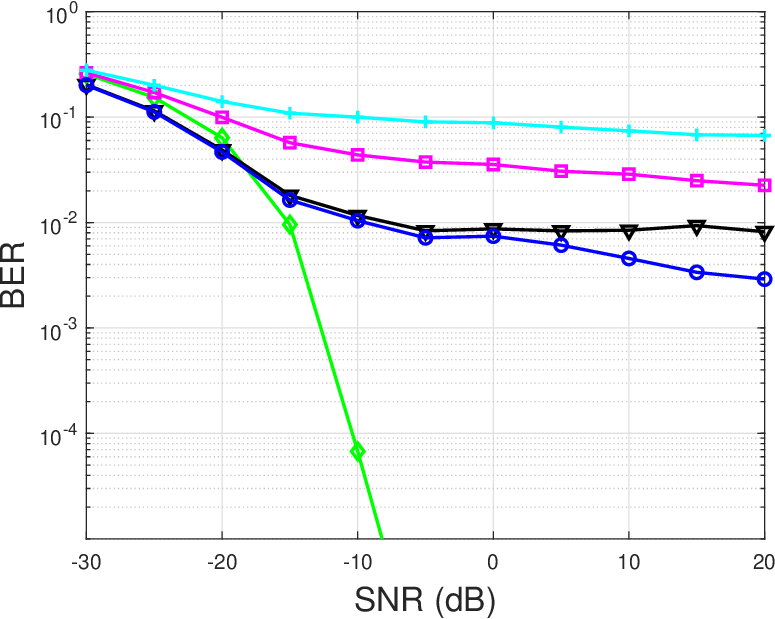}}
        \caption{ $M\!=\!4, d = \lambda/2$}
     \end{subfigure}\hfill 
         \begin{subfigure}[t]{0.245\textwidth}     \centerline{\includegraphics[width=\textwidth]{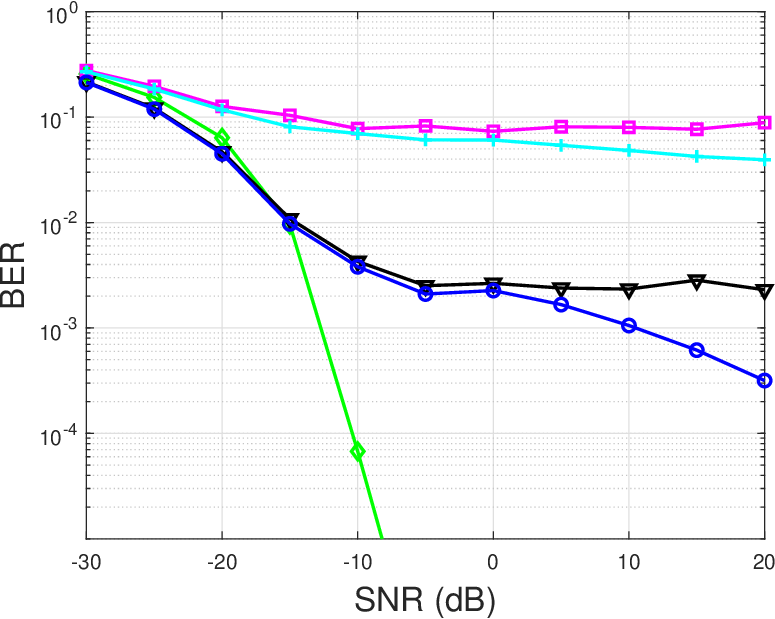}}
        \caption{ $M\!=\!5, d = \lambda/2$}
     \end{subfigure}\hfill 
\begin{subfigure}[t]{0.245\textwidth}     \centerline{\includegraphics[width=\textwidth]{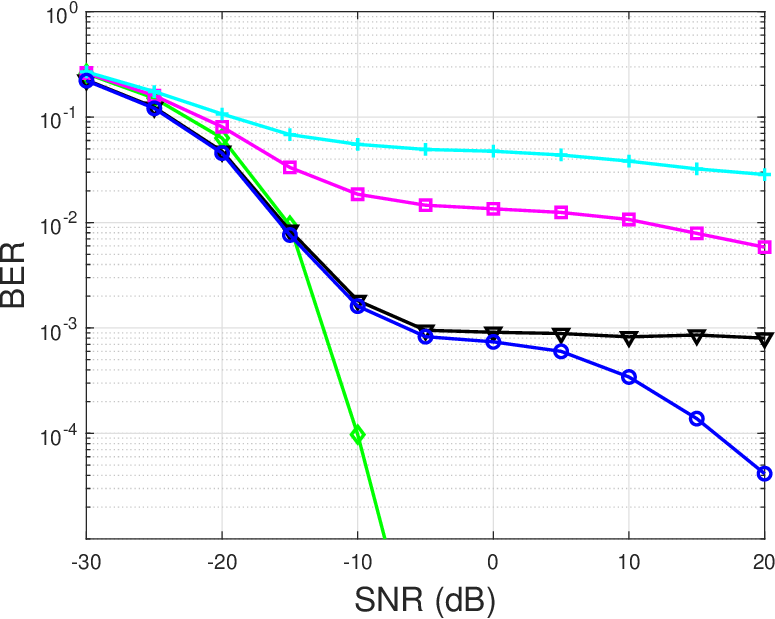}}
        \caption{ $M\!=\!6, d = \lambda/2$}
     \end{subfigure}\hfill 
     \begin{subfigure}[t]{0.245\textwidth}     \centerline{\includegraphics[width=\textwidth]{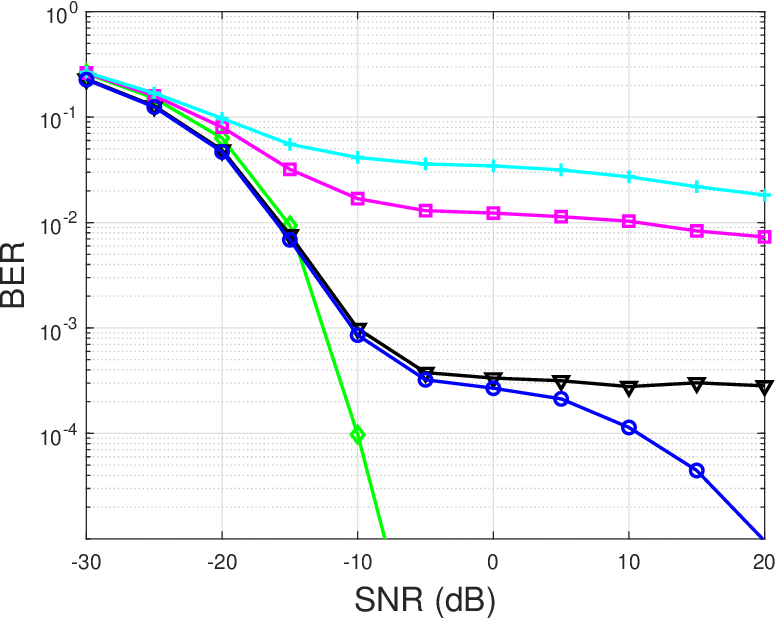}}
        \caption{ $M\!=\!7, d = \lambda/2$}
     \end{subfigure}\hfill \\\vspace{6pt}
    \begin{subfigure}[t]{0.245\textwidth}
        \centerline{\includegraphics[width=\textwidth]{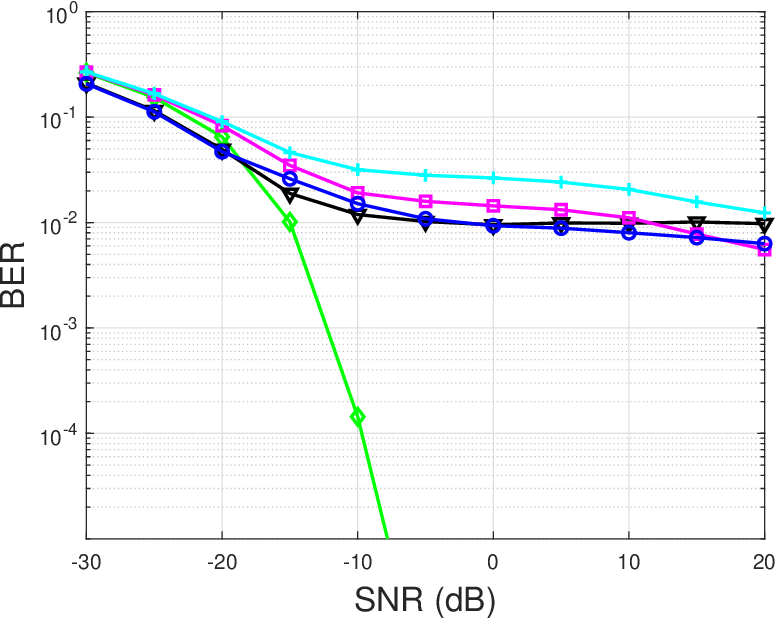}}
        \caption{ $M\!=\!4, d = \lambda/4$}
     \end{subfigure}\hfill 
         \begin{subfigure}[t]{0.245\textwidth}     \centerline{\includegraphics[width=\textwidth]{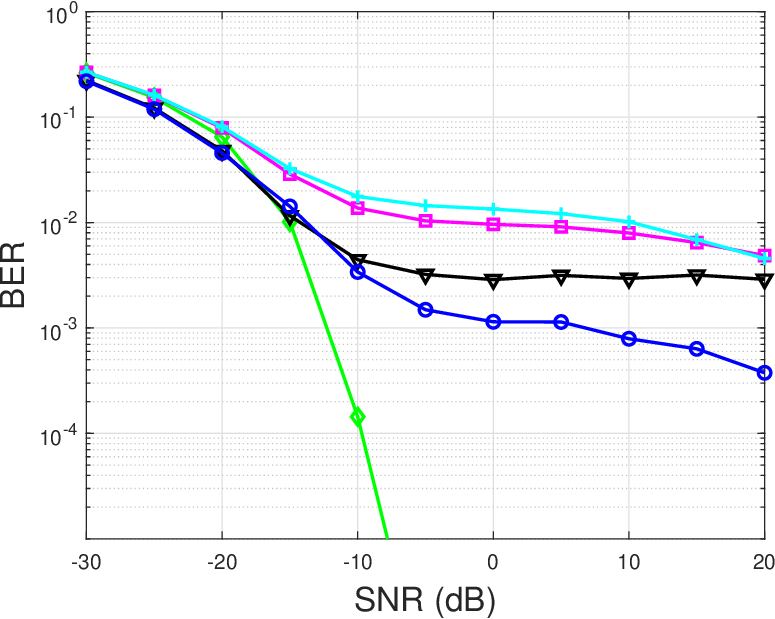}}
        \caption{ $M\!=\!5, d = \lambda/4$}
     \end{subfigure}\hfill 
\begin{subfigure}[t]{0.245\textwidth}     \centerline{\includegraphics[width=\textwidth]{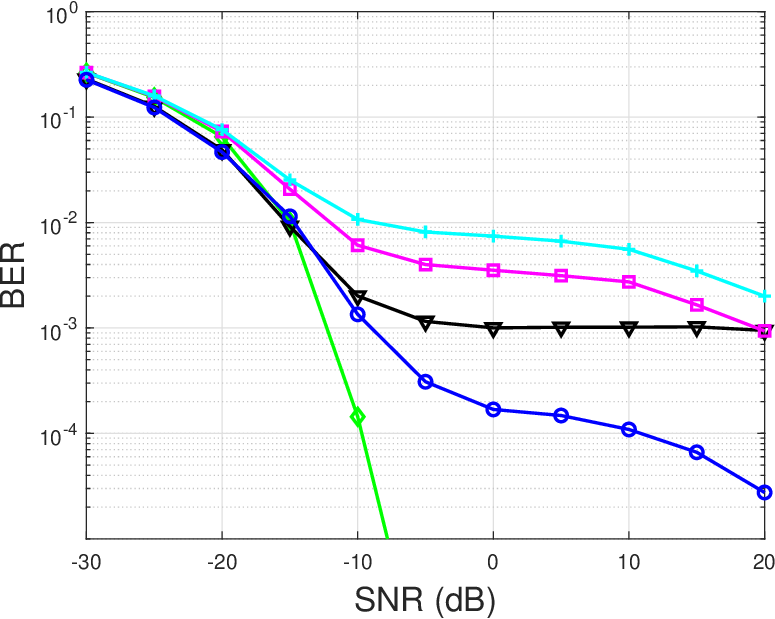}}
        \caption{ $M\!=\!6, d = \lambda/4$}
     \end{subfigure}\hfill 
     \begin{subfigure}[t]{0.245\textwidth}     \centerline{\includegraphics[width=\textwidth]{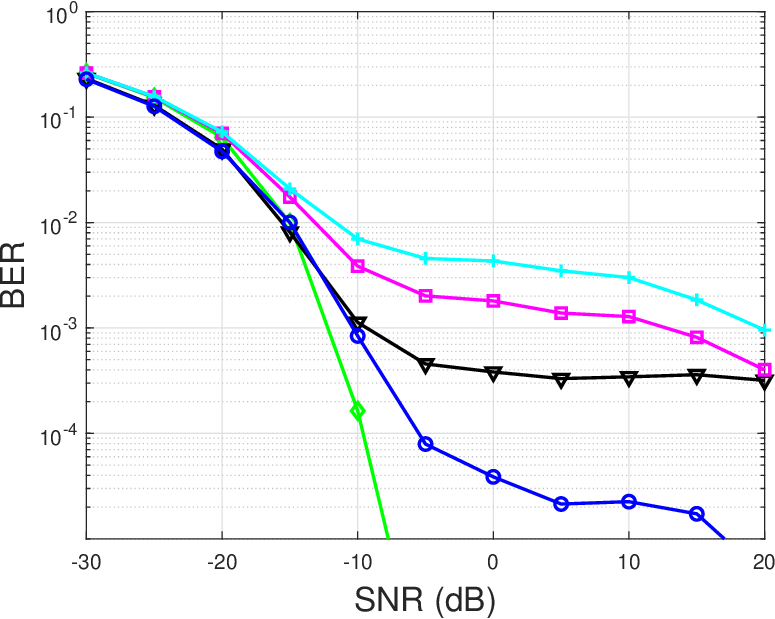}}
        \caption{ $M\!=\!7, d = \lambda/4$}
     \end{subfigure}\hfill \\\vspace{6pt}
    \begin{subfigure}[t]{0.245\textwidth}
        \centerline{\includegraphics[width=\textwidth]{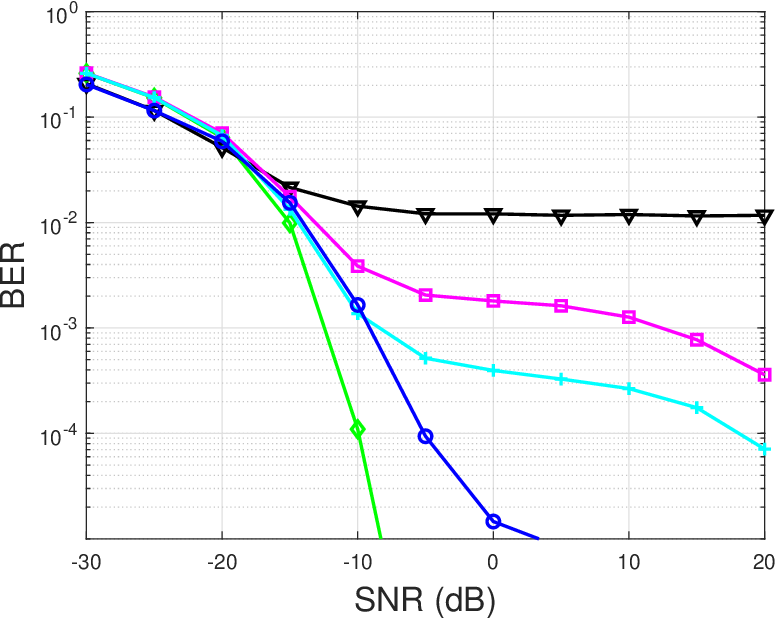}}
        \caption{ $M\!=\!4, d = \lambda/8$}
     \end{subfigure}\hfill 
         \begin{subfigure}[t]{0.245\textwidth}     \centerline{\includegraphics[width=\textwidth]{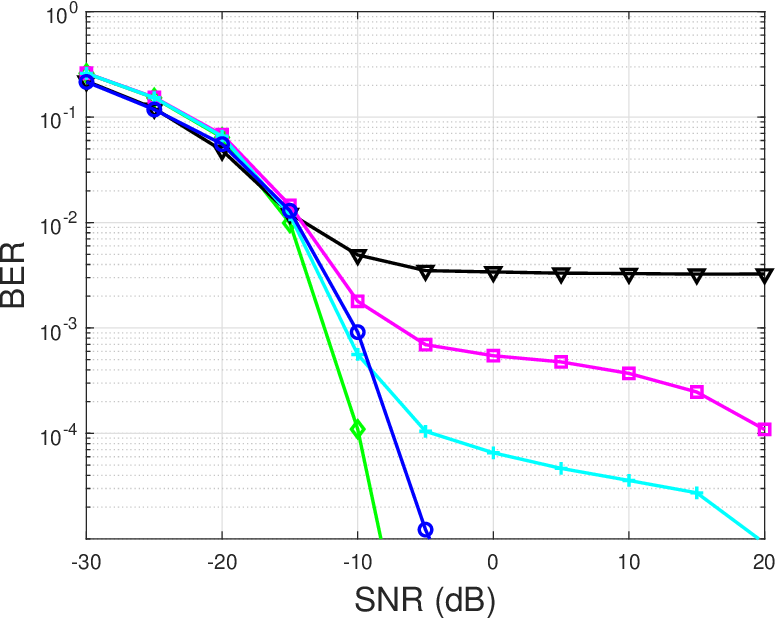}}
        \caption{ $M\!=\!5, d = \lambda/8$}
     \end{subfigure}\hfill 
\begin{subfigure}[t]{0.245\textwidth}     \centerline{\includegraphics[width=\textwidth]{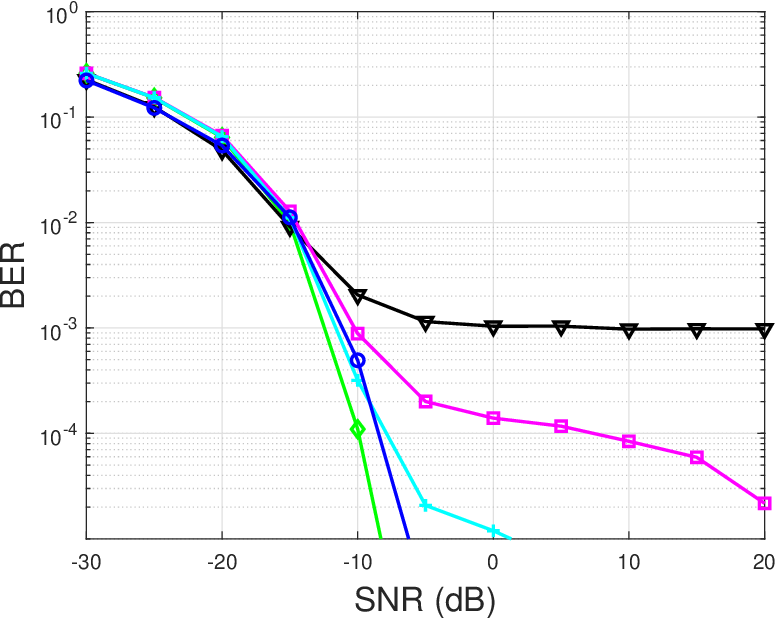}}
        \caption{ $M\!=\!6, d = \lambda/8$}
     \end{subfigure}\hfill 
     \begin{subfigure}[t]{0.245\textwidth}     \centerline{\includegraphics[width=\textwidth]{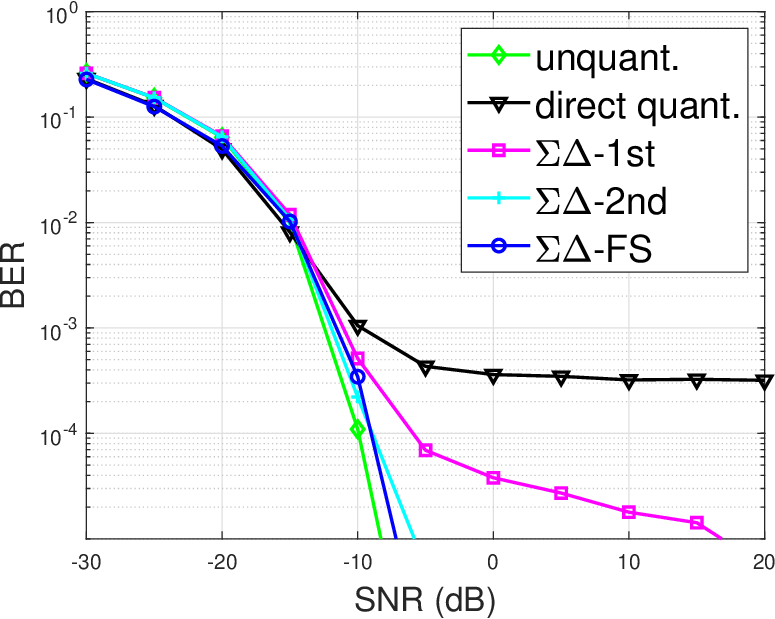}}
        \caption{ $M\!=\!7, d = \lambda/8$}
     \end{subfigure}\hfill 
\end{minipage}
\caption{BER performance of the fixed-sector optimized $\Sigma\Delta$ modulation scheme for various values of the inter-antenna spacing $d$. $N = 1024$, $K=8$, $L = 16$, $[\theta_l, \theta_u] = [-75^\circ, 75^\circ]$.
See the caption of Figure \ref{fig:sectorpass-baseband} for a description of the legend labels.
}
\label{fig:different-d-M}
\end{figure*}

We are also interested in how the performance changes with the inter-antenna spacing $d$.
As discussed, the $\Sigma\Delta$ notion suggests that we want $d$ to be as small as possible, but physical limitations disallow us from making $d$ too small.
Figure \ref{fig:different-d-M} shows the results for various values of $d$, wherein we set the angle sector  as $[\theta_l, \theta_u] = [-75^\circ, 75^\circ]$ which is relatively wide.
We see that a smaller $d$ leads to better performance for all the $\Sigma\Delta$ schemes,
while a larger $d$ requires us to use a larger quantization level number $M$ to get reasonable performance.
This simulation result, together with the previous results, indicate a tradeoff---if we want to have a wider angle sector and/or a larger inter-antenna spacing, the $\Sigma\Delta$ noise shaping problem becomes harder and we need a greater number of quantization levels to meet the challenge. 

\subsection{User-Targeted Design}

\begin{figure*}[p!]
\begin{subfigure}[t]{0.245\textwidth}
       \centerline{\includegraphics[width=\textwidth]{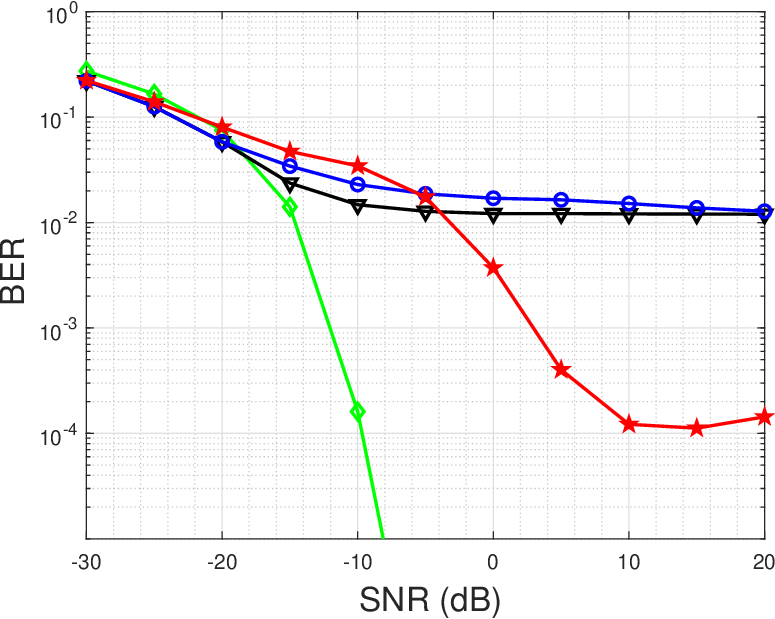}}\vspace{-0pt}
       \caption{\footnotesize $d = \frac{\lambda}{2}, |\theta|\leq 30^\circ, M = 4$}
\end{subfigure}\hfill
     \begin{subfigure}[t]{0.245\textwidth}
       \centerline{\includegraphics[width=\textwidth]{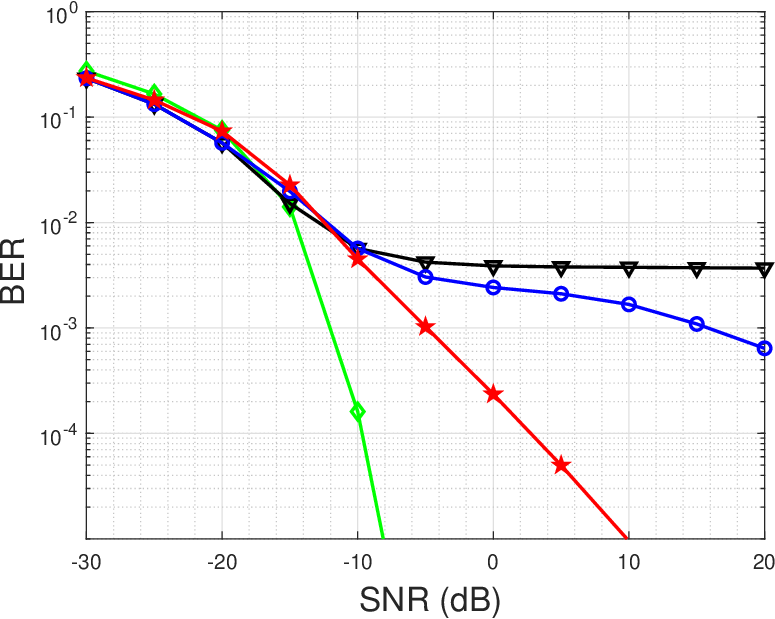}}\vspace{-0pt}
       \caption{\footnotesize $d = \frac{\lambda}{2}, |\theta|\leq 30^\circ, M = 5$}
\end{subfigure}\hfill
     \begin{subfigure}[t]{0.245\textwidth}
       \centerline{\includegraphics[width=\textwidth]{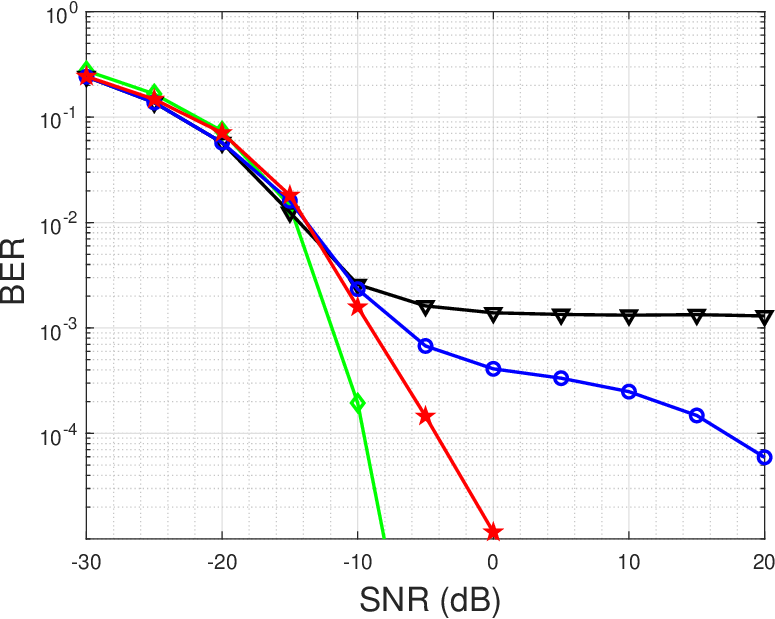}}\vspace{-0pt}
       \caption{\footnotesize $d = \frac{\lambda}{2}, |\theta|\leq 30^\circ, M = 6$}
\end{subfigure}\hfill
     \begin{subfigure}[t]{0.245\textwidth}
       \centerline{\includegraphics[width=\textwidth]{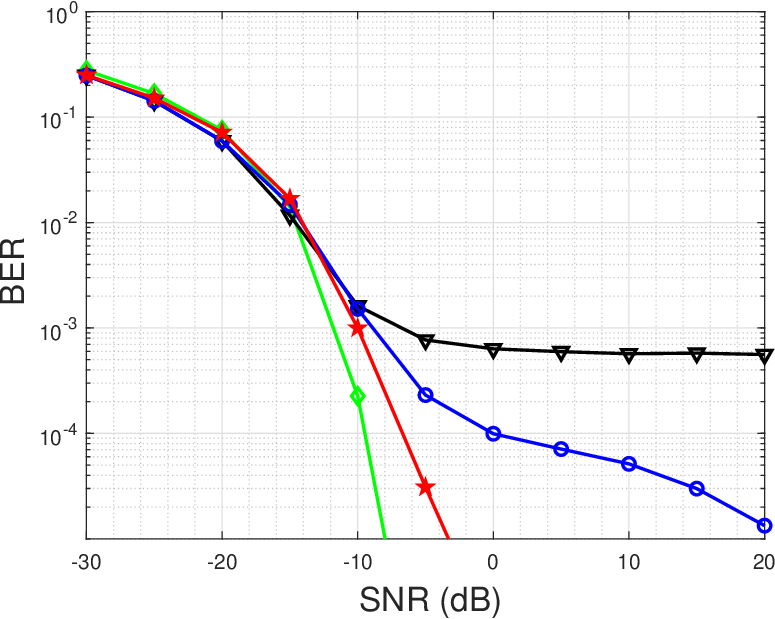}}\vspace{-0pt}
       \caption{\footnotesize $d = \frac{\lambda}{2}, |\theta|\leq 30^\circ, M = 7$}
\end{subfigure}\hfill
\\\vspace{10pt}

\begin{subfigure}[t]{0.245\textwidth}
       \centerline{\includegraphics[width=\textwidth]{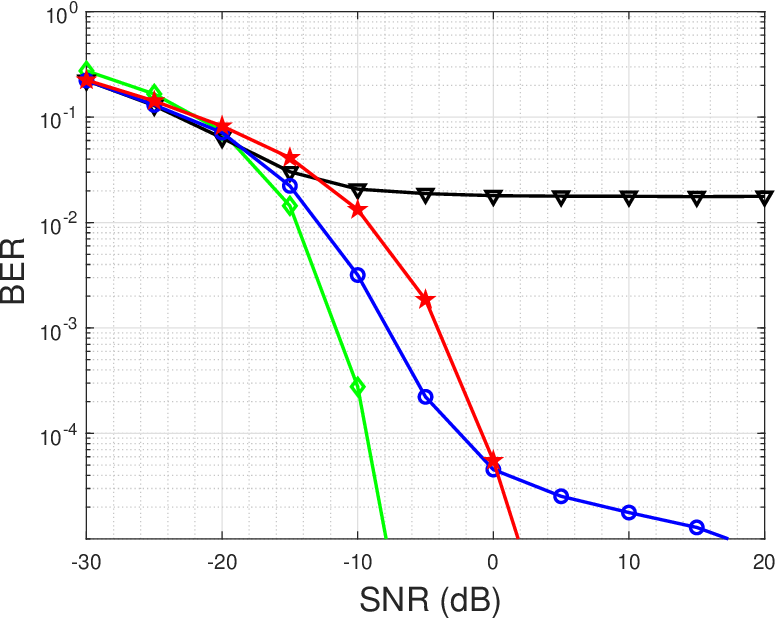}}\vspace{-0pt}
       \caption{\footnotesize $d = \frac{\lambda}{4}, |\theta|\leq 30^\circ, M = 4$}
\end{subfigure}\hfill
     \begin{subfigure}[t]{0.245\textwidth}
       \centerline{\includegraphics[width=\textwidth]{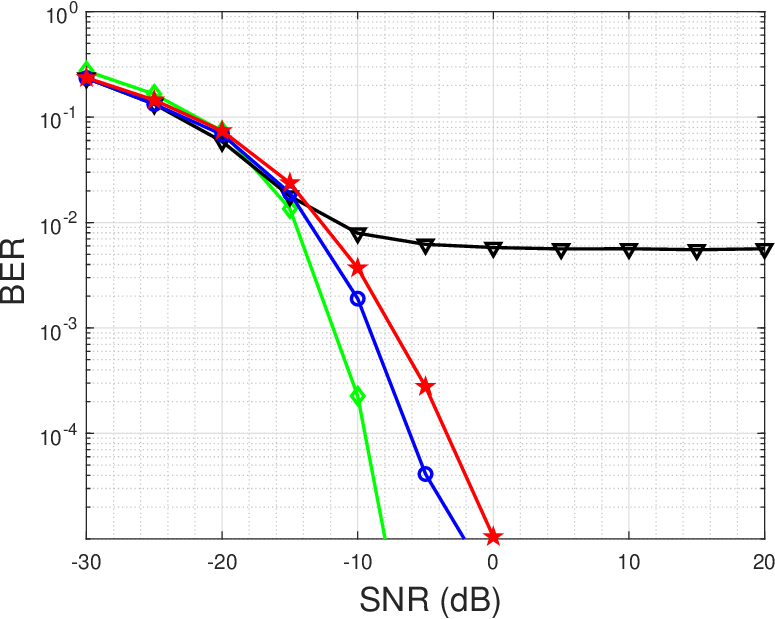}}\vspace{-0pt}
       \caption{\footnotesize $d = \frac{\lambda}{4}, |\theta|\leq 30^\circ, M = 5$}
\end{subfigure}\hfill
     \begin{subfigure}[t]{0.245\textwidth}
       \centerline{\includegraphics[width=\textwidth]{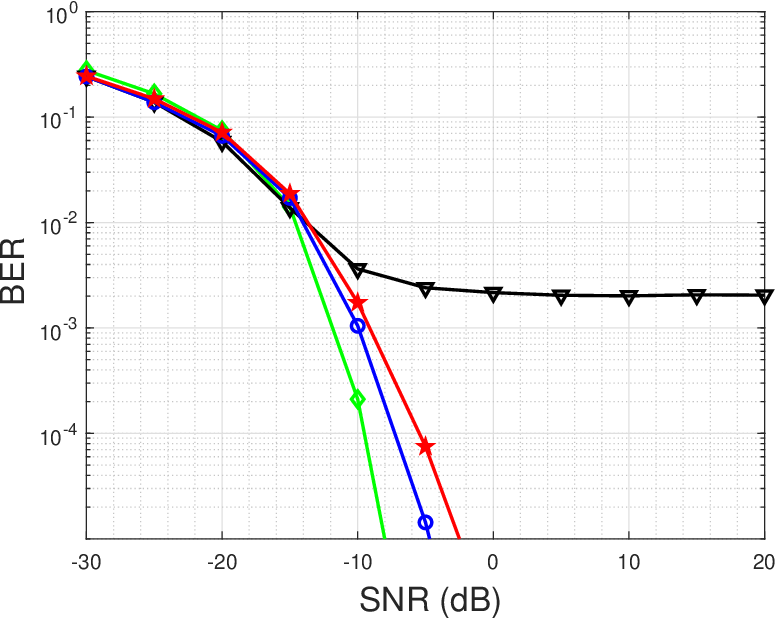}}\vspace{-0pt}
       \caption{\footnotesize $d = \frac{\lambda}{4}, |\theta|\leq 30^\circ, M = 6$}
\end{subfigure}\hfill
     \begin{subfigure}[t]{0.245\textwidth}
       \centerline{\includegraphics[width=\textwidth]{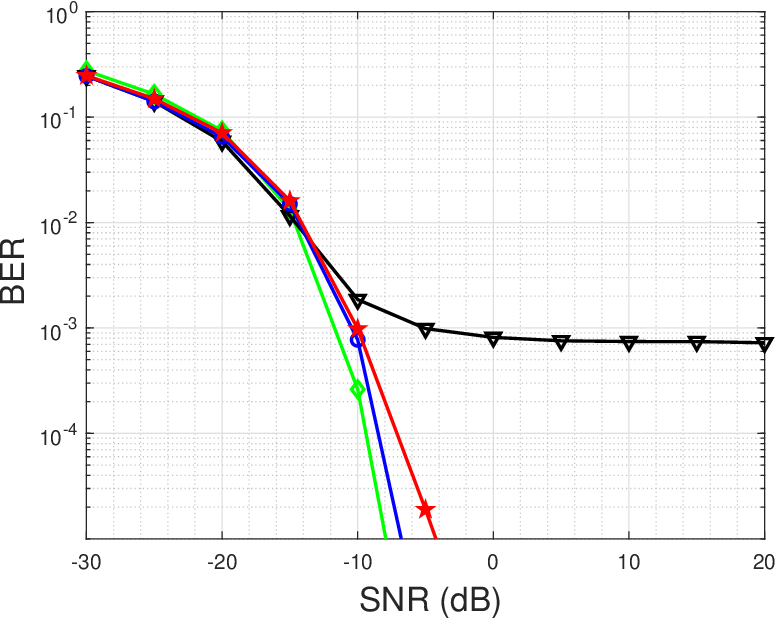}}\vspace{-0pt}
       \caption{\footnotesize $d = \frac{\lambda}{4}, |\theta|\leq 30^\circ, M = 7$}
\end{subfigure}\hfill
\\\vspace{20pt}

\begin{subfigure}[t]{0.245\textwidth}
       \centerline{\includegraphics[width=\textwidth]{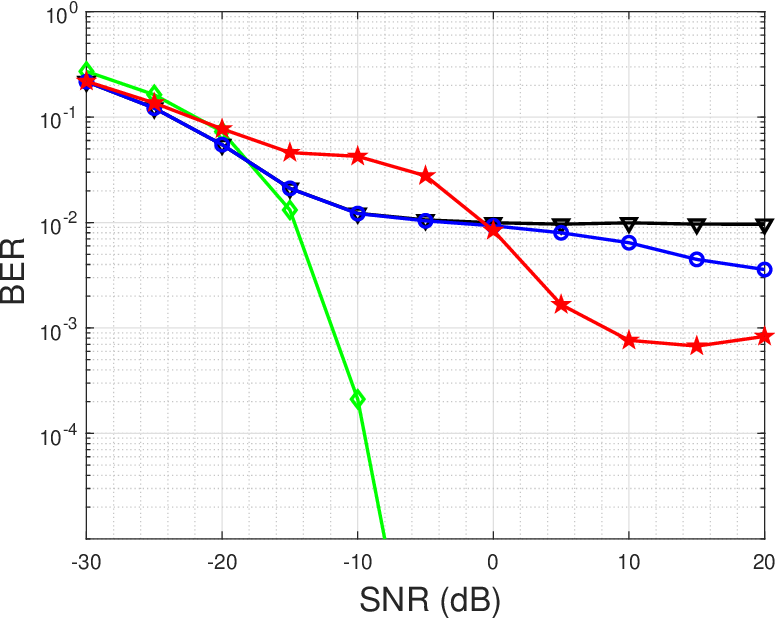}}\vspace{-0pt}
       \caption{\footnotesize $d = \frac{\lambda}{2}, |\theta|\leq 80^\circ, M = 4$}
\end{subfigure}\hfill
     \begin{subfigure}[t]{0.245\textwidth}
       \centerline{\includegraphics[width=\textwidth]{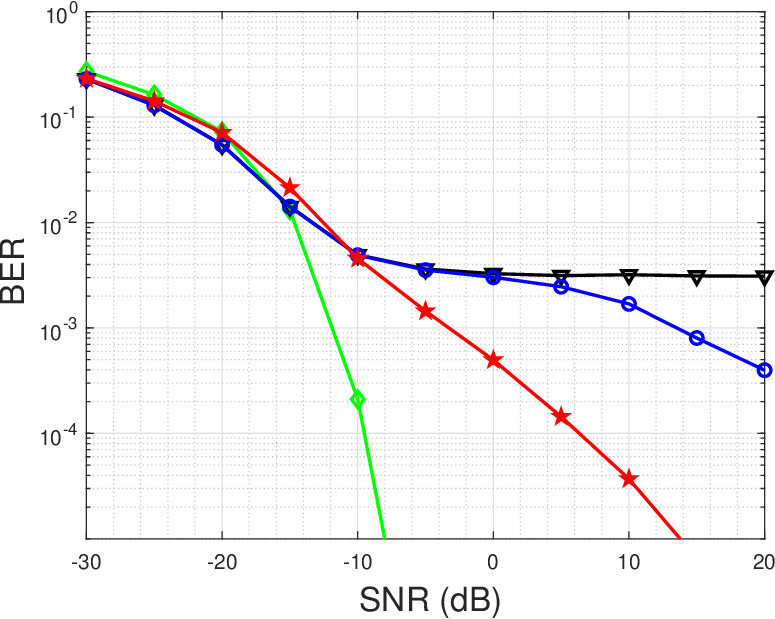}}\vspace{-0pt}
       \caption{\footnotesize $d = \frac{\lambda}{2}, |\theta|\leq 80^\circ, M = 5$}
\end{subfigure}\hfill
     \begin{subfigure}[t]{0.245\textwidth}
       \centerline{\includegraphics[width=\textwidth]{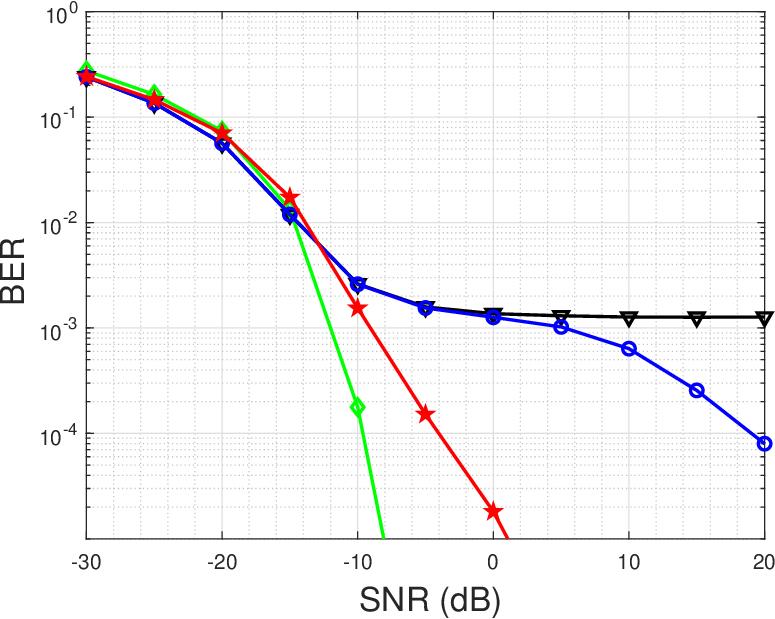}}\vspace{-0pt}
       \caption{\footnotesize $d = \frac{\lambda}{2}, |\theta|\leq 80^\circ, M = 6$}
\end{subfigure}\hfill
     \begin{subfigure}[t]{0.245\textwidth}
       \centerline{\includegraphics[width=\textwidth]{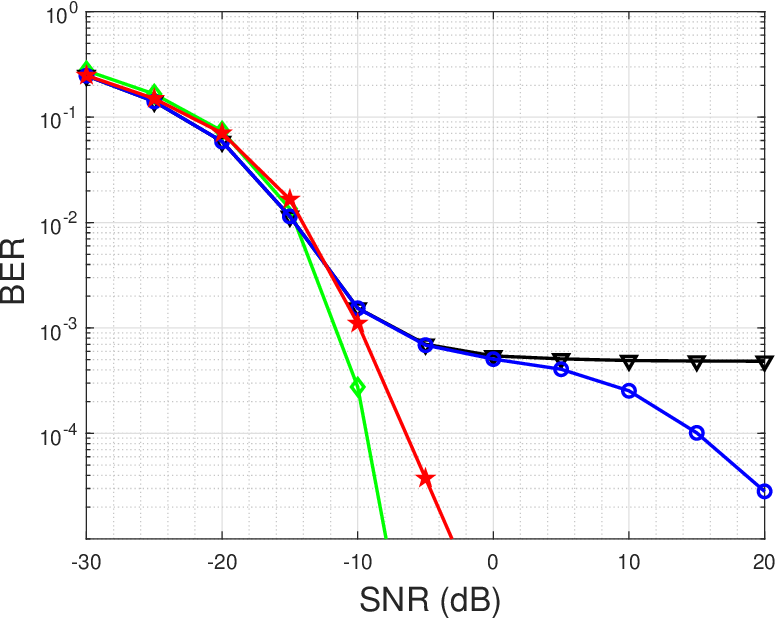}}\vspace{-0pt}
       \caption{\footnotesize $d = \frac{\lambda}{2}, |\theta|\leq 80^\circ, M = 7$}
\end{subfigure}\hfill
\\ \vspace{10pt}

\begin{subfigure}[t]{0.245\textwidth}
       \centerline{\includegraphics[width=\textwidth]{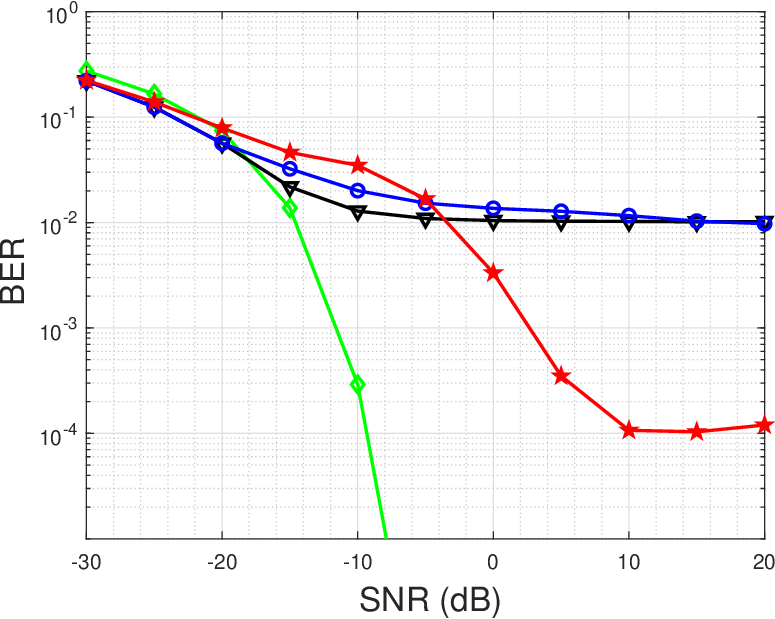}}\vspace{-0pt}
       \caption{\footnotesize $d = \frac{\lambda}{4}, |\theta|\leq 80^\circ, M = 4$}
\end{subfigure}\hfill
     \begin{subfigure}[t]{0.245\textwidth}
       \centerline{\includegraphics[width=\textwidth]{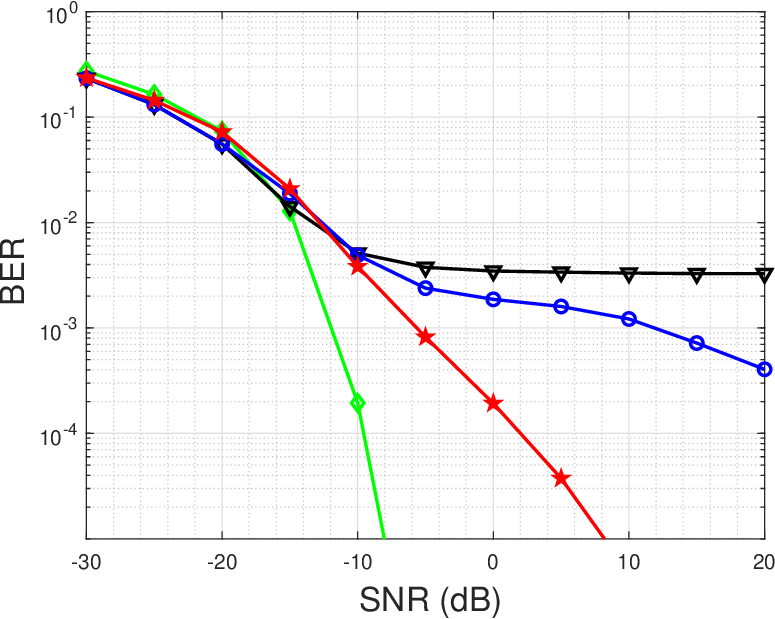}}\vspace{-0pt}
       \caption{\footnotesize $d = \frac{\lambda}{4}, |\theta|\leq 80^\circ, M = 5$}
\end{subfigure}\hfill
     \begin{subfigure}[t]{0.245\textwidth}
       \centerline{\includegraphics[width=\textwidth]{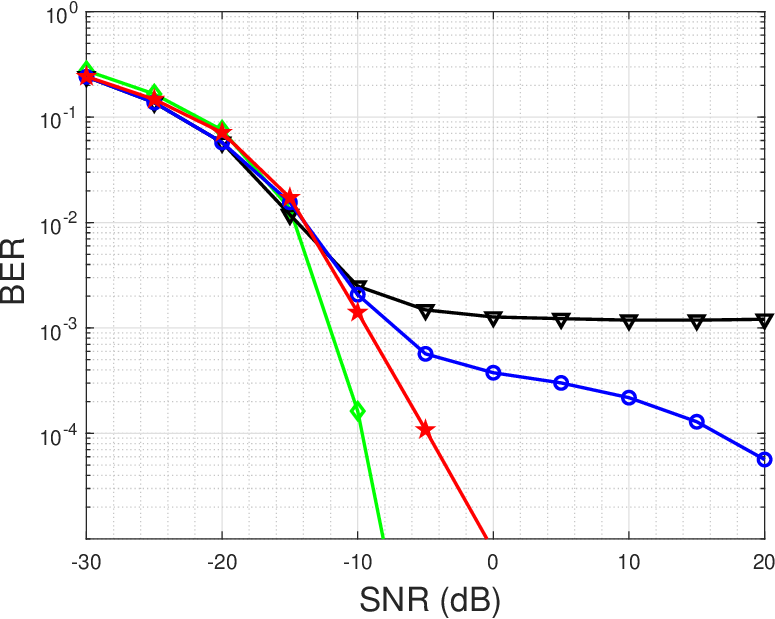}}\vspace{-0pt}
       \caption{\footnotesize $d = \frac{\lambda}{4}, |\theta|\leq 80^\circ, M = 6$}
\end{subfigure}\hfill
     \begin{subfigure}[t]{0.245\textwidth}
       \centerline{\includegraphics[width=\textwidth]{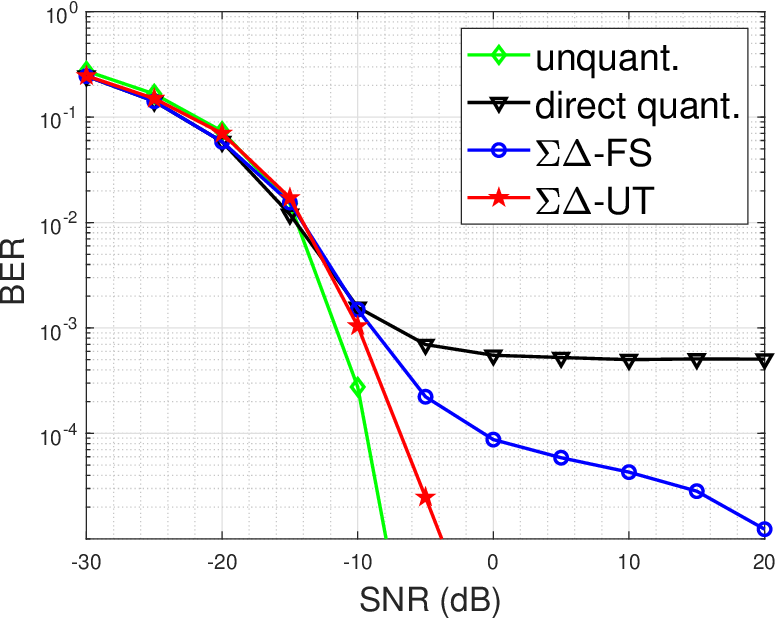}}\vspace{-0pt}
       \caption{\footnotesize $d = \frac{\lambda}{4}, |\theta|\leq 80^\circ, M = 7$}
\end{subfigure}\hfill
\caption{BER performance of the user-targeted $\Sigma\Delta$ modulation scheme for $K= 9$.
$N = 1024$, 
$L = 24$, $|\theta| \leq \vartheta$ means that the angle sector is $[\theta_l,\theta_u]= [-\vartheta,\vartheta]$.
$\Sigma\Delta$-UT: the user-targeted $\Sigma\Delta$ modulation scheme.
See the caption of Figure \ref{fig:sectorpass-baseband} for a description of the other legend labels.
}
\label{fig:UT-1024x9}
\end{figure*}

We turn our interest to the user-targeted design in  Section~\ref{sect:main_meat}-\ref{sect:SQNR_max}.
The settings are identical to those in the last subsection, except that the filter order of our optimization-based $\Sigma\Delta$ modulator is $L=24$. 
Figure \ref{fig:UT-1024x9}  displays 
a set of BER-versus-SNR plots
when the user number is $K=9$.
We see that the user-targeted $\Sigma\Delta$ modulation scheme can improve upon the fixed-sector optimized $\Sigma\Delta$ modulation scheme,
and the improvement is significant for larger values of the quantization level number $M$.
For instance, for $d= \lambda/2$ and $[\theta_l,\theta_u] = [-80^\circ,80^\circ]$, which is a challenging setting, Figure \ref{fig:UT-1024x9}(l) shows that the user-targeted $\Sigma\Delta$ modulation scheme can lead to BER performance close to the unquantized performance baseline.
Also we see that if $d$ is larger and/or the angle sector width is larger, the user-targeted $\Sigma\Delta$ modulation scheme requires a larger $M$ to provide good performance.
This is in agreement with our observation with the fixed-sector $\Sigma\Delta$ modulation scheme in the last subsection.

\begin{figure*}[p!]
\begin{subfigure}[t]{0.245\textwidth}
       \centerline{\includegraphics[width=\textwidth]{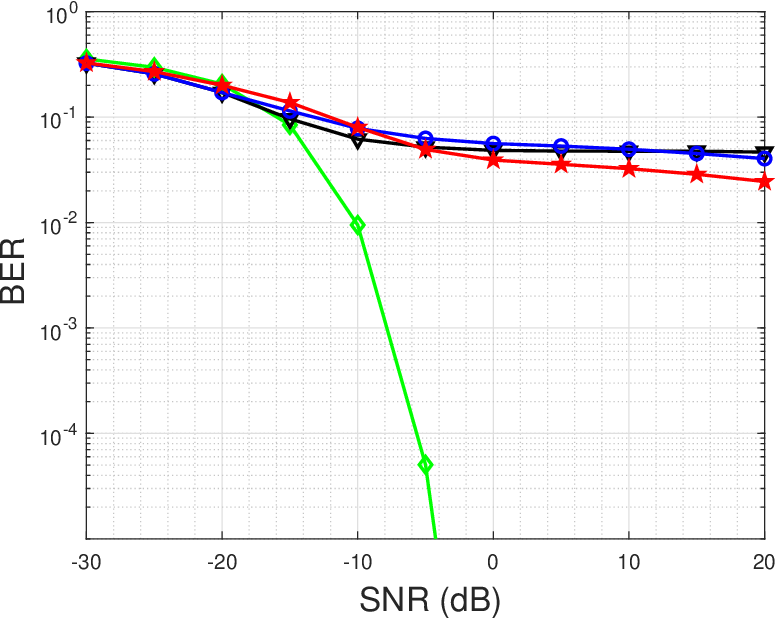}}\vspace{-0pt}
       \caption{\footnotesize $d = \frac{\lambda}{2}, |\theta|\leq 30^\circ, M = 4$}
\end{subfigure}\hfill
     \begin{subfigure}[t]{0.245\textwidth}
       \centerline{\includegraphics[width=\textwidth]{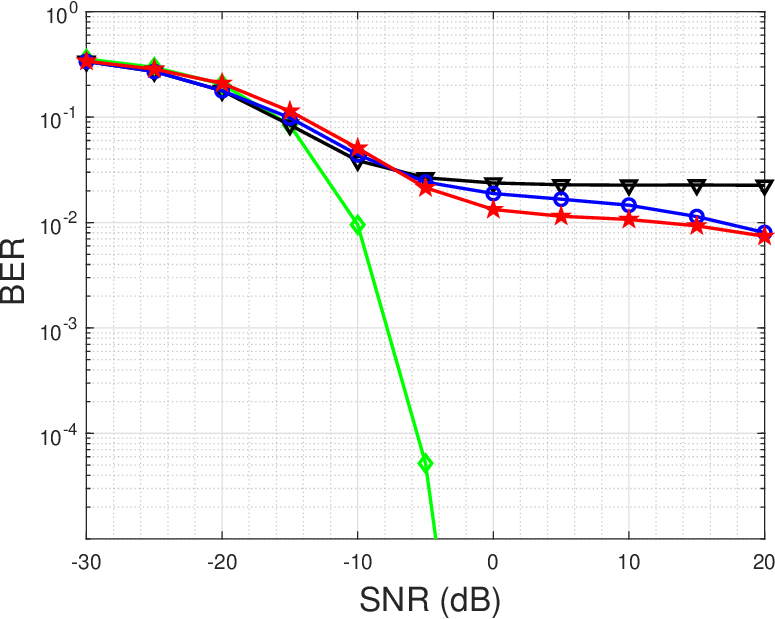}}\vspace{-0pt}
       \caption{\footnotesize $d = \frac{\lambda}{2}, |\theta|\leq 30^\circ, M = 5$}
\end{subfigure}\hfill
     \begin{subfigure}[t]{0.245\textwidth}
       \centerline{\includegraphics[width=\textwidth]{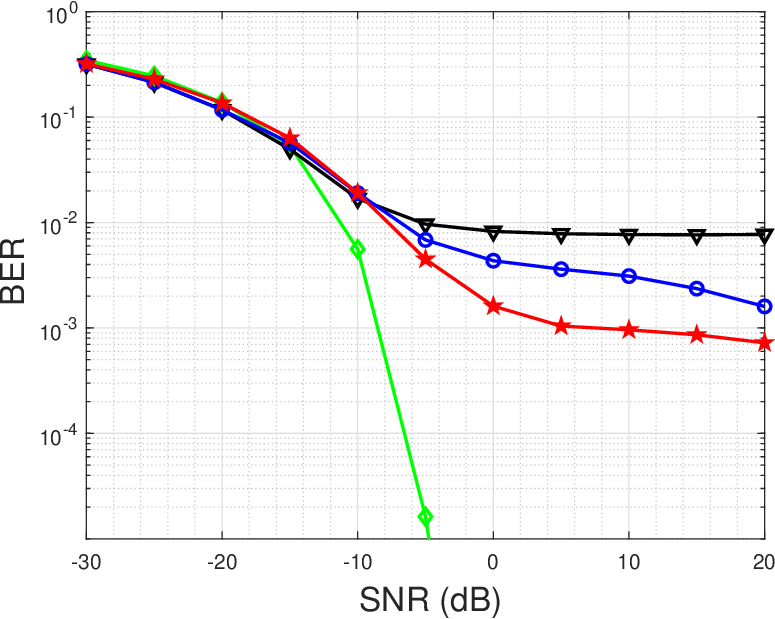}}\vspace{-0pt}
       \caption{\footnotesize $d = \frac{\lambda}{2}, |\theta|\leq 30^\circ, M = 6$}
\end{subfigure}\hfill
     \begin{subfigure}[t]{0.245\textwidth}
       \centerline{\includegraphics[width=\textwidth]{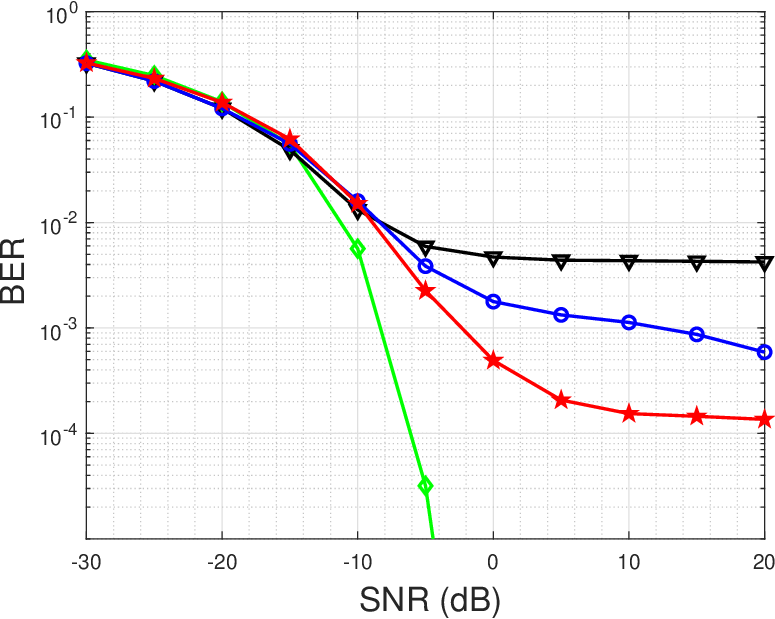}}\vspace{-0pt}
       \caption{\footnotesize $d = \frac{\lambda}{2}, |\theta|\leq 30^\circ, M = 7$}
\end{subfigure}\hfill
\\\vspace{10pt}

\begin{subfigure}[t]{0.245\textwidth}
       \centerline{\includegraphics[width=\textwidth]{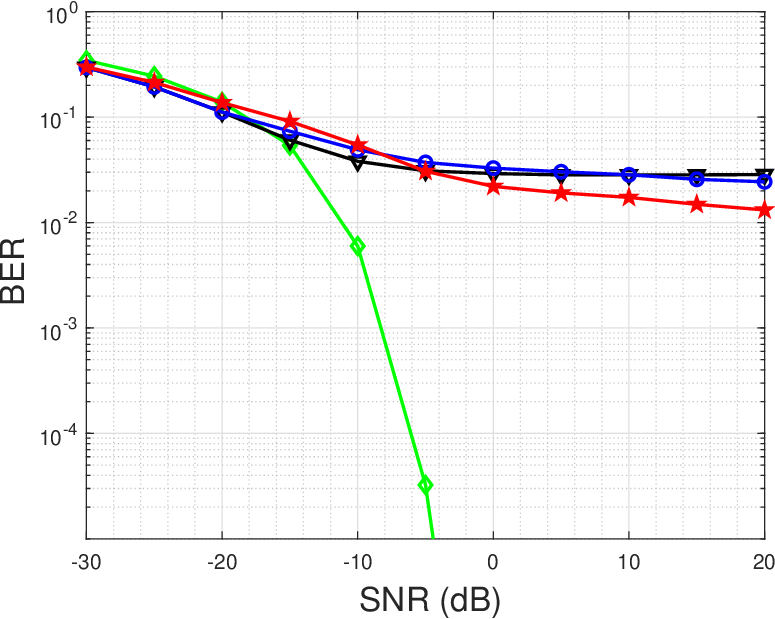}}\vspace{-0pt}
       \caption{\footnotesize $d = \frac{\lambda}{4}, |\theta|\leq 30^\circ, M = 4$}
\end{subfigure}\hfill
     \begin{subfigure}[t]{0.245\textwidth}
       \centerline{\includegraphics[width=\textwidth]{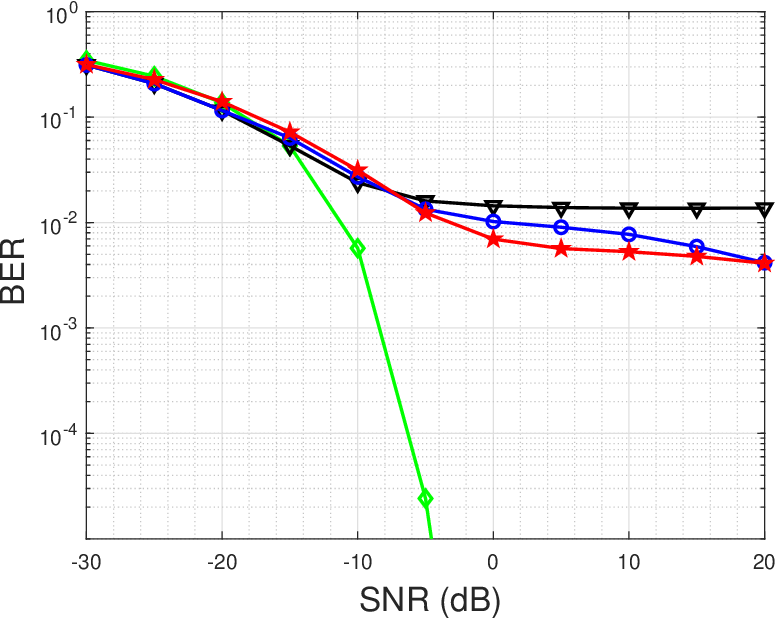}}\vspace{-0pt}
       \caption{\footnotesize $d = \frac{\lambda}{4}, |\theta|\leq 30^\circ, M = 5$}
\end{subfigure}\hfill
     \begin{subfigure}[t]{0.245\textwidth}
       \centerline{\includegraphics[width=\textwidth]{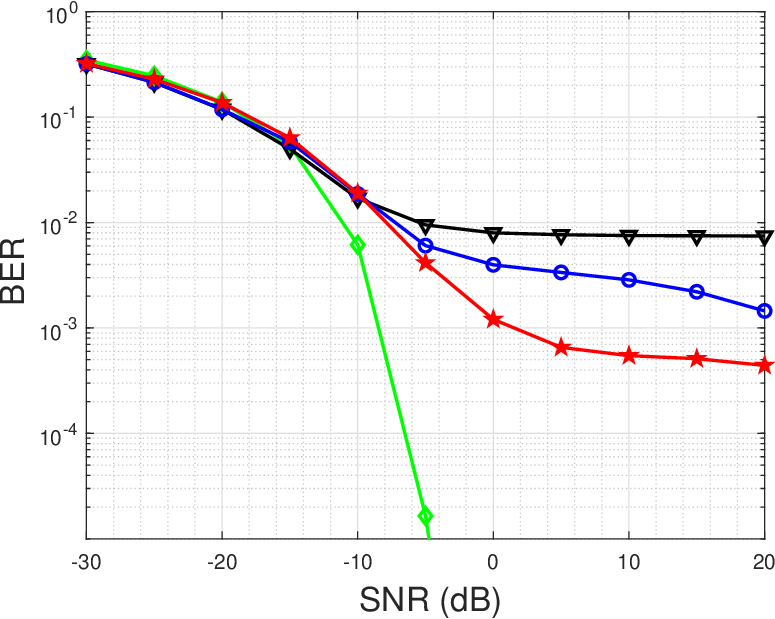}}\vspace{-0pt}
       \caption{\footnotesize $d = \frac{\lambda}{4}, |\theta|\leq 30^\circ, M = 6$}
\end{subfigure}\hfill
     \begin{subfigure}[t]{0.245\textwidth}
       \centerline{\includegraphics[width=\textwidth]{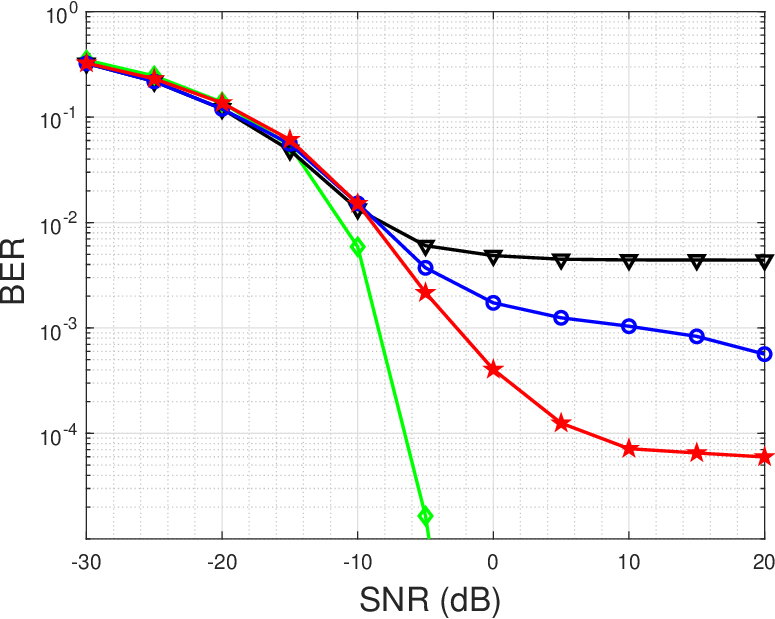}}\vspace{-0pt}
       \caption{\footnotesize $d = \frac{\lambda}{4}, |\theta|\leq 30^\circ, M = 7$}
\end{subfigure}\hfill
\\\vspace{20pt}

\begin{subfigure}[t]{0.245\textwidth}
       \centerline{\includegraphics[width=\textwidth]{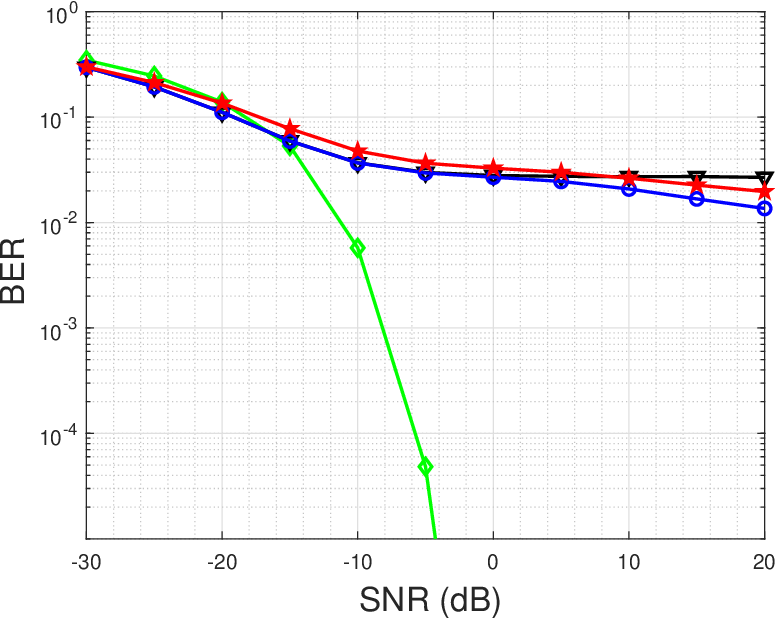}}\vspace{-0pt}
       \caption{\footnotesize $d = \frac{\lambda}{2}, |\theta|\leq 80^\circ, M = 4$}
\end{subfigure}\hfill
     \begin{subfigure}[t]{0.245\textwidth}
       \centerline{\includegraphics[width=\textwidth]{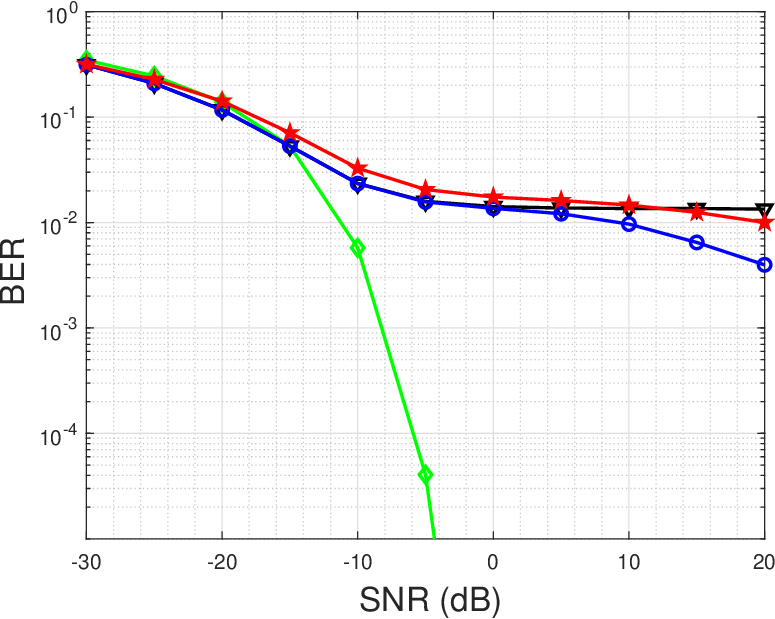}}\vspace{-0pt}
       \caption{\footnotesize $d = \frac{\lambda}{2}, |\theta|\leq 80^\circ, M = 5$}
\end{subfigure}\hfill
     \begin{subfigure}[t]{0.245\textwidth}
       \centerline{\includegraphics[width=\textwidth]{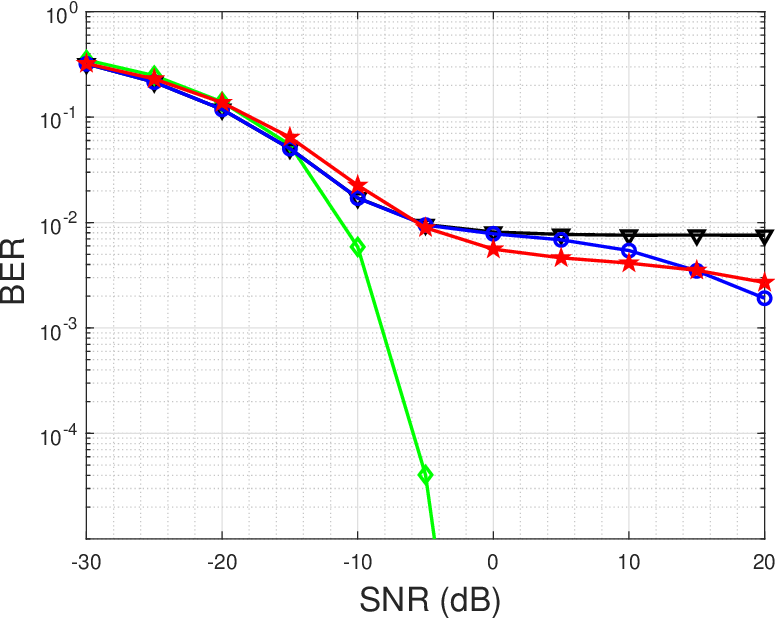}}\vspace{-0pt}
       \caption{\footnotesize $d = \frac{\lambda}{2}, |\theta|\leq 80^\circ, M = 6$}
\end{subfigure}\hfill
     \begin{subfigure}[t]{0.245\textwidth}
       \centerline{\includegraphics[width=\textwidth]{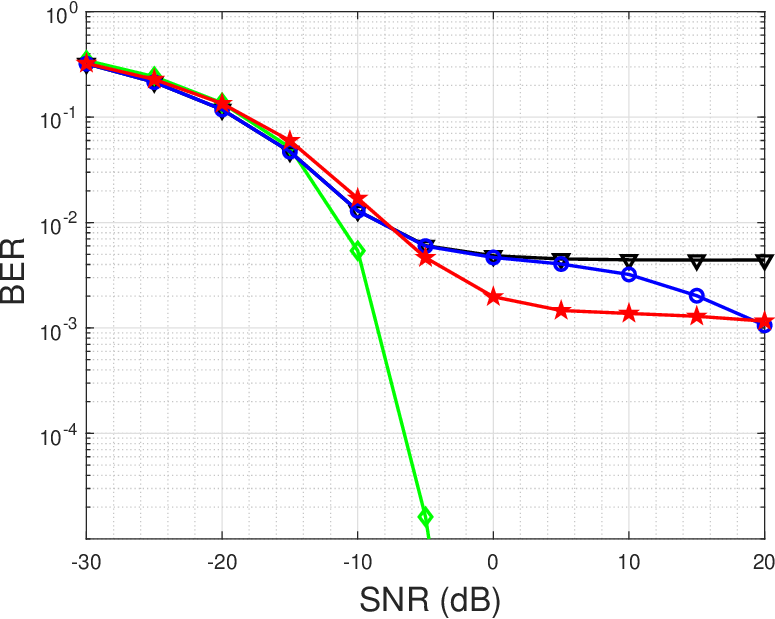}}\vspace{-0pt}
       \caption{\footnotesize $d = \frac{\lambda}{2}, |\theta|\leq 80^\circ, M = 7$}
\end{subfigure}\hfill
\\ \vspace{10pt}

\begin{subfigure}[t]{0.245\textwidth}
       \centerline{\includegraphics[width=\textwidth]{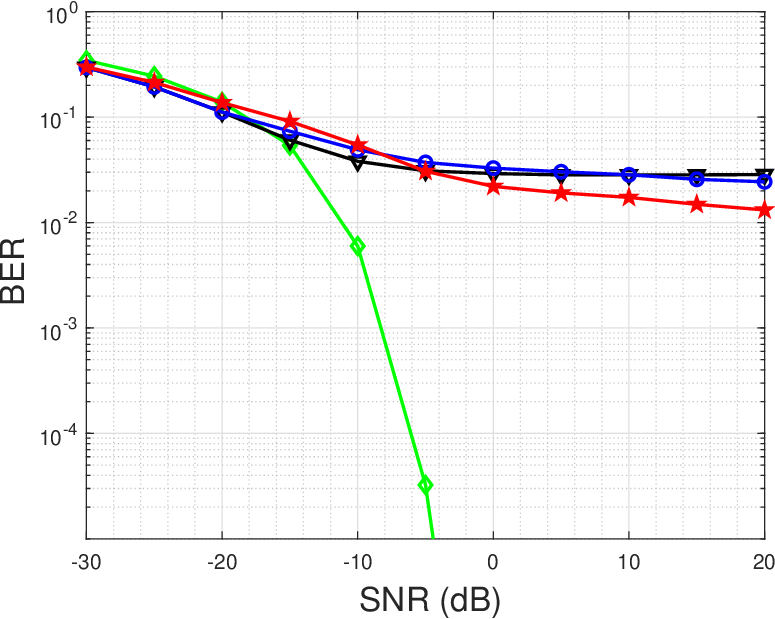}}\vspace{-0pt}
       \caption{\footnotesize $d = \frac{\lambda}{4}, |\theta|\leq 80^\circ, M = 4$}
\end{subfigure}\hfill
     \begin{subfigure}[t]{0.245\textwidth}
       \centerline{\includegraphics[width=\textwidth]{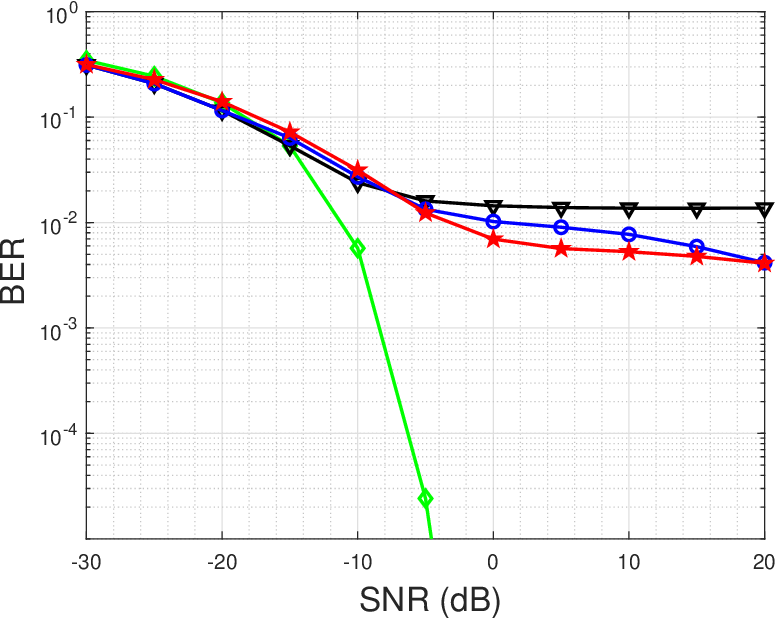}}\vspace{-0pt}
       \caption{\footnotesize $d = \frac{\lambda}{4}, |\theta|\leq 80^\circ, M = 5$}
\end{subfigure}\hfill
     \begin{subfigure}[t]{0.245\textwidth}
       \centerline{\includegraphics[width=\textwidth]{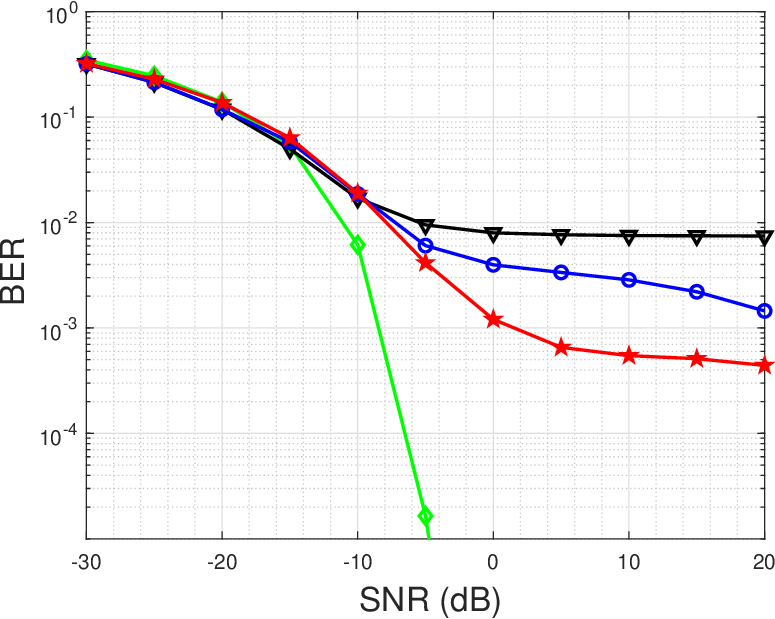}}\vspace{-0pt}
       \caption{\footnotesize $d = \frac{\lambda}{4}, |\theta|\leq 80^\circ, M = 6$}
\end{subfigure}\hfill
     \begin{subfigure}[t]{0.245\textwidth}
       \centerline{\includegraphics[width=\textwidth]{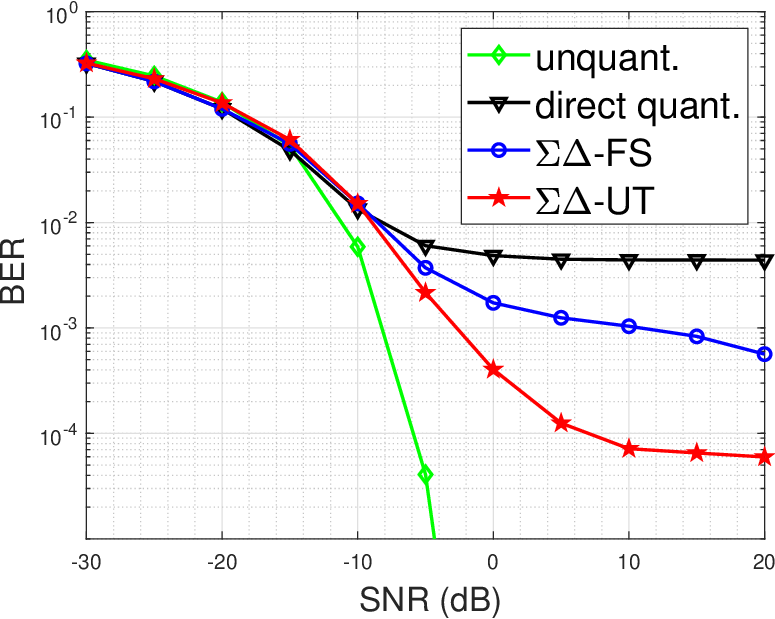}}\vspace{-0pt}
       \caption{\footnotesize $d = \frac{\lambda}{4}, |\theta|\leq 80^\circ, M = 7$}
\end{subfigure}\hfill
\caption{BER performance of the user-targeted $\Sigma\Delta$ modulation scheme for $K= 18$.
The settings are identical to those in Figure \ref{fig:UT-1024x9}.}
\label{fig:UT-1024x18}
\end{figure*}

Figure \ref{fig:UT-1024x18} shows another set of plots wherein we increase the user number to $K=18$.
Comparing this result with the previous result ($K=9$), we observe that (i) the performance behaviors of the current result appears to be consistent with those of the previous; (ii) the performance sees degradation as the user number increases. 
We argue that the second observation is an inevitable limitation, as alluded to by Proposition \ref{prop:zero_qe} which suggests that achieving zero quantization noise would require the quantization level number $M$ to increase exponentially with the user number $K$.

\subsection{2D Spatial $\Sigma\Delta$ Modulation for Uniform Planar Arrays}

\begin{figure*}[t]
    \begin{subfigure}[t]{0.245\textwidth}
       \centerline{\includegraphics[width=\textwidth]{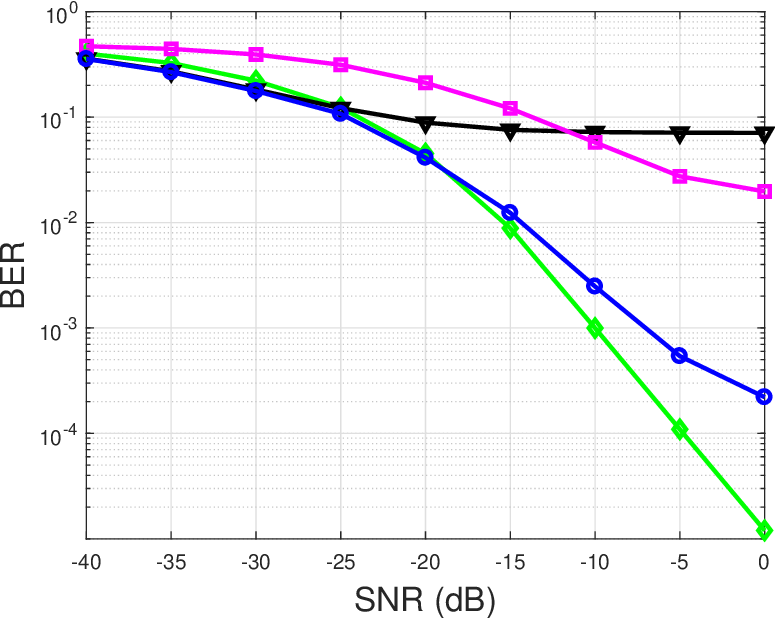}}\vspace{-0pt}
       \caption{\footnotesize $N_1 = N_2 = 40, M = 4$}
\end{subfigure}\hfill
     \begin{subfigure}[t]{0.245\textwidth}
       \centerline{\includegraphics[width=\textwidth]{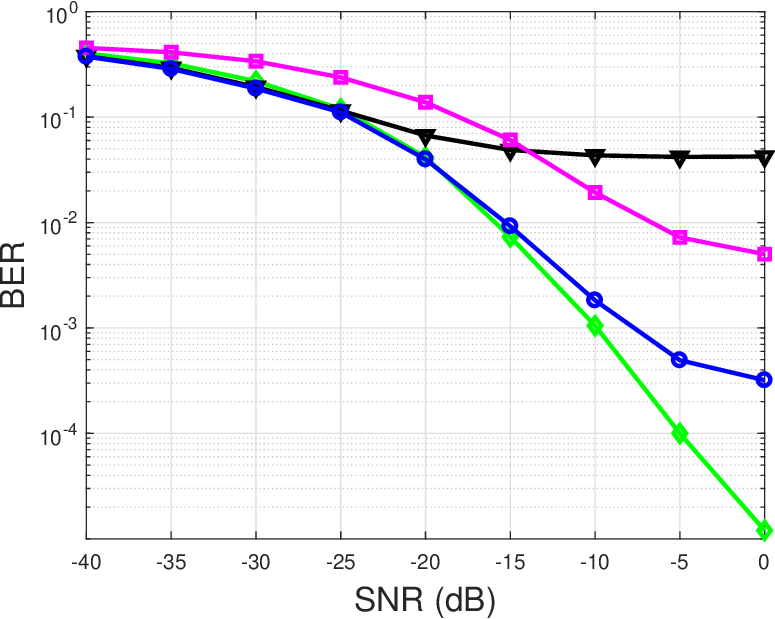}}\vspace{-0pt}
       \caption{\footnotesize $N_1 = N_2 = 40, M = 5$}
\end{subfigure}\hfill
     \begin{subfigure}[t]{0.245\textwidth}
       \centerline{\includegraphics[width=\textwidth]{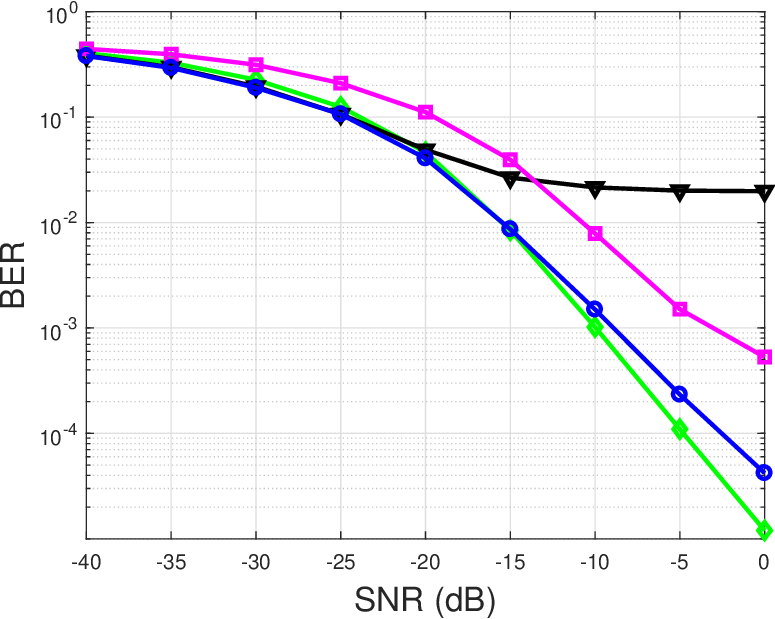}}\vspace{-0pt}
       \caption{\footnotesize $N_1 = N_2 = 40, M = 6$}
\end{subfigure}\hfill
     \begin{subfigure}[t]{0.245\textwidth}
       \centerline{\includegraphics[width=\textwidth]{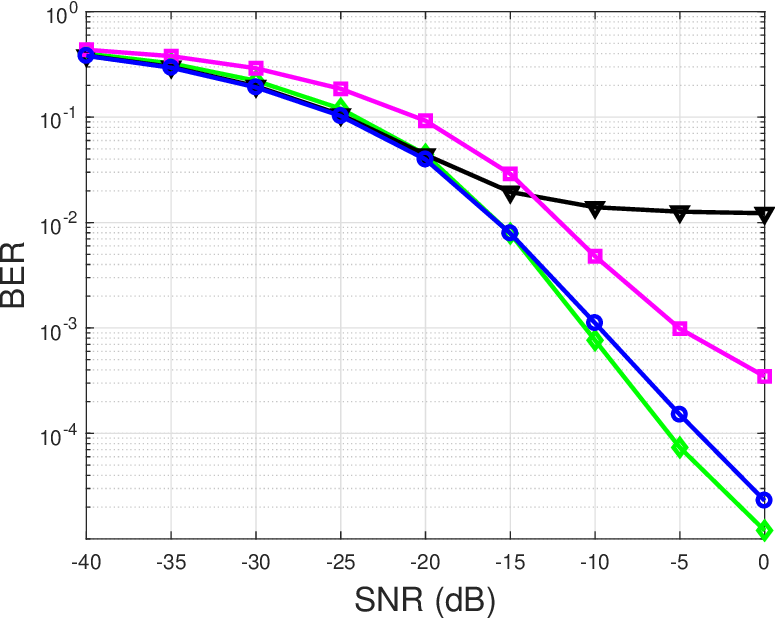}}\vspace{-0pt}
       \caption{\footnotesize $N_1 = N_2 = 40, M = 7$}
\end{subfigure}\vspace{10pt} \hfill
\\ 
    \begin{subfigure}[t]{0.245\textwidth}
       \centerline{\includegraphics[width=\textwidth]{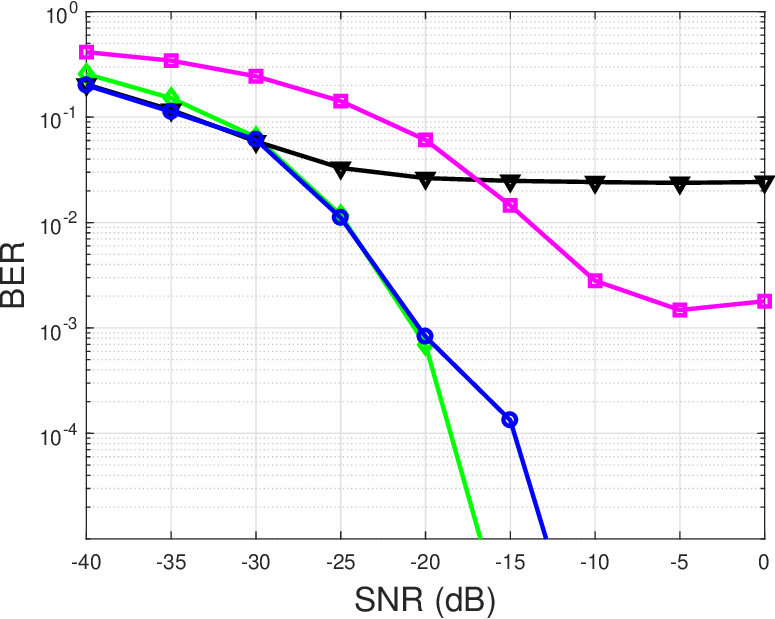}}\vspace{-0pt}
       \caption{\footnotesize $N_1 = N_2 = 60, M = 4$}
\end{subfigure}\hfill
     \begin{subfigure}[t]{0.245\textwidth}
       \centerline{\includegraphics[width=\textwidth]{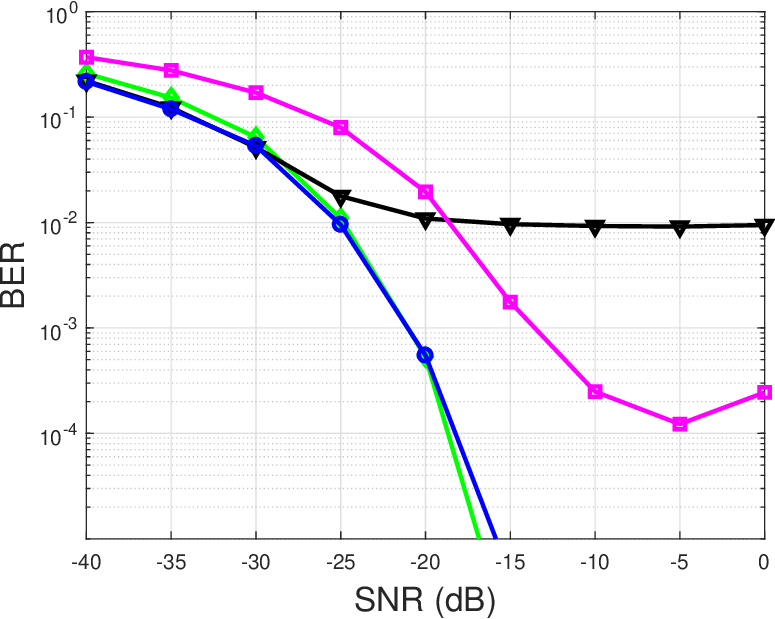}}\vspace{-0pt}
       \caption{\footnotesize $N_1 = N_2 = 60, M = 5$}
\end{subfigure}\hfill
     \begin{subfigure}[t]{0.245\textwidth}
       \centerline{\includegraphics[width=\textwidth]{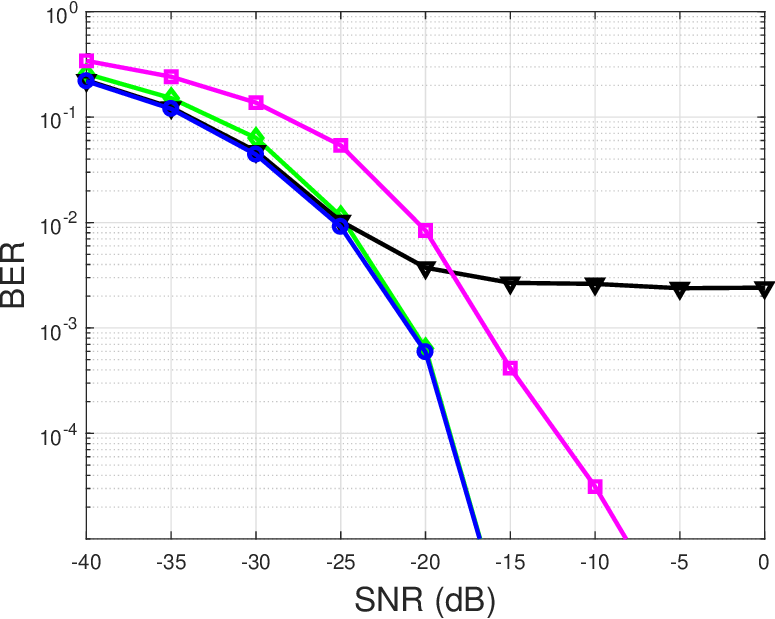}}\vspace{-0pt}
       \caption{\footnotesize $N_1 = N_2 = 60, M = 6$}
\end{subfigure}\hfill
     \begin{subfigure}[t]{0.245\textwidth}
       \centerline{\includegraphics[width=\textwidth]{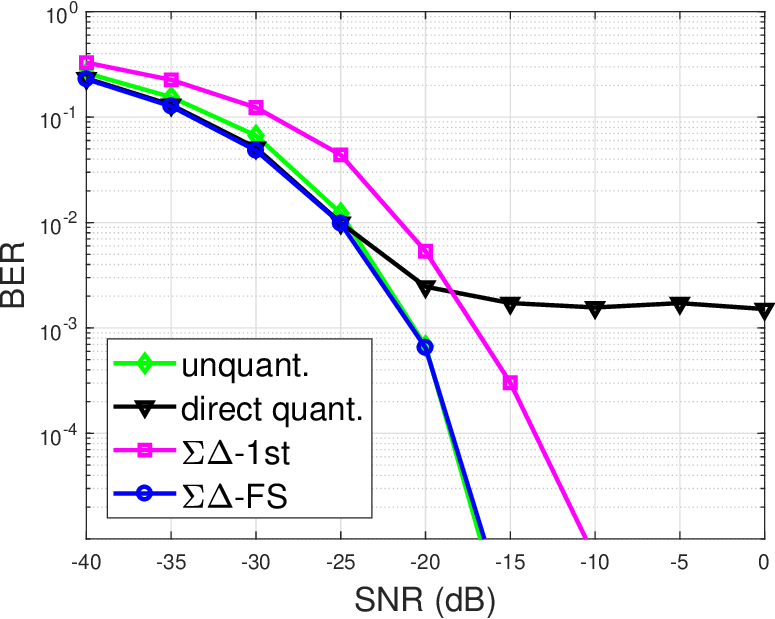}}\vspace{-0pt}
       \caption{\footnotesize $N_1 = N_2 = 60, M = 7$}
\end{subfigure}\hfill
\caption{BER performance of the 2D fixed-sector optimized $\Sigma\Delta$ modulation scheme.  
$d_1 = d_2 = \lambda/4$, $K=8$, $L_1 = L_2 = 5$, $[\theta_l, \theta_u] \times [\phi_l, \phi_u] = [-30^\circ, 30^\circ] \times [0, 20^\circ]$.
See the caption of Figure \ref{fig:sectorpass-baseband} for a description of the legend labels.}
\label{fig:2D-FS}
\end{figure*}

We consider the 2D spatial $\Sigma\Delta$ modulation schemes for uniform planar arrays, described in Section~\ref{sect:2D_sigdel}.
The simulation workflow is identical to the above.
The simulation settings are as follows:
the user number is $K=8$;
the inter-antenna spacings are $d_1 = d_2 = \lambda/4$;
the angle sector is $[\theta_l, \theta_u] \times [\phi_l, \phi_u] = [-30^\circ, 30^\circ]\times[0^\circ, 20^\circ]$;
the filter order of our optimized $\Sigma\Delta$ modulator is $(L_1,L_2)= (5,5)$.
We consider the fixed-sector design, and we use the 2D first-order $\Sigma\Delta$ modulator (cf. \eqref{eq:1st-2D-noise-shaping}) as our main benchmark.
Figure \ref{fig:2D-FS} shows the results for two different settings of the transmit antenna size $(N_1,N_2)$.
The results demonstrate that the 2D spatial $\Sigma\Delta$ modulation schemes are viable.
We should remark that, to the best of our knowledge, 2D spatial $\Sigma\Delta$ modulation for coarsely quantized MIMO precoding with uniform planar arrays was not attempted before;
even the 2D first-order $\Sigma\Delta$ modulation scheme is a new attempt.

\section{Conclusion}
\label{sect:con}

To summarize, we developed a spatial  $\Sigma\Delta$ modulator design framework for coarsely quantized massive MIMO downlink precoding.
Our framework is flexible. It can handle any $\Sigma\Delta$ filter order and any number of quantization levels. It can deal with various SQNR requirements, such as max-min-fair SQNR enhancement over a prescribed angle sector, or SQNR enhancement in accordance with the user angles in an instantaneous fashion. It can also be extended to 2D uniform planar arrays. Our design framework is based on convex optimization. Numerical results showed that $\Sigma\Delta$ modulators designed under our framework outperform the existing $\Sigma\Delta$ modulators,
and may lead to near-ideal (unquantized) performance under certain operating conditions.

%% file: main-arXiv.bbl
\begin{thebibliography}{10}
\providecommand{\url}[1]{#1}
\csname url@samestyle\endcsname
\providecommand{\newblock}{\relax}
\providecommand{\bibinfo}[2]{#2}
\providecommand{\BIBentrySTDinterwordspacing}{\spaceskip=0pt\relax}
\providecommand{\BIBentryALTinterwordstretchfactor}{4}
\providecommand{\BIBentryALTinterwordspacing}{\spaceskip=\fontdimen2\font plus
\BIBentryALTinterwordstretchfactor\fontdimen3\font minus \fontdimen4\font\relax}
\providecommand{\BIBforeignlanguage}[2]{{%
\expandafter\ifx\csname l@#1\endcsname\relax
\typeout{** WARNING: IEEEtran.bst: No hyphenation pattern has been}%
\typeout{** loaded for the language `#1'. Using the pattern for}%
\typeout{** the default language instead.}%
\else
\language=\csname l@#1\endcsname
\fi
#2}}
\providecommand{\BIBdecl}{\relax}
\BIBdecl

\bibitem{marzetta_larsson_yang_ngo_2016}
T.~L. Marzetta, E.~G. Larsson, H.~Yang, and H.~Q. Ngo, \emph{Fundamentals of Massive {MIMO}}.\hskip 1em plus 0.5em minus 0.4em\relax Cambridge University Press, 2016.

\bibitem{risi2014massive}
C.~Risi, D.~Persson, and E.~G. Larsson, ``{Massive MIMO with 1-bit ADC},'' \emph{arXiv preprint arXiv:1404.7736}, 2014.

\bibitem{choi-mo-heath-2016}
J.~Choi, J.~Mo, and R.~W. Heath, ``Near maximum-likelihood detector and channel estimator for uplink multiuser massive {MIMO} systems with one-bit {ADCs},'' \emph{IEEE Trans. Commun.}, vol.~64, no.~5, pp. 2005--2018, 2016.

\bibitem{MollenCLH17}
C.~Moll\'{e}n, J.~Choi, E.~G. Larsson, and R.~W. Heath, ``Uplink performance of wideband massive {MIMO} with one-bit {ADCs},'' \emph{IEEE Trans. Wireless Commun.}, vol.~16, no.~1, pp. 87--100, Jan 2017.

\bibitem{li-tao-gonzalo-2017}
Y.~Li, C.~Tao, G.~Seco-Granados, A.~Mezghani, A.~L. Swindlehurst, and L.~Liu, ``Channel estimation and performance analysis of one-bit massive {MIMO} systems,'' \emph{IEEE Trans. Signal Process.}, vol.~65, no.~15, pp. 4075--4089, 2017.

\bibitem{saxena-fijalkow-swindlehurst-2017}
A.~K. Saxena, I.~Fijalkow, and A.~L. Swindlehurst, ``Analysis of one-bit quantized precoding for the multiuser massive {MIMO} downlink,'' \emph{IEEE Trans. Signal Process.}, vol.~65, no.~17, pp. 4624--4634, 2017.

\bibitem{swindlehurst-saxena-mezghani-2017}
A.~Swindlehurst, A.~Saxena, A.~Mezghani, and I.~Fijalkow, ``Minimum probability-of-error perturbation precoding for the one-bit massive {MIMO} downlink,'' in \emph{Proc. IEEE Int. Conf. Acoust., Speech, Signal Process. (ICASSP)}, 2017, pp. 6483--6487.

\bibitem{jacobsson-durisi-coldrey-2017}
S.~Jacobsson, G.~Durisi, M.~Coldrey, T.~Goldstein, and C.~Studer, ``Quantized precoding for massive {MU-MIMO},'' \emph{IEEE Trans. Commun.}, vol.~65, no.~11, pp. 4670--4684, 2017.

\bibitem{studer2016quantized}
C.~Studer and G.~Durisi, ``Quantized massive {MU-MIMO-OFDM} uplink,'' \emph{IEEE Trans. Commun.}, vol.~64, no.~6, pp. 2387--2399, 2016.

\bibitem{shao2022accelerated}
M.~Shao, W.-K. Ma, J.~Liu, and Z.~Huang, ``Accelerated and deep expectation maximization for one-bit {MIMO-OFDM} detection,'' \emph{arXiv preprint arXiv:2210.03888}, 2022.

\bibitem{li-tao-swindlehurst-2017}
Y.~Li, C.~Tao, A.~Lee~Swindlehurst, A.~Mezghani, and L.~Liu, ``Downlink achievable rate analysis in massive {MIMO} systems with one-bit {DACs},'' \emph{IEEE Commun. Lett.}, vol.~21, no.~7, pp. 1669--1672, 2017.

\bibitem{mezghani-ghiat-nossek-2009}
A.~Mezghani, R.~Ghiat, and J.~A. Nossek, ``Transmit processing with low resolution {D/A}-converters,'' in \emph{Proc. 16th IEEE Int. Conf. Electron., Circuits, Syst.}, Dec 2009, pp. 683--686.

\bibitem{Sohrabi2018}
F.~Sohrabi, Y.-F. Liu, and W.~Yu, ``One-bit precoding and constellation range design for massive {MIMO} with {QAM} signaling,'' \emph{IEEE J. Sel. Topics Signal Process.}, vol.~12, no.~3, pp. 557--570, 2018.

\bibitem{shao2018framework}
M.~Shao, Q.~Li, W.-K. Ma, and A.~M.-C. So, ``{A framework for one-bit and constant-envelope precoding over multiuser massive {MISO} channels},'' \emph{IEEE Trans. Signal Process.}, vol.~67, no.~20, pp. 5309--5324, 2019.

\bibitem{castaneda-jacobsson-durisi-2017}
O.~Castañeda, S.~Jacobsson, G.~Durisi, M.~Coldrey, T.~Goldstein, and C.~Studer, ``1-bit massive {MU-MIMO} precoding in {VLSI},'' \emph{IEEE J. Emerging and Sel. Topics in Circuits and Syst.}, vol.~7, no.~4, pp. 508--522, 2017.

\bibitem{li-masouros-liu-2018}
A.~Li, C.~Masouros, F.~Liu, and A.~L. Swindlehurst, ``Massive {MIMO} 1-bit {DAC} transmission: A low-complexity symbol scaling approach,'' \emph{IEEE Trans. Wireless Commun.}, vol.~17, no.~11, pp. 7559--7575, 2018.

\bibitem{jedda-mezghani-swindlehurst-2018}
H.~Jedda, A.~Mezghani, A.~L. Swindlehurst, and J.~A. Nossek, ``Quantized constant envelope precoding with {PSK} and {QAM} signaling,'' \emph{IEEE Trans. Wireless Commun.}, vol.~17, no.~12, pp. 8022--8034, 2018.

\bibitem{kazemi-aghaeinia-2017}
M.~Kazemi, H.~Aghaeinia, and T.~M. Duman, ``Discrete-phase constant envelope precoding for massive {MIMO} systems,'' \emph{IEEE Trans. Commun.}, vol.~65, no.~5, pp. 2011--2021, 2017.

\bibitem{wu-liu-jiang-dai-2023}
Z.~Wu, Y.-F. Liu, B.~Jiang, and Y.-H. Dai, ``Efficient quantized constant envelope precoding for multiuser downlink massive {MIMO} systems,'' in \emph{Proc. IEEE Int. Conf. Acoust., Speech, Signal Process. (ICASSP)}, 2023, pp. 1--5.

\bibitem{shao2019one}
M.~Shao, W.-K. Ma, Q.~Li, and A.~L. Swindlehurst, ``One-bit {Sigma-Delta MIMO} precoding,'' \emph{IEEE J. Sel. Topics Signal Process.}, vol.~13, no.~5, pp. 1046--1061, 2019.

\bibitem{shao2020multiuser}
M.~Shao, W.-K. Ma, and L.~Swindlehurst, ``Multiuser massive {MIMO} downlink precoding using second-order spatial {Sigma-Delta} modulation,'' in \emph{Proc. IEEE Int. Conf. Acoust., Speech, Signal Process. (ICASSP)}, 2020, pp. 8966--8970.

\bibitem{schreier2005understanding}
R.~Schreier and G.~C. Temes, \emph{Understanding {Delta-Sigma} Data Converters}.\hskip 1em plus 0.5em minus 0.4em\relax IEEE Press, Piscataway, NJ, 2005, vol.~74.

\bibitem{baracspatial}
D.~Barac and E.~Lindqvist, ``Spatial {Sigma-Delta} modulation in a massive {MIMO} cellular system,'' Master's thesis, Department of Computer Science and Engineering, Chalmers University of Technology, 2016.

\bibitem{Corey_Sig}
R.~M. Corey and A.~C. Singer, ``Spatial sigma-delta signal acquisition for wideband beamforming arrays,'' in \emph{Proc. Int. ITG Workshop Smart Antennas (WSA)}, March 2016.

\bibitem{pirzadeh2020spectral}
H.~Pirzadeh, G.~Seco-Granados, S.~Rao, and A.~L. Swindlehurst, ``Spectral efficiency of one-bit sigma-delta massive {MIMO},'' \emph{IEEE J. Sel. Areas Commun.}, vol.~38, no.~9, pp. 2215--2226, 2020.

\bibitem{rao-seco-pirzadeh-2021}
S.~Rao, G.~Seco-Granados, H.~Pirzadeh, J.~A. Nossek, and A.~L. Swindlehurst, ``Massive {MIMO} channel estimation with low-resolution spatial {Sigma-Delta ADCs},'' \emph{IEEE Access}, vol.~9, pp. 109\,320--109\,334, 2021.

\bibitem{sankar-chepuri-2022}
R.~P. Sankar and S.~P. Chepuri, ``Channel estimation in {MIMO} systems with one-bit spatial {Sigma-Delta ADCs},'' \emph{IEEE Trans. Signal Process.}, vol.~70, pp. 4681--4696, 2022.

\bibitem{nagahara2012frequency}
M.~Nagahara and Y.~Yamamoto, ``Frequency domain min-max optimization of noise-shaping {Delta-Sigma} modulators,'' \emph{IEEE Trans. Signal Process.}, vol.~60, no.~6, pp. 2828--2839, 2012.

\bibitem{gray1990quantization}
R.~M. Gray, ``Quantization noise spectra,'' \emph{IEEE Trans. Inf. Theory}, vol.~36, no.~6, pp. 1220--1244, 1990.

\bibitem{daubechies2003approximating}
I.~Daubechies and R.~DeVore, ``Approximating a bandlimited function using very coarsely quantized data: A family of stable sigma-delta modulators of arbitrary order,'' \emph{Ann. Math.}, vol. 158, no.~2, pp. 679--710, 2003.

\bibitem{gunturk2003one}
C.~S. G{\"u}nt{\"u}rk, ``One-bit {Sigma-Delta} quantization with exponential accuracy,'' \emph{Commun. Pure Appl. Math.}, vol.~56, no.~11, pp. 1608--1630, 2003.

\bibitem{tse2005fundamentals}
D.~Tse and P.~Viswanath, \emph{Fundamentals of Wireless Communication}.\hskip 1em plus 0.5em minus 0.4em\relax Cambridge University Press, 2005.

\bibitem{schreier1991stability}
R.~Schreier and M.~Snelgrove, ``Stability in a general {Sigma Delta} modulator,'' in \emph{Proc. IEEE Int. Conf. Acoust., Speech, Signal Process. (ICASSP)}, 1991, pp. 1769--1772.

\bibitem{charnes1962programming}
A.~Charnes and W.~W. Cooper, ``Programming with linear fractional functionals,'' \emph{Naval Res. Logist. Quarter.,}, vol.~9, no. 3-4, pp. 181--186, 1962.

\bibitem{cvx}
M.~Grant and S.~Boyd, ``{CVX}: Matlab software for disciplined convex programming, version 2.1,'' \url{http://cvxr.com/cvx}, Mar. 2014.

\bibitem{kite1997digital}
T.~D. Kite, B.~L. Evans, A.~C. Bovik, and T.~L. Sculley, ``Digital halftoning as {2-D} {Delta-Sigma} modulation,'' in \emph{Proc. IEEE Int. Conf. Image Process. (ICIP)}, vol.~1, 1997, pp. 799--802.

\bibitem{balanis2015antenna}
C.~A. Balanis, \emph{Antenna Theory: Analysis and Design}.\hskip 1em plus 0.5em minus 0.4em\relax John Wiley \& Sons, 2015.

\end{thebibliography}
